\documentclass{aa}

\usepackage{txfonts}
\usepackage{graphicx}
\begin{document}

\title{MHD wave propagation in the neighbourhood \\ of a two-dimensional null point }
\author{J.~A. McLaughlin \and A.~W. Hood}
\institute{School of Mathematics and Statistics, University of St
Andrews,
KY16 9SS, UK}
\offprints{J.~A. McLaughlin,
\email{\emph{james@mcs.st-and.ac.uk}}}
\date{Received 19 December 2003 / Accepted 12 March 2004 }

\titlerunning{MHD Waves in the Neighbourhood of a Null}
\authorrunning{J. McLaughlin \& A. Hood}

% ----------------------------------------------------------------
\abstract{The nature of fast magnetoacoustic and Alfv\'en waves is investigated in a zero $\beta$ plasma. This gives an indication of wave propagation in the low $\beta$ solar corona. It is found
that for a two-dimensional null point, the fast wave is attracted to
that point and the front of the wave slows down as it approaches the null point, causing the current density to accumulate there and rise rapidly.
Ohmic dissipation will extract the energy in the wave at this
point. This illustrates that null points play an important
role in the rapid dissipation of fast magnetoacoustic waves and suggests the location where wave heating will occur in the corona. The
Alfv\'en wave behaves in a different manner in
that the wave energy is dissipated along the separatrices. For Alfv\'en waves that are decoupled from fast waves, the value of the plasma $\beta$ is unimportant. However, the phenomenon of dissipating the majority of the wave energy at a specific place is a feature of both wave types.
}

\maketitle

% ----------------------------------------------------------------

\section{Introduction}

%Null points in corona. Mention potential field models imply existence of coronal null points.
The coronal heating problem remains a key unsolved problem in solar physics. While
the coronal magnetic field is ultimately involved, there are many
rival theories ranging from reconnection models involving
\emph{nanoflares} and wave heating models involving
\emph{phase mixing} and \emph{resonant absorption}. The
reconnection models either require the formation of many current
sheets, due to random photospheric boundary motions that braid the
magnetic field, or the collapse of null points. The wave heating
models rely on the generation of small length scale wave motions
in the corona. There is clear evidence from SOHO and TRACE
observations of slow MHD waves (Berghmans \& Clette, 1999; De Moortel \emph{et al.}, 2000), fast MHD waves
(Nakariakov \emph{et al.}, 1999) and non-thermal line broadening due to Alfv\'en waves
(Harrison \emph{et al.}, 2002). While there may be insufficient energy in these waves to heat the whole corona, their dissipation will contribute to the overall energy budget. This paper is concerned with the propagation of MHD waves in the neighbourhood of null points in a zero $\beta$ plasma, giving an indication of how MHD waves behave in the low $\beta$ plasma of the solar corona.

The existence of null points is predicted on theoretical grounds
and their importance lies in the fact that the Alfv\'en speed is
actually zero at that point. This important consequence will be
utilised later. Potential field extrapolations, using photospheric
magnetograms to provide the field distribution on the lower
boundary, suggest that there are always likely to be null points
in the corona. The number of such points does depend on the
magnetic complexity of the photospheric flux distribution.
Detailed investigations of the coronal magnetic field, using such
potential field calculations, can be found in \cite{Beveridge2002} and \cite{Brown2001}.

%Null point collapse (Craig plus others) Mention induce collapse through moving and holding boundaries.
The fact that null points are a weakness in the magnetic field has
been used to investigate how they collapse in response to boundary
motions. This has been investigated by Craig and co-workers and other authors using analytical and numerical approaches (Craig \& McClymont, 1991; Hassam, 1992). The basic aim is to
move the field lines passing through the boundary in a particular
manner in order to perturb the field. The resulting field
perturbations cause the null point to collapse to form a current
sheet in which reconnection can release magnetic energy. In these
models the boundary motions move the field lines but do not return
them to their original positions. Thus, the Poynting flux induced
by the imposed motion (and then fixing the field after the motion
is complete) accumulates at the resulting current sheet and
provides the energy released in the reconnection. However, if the
boundary motions are simply due to the passing of incoming waves
through the boundary, then it is not clear that the null point
need collapse and form a current sheet. If this is the case, then
it is not clear if the energy in the wave, again due to the
Poynting flux through the boundary, will dissipate or simply pass
through one of the other boundaries.

%Null point waves (Craig? radial waves)

Waves in the neighbourhood of 2D null points have been investigated
by various authors. Bulanov and Syrovatskii (1980) provided a
detailed discussion of the propagation of fast and Alfv\'en waves
using cylindrical symmetry. In their paper, harmonic fast waves are generated and
these propagate towards the null point. However, the assumed
cylindrical symmetry means that the disturbances can only propagate
either towards or away from the null point. Craig and Watson (1992)
mainly consider the radial propagation of the $m=0$ mode (where $m$ is the
azimuthal wavenumber) using a mixture of analytical and numerical
solutions. In their investigation, the outer radial boundary is held fixed so that any
outgoing waves will be reflected back towards the null point. This
means that all the energy in the wave motions is contained within a
fixed region. They show that the propagation of the $m=0$ wave
towards the null point generates an exponentially large increase in
the current density and that magnetic resistivity dissipates this current in
a time related to $\log { \eta }$. Their initial disturbance is given as a
function of radius. In this paper, we are interested in generating
the disturbances at the boundary rather than internally. Craig and
McClymont (1991, 1993) investigate the normal mode solutions for both
$m=0$ and $m\ne 0$ modes with resistivity included. Again they
emphasise that the current builds up as the inverse square of the
radial distance from the null point. However, attention has been
restricted to a circular reflecting boundary. This paper will
investigate wave propagation in a zero $\beta$ plasma in the
neighbourhood of a simple 2D null point but for more general
disturbances, more general boundary conditions and single wave
pulses. This allows us to concentrate on the transient features that
are not always apparent on using normal mode analysis. Galsgaard \emph{et al.} (2003) looked at a particular type of wave disturbance for a symmetrical 3D null. They investigated the effect of rotating the field lines around the spine and found that a twist wave (essentially an Alfv\'en wave) propagates in towards the null. They found that while the helical Alfv\'en wave spreads out, coupling due to the field geometry generates a  fast wave that focuses on the null and wraps around it. This wrapping effect is analysed in more detail here for the simpler 2D null.

%2D analysis in this paper. Investigate basic process. Investigate fast and Alfv\'en uncoupled. Same behaviour in 3D but waves coupled together.
The propagation of fast magnetoacoustic waves in an inhomogeneous
coronal plasma has been investigated by \cite{Nakariakov1995},
who showed how the waves are refracted into regions of low Alfv\'en
speed. In the case of null points, it is the aim of this paper to
see how this refraction proceeds when the Alfv\'en speed actually
drops to zero.

The paper has the following outline. In Section \ref{sec:2} the
basic equations are described. The results for an uncoupled fast
magnetoacoustic wave are presented in Section \ref{sec:3}. Some
simple analytical results are also discussed as a verification to and interpretation of the
numerical simulation. Section \ref{sec:4} discusses the
propagation of Alfv\'en waves and the conclusions are given in Section \ref{sec:5}.

\section{Basic Equations and Numerical Method}\label{sec:2}

The usual MHD equations for a low $\beta$ plasma appropriate to
the solar corona are used. Hence,
\begin{eqnarray}
\qquad  \rho \left( {\partial {\bf{v}}\over \partial t} + \left( {\bf{v}}\cdot\nabla \right) {\bf{v}} \right) &=& {1\over \mu}\left(\nabla \times {\bf{B}}\right)\times {\bf{B}},\label{eq:2.1a} \\
  {\partial {\bf{B}}\over \partial t} &=& \nabla \times \left
  ({\bf{v}}\times {\bf{B}}\right ) + \eta \nabla^2
  {\bf{B}},\label{eq:2.1b} \\
  {\partial \rho\over \partial t} + \nabla \cdot \left (\rho {\bf{v}}\right )
  &=& 0, \label{eq:2.1c}
\end{eqnarray}
where $\rho$ is the mass density, ${\bf{v}}$ is the plasma
velocity, ${\bf{B}}$ the magnetic induction (usually called the
magnetic field), $ \mu = 4 \pi \times 10^{-7} \/\mathrm{Hm^{-1}}$  the magnetic
permeability, $\eta = 1/\mu\sigma$ is the magnetic diffusivity $\left( \mathrm{m}^2\mathrm{s}^{-1}\right)$,
and $\sigma$ the electrical conductivity. The gas pressure and the
adiabatic energy equation are neglected in the low $\beta$ approximation.

\subsection{Basic equilibrium}\label{sec:2.1}

\begin{figure}[hb]
\begin{center}
\includegraphics[width=2.8in]{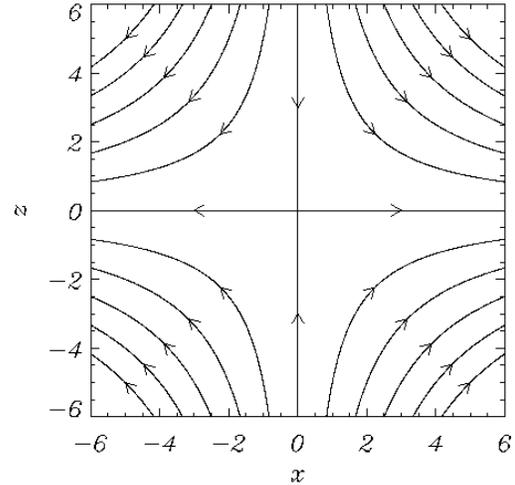}
\caption{Our choice of equilibrium magnetic field.}
\label{figureone}
\end{center}
\end{figure}

The basic magnetic field structure is taken as a simple 2D
X-type neutral point. The aim of studying waves in a 2D
configuration is one of simplicity. The individual effects are
much easier to identify when there is no coupling between the fast
and Alfv\'en modes. However, the extension to 3D is relatively
straightforward. The modes will become coupled but their evolution
is predictable from the 2D case. Therefore, the magnetic field is taken as
\begin{equation}\label{eq:2.2}
\qquad {\bf{B}}_0 = B_0 \left({x\over a}, 0, -{z\over a}\right),
\end{equation}
where $B_0$ is a characteristic field strength and $a$ is the
length scale for magnetic field variations. This magnetic field can be seen in Figure \ref{figureone}. Obviously this
particular choice of magnetic field is only valid in the
neighbourhood of the null point located at $x=0, z=0$.

\subsection{Linearised equations}\label{sec:2.2}

To study the nature of wave propagation near null points, the
linearised MHD equations are used. Using subscripts of $0$ for equilibrium quantities and $1$ for
perturbed quantities, the linearised equation of motion becomes
\begin{equation}\label{eq:2.3}
\qquad  \rho_0 \frac{\partial \mathbf{v}_1}{\partial t} = \left(\frac{ \nabla \times \mathbf{B}_1}{\mu} \right) \times \mathbf{B}_0 \; ,
\end{equation}
the linearised induction equation
\begin{equation}\label{eq:2.5}
\qquad    {\partial {\bf{B}}_1\over \partial t} = \nabla \times
    ({\bf{v}}_1 \times {\bf{B}}_0) + \eta \nabla^2 {\bf{B}}_1 \; ,
\end{equation}
and the linearised equation of mass continuity
\begin{equation}\label{eq:2.4}
\qquad \frac{\partial \rho_1} {\partial t} + \nabla\cdot\left( \rho_0 \mathbf{v} _1 \right) =0 \; .
\end{equation}
We will not discuss equation (\ref{eq:2.4}) further as it can be solved once we know $\mathbf{v} _1$. In fact, it has no influence on the momentum equation (in the low $\beta$ approximation) and so in effect the plasma is arbitrarily compressible (Craig \& Watson, 1992). We assume the background gas density is uniform and label it as $\rho_0$. A spatial variation in $\rho _0$ can cause phase mixing (Heyvaerts \& Priest, 1983).

We now consider a change of scale to non-dimensionalise; let ${\mathbf{\mathrm{v}}}_1 = \bar{\rm{v}} {\mathbf{v}}_1^*$, ${\mathbf{B}}_0 = B_0 {\mathbf{B}}_0^*$, ${\mathbf{B}}_1 = B_0 {\mathbf{B}}_1^*$, $x = a x^*$, $z=az^*$, $\nabla = \frac{1}{a}\nabla^*$ and $t=\bar{t}t^*$, where we let * denote a dimensionless quantity and $\bar{\rm{v}}$, $B_0$, $a$ and $\bar{t}$ are constants with the dimensions of the variable they are scaling. We then set $\frac {B_0}{\sqrt{\mu \rho _0 } } =\bar{\rm{v}}$ and $\bar{\rm{v}} =  {a} / {\bar{t}}$ (this sets $\bar{\rm{v}}$ as a sort of constant background Alfv\'{e}n speed). This process non-dimensionalises equations (\ref{eq:2.3}) and (\ref{eq:2.5}), and under these scalings, $t^*=1$ (for example) refers to $t=\bar{t}=  {a} / {\bar{\rm{v}}}$; i.e. the (background) Alfv\'en time taken to travel a distance $a$. For the rest of this paper, we drop the star indices; the fact that they are now non-dimensionalised is understood.

The ideal linearised MHD equations naturally decouple into two equations for the fast
MHD wave and the Alfv\'en wave. The slow MHD wave is absent in
this limit and there is no velocity component along the background
magnetic field (as can be seen by taking the scalar product of equation (\ref{eq:2.3}) with ${\mathbf{B}} _0$. The magnetic resistivity, $\eta$, in equation (\ref{eq:2.5}) will be neglected in the numerical simulations but is included for discussion in the conclusions.

The linearised equations for the fast magnetoacoustic wave are:
\begin{eqnarray}
\qquad \frac{\partial V}{\partial t} &=& v_A^2 \left( x,z \right) \left( \frac{\partial b_z}{\partial x} - \frac{\partial b_x}{\partial z}  \right) \nonumber \\
\frac{\partial b_x}{\partial t} &=& -\frac{\partial V}{\partial z} \; , \; \frac{\partial b_z}{\partial t} =   \frac{\partial V}{\partial x} \; \label{fastalpha},
\end{eqnarray}
where the Alfv\'{e}n speed, $v_A \left( x,z \right)$, is equal to $ \sqrt{x^2+z^2}$, $ { \mathbf{B} } _1 = \left( b_x,0,b_z \right) $ and the variable $ V $ is related to the perpendicular velocity; $ V = \left[ \left( \mathbf{v} _1 \times { \mathbf{B} } _0 \right) \cdot {\hat{\mathbf{e}} }_y \right] $.
These equations can be combined to form a single wave equation: 
\begin{eqnarray}
\qquad \frac{\partial ^2 V}{\partial t^2} = v_A^2 \left( x,z \right) \left( \frac{\partial^2 V}{\partial x^2} + \frac{\partial ^2 V}{\partial z^2}  \right) \; \label{fastbeta},  
\end{eqnarray}

The linearised equations for the Alfv\'en wave, with ${\mathbf{v}} _1 = \left( 0,v_y,0 \right)$ and ${\mathbf{B}} _1 = \left( 0,b_y,0 \right)$ are:
\begin{eqnarray}
\qquad \frac {\partial v_y }{\partial t} = x \frac {\partial b_y }{\partial x} - z\frac {\partial b_y }{\partial z} \; , \quad \frac {\partial b_y }{\partial t} = x \frac {\partial v_y }{\partial x} - z\frac {\partial v_y }{\partial z} \; \label{alfvenalpha} , 
\end{eqnarray}
which can be combined to form a single wave equation:
\begin{eqnarray}
\qquad \frac {\partial^2 v_y }{\partial t^2} = \left(x \frac {\partial }{\partial x} - z\frac {\partial }{\partial z} \right) ^2 v_y \; .
\end{eqnarray}

\section{Fast waves}\label{sec:3}

\begin{figure*}[ht]
\begin{center}
\includegraphics[width=2.0in]{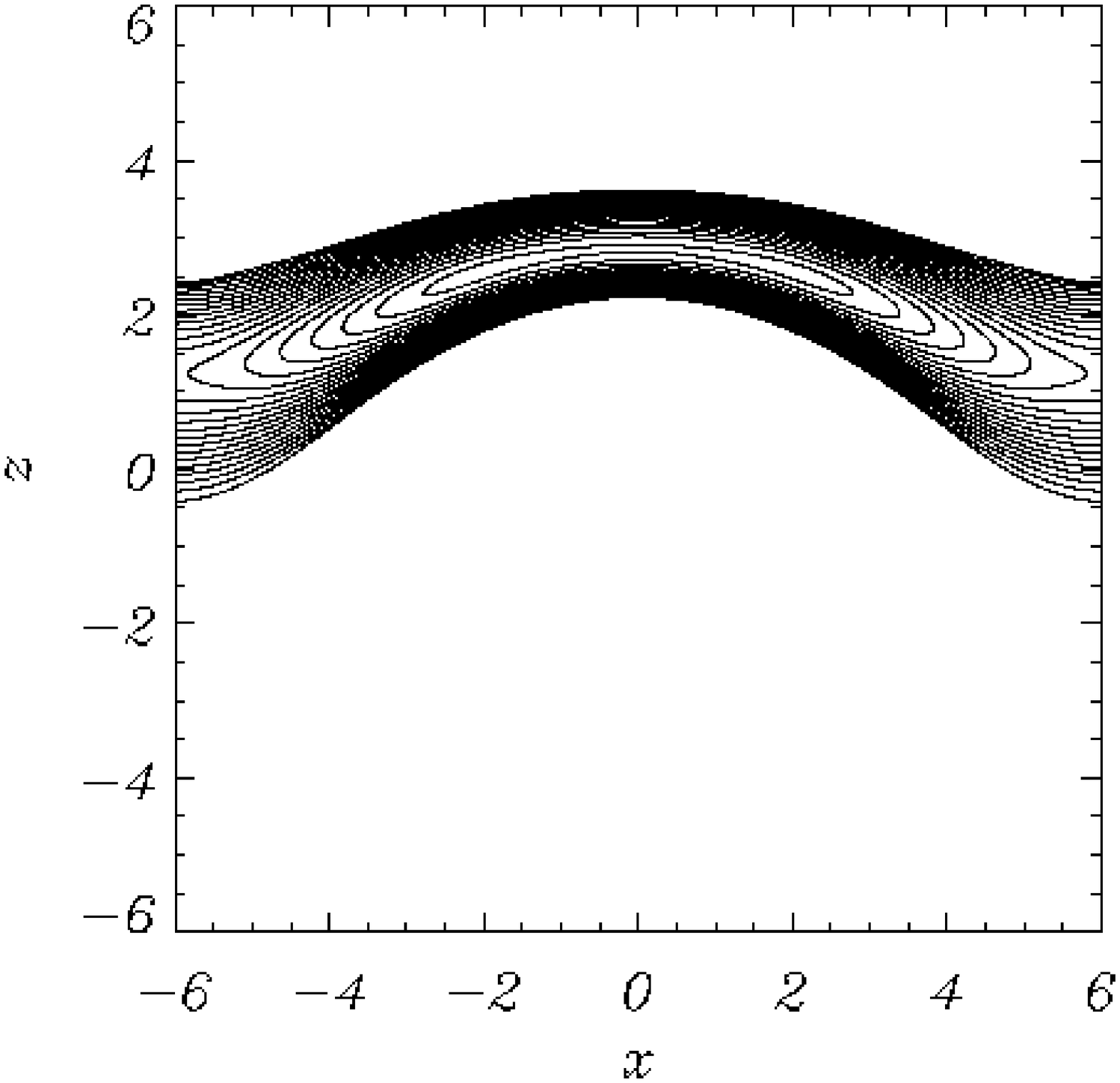}
\hspace{0.15in}
\includegraphics[width=2.0in]{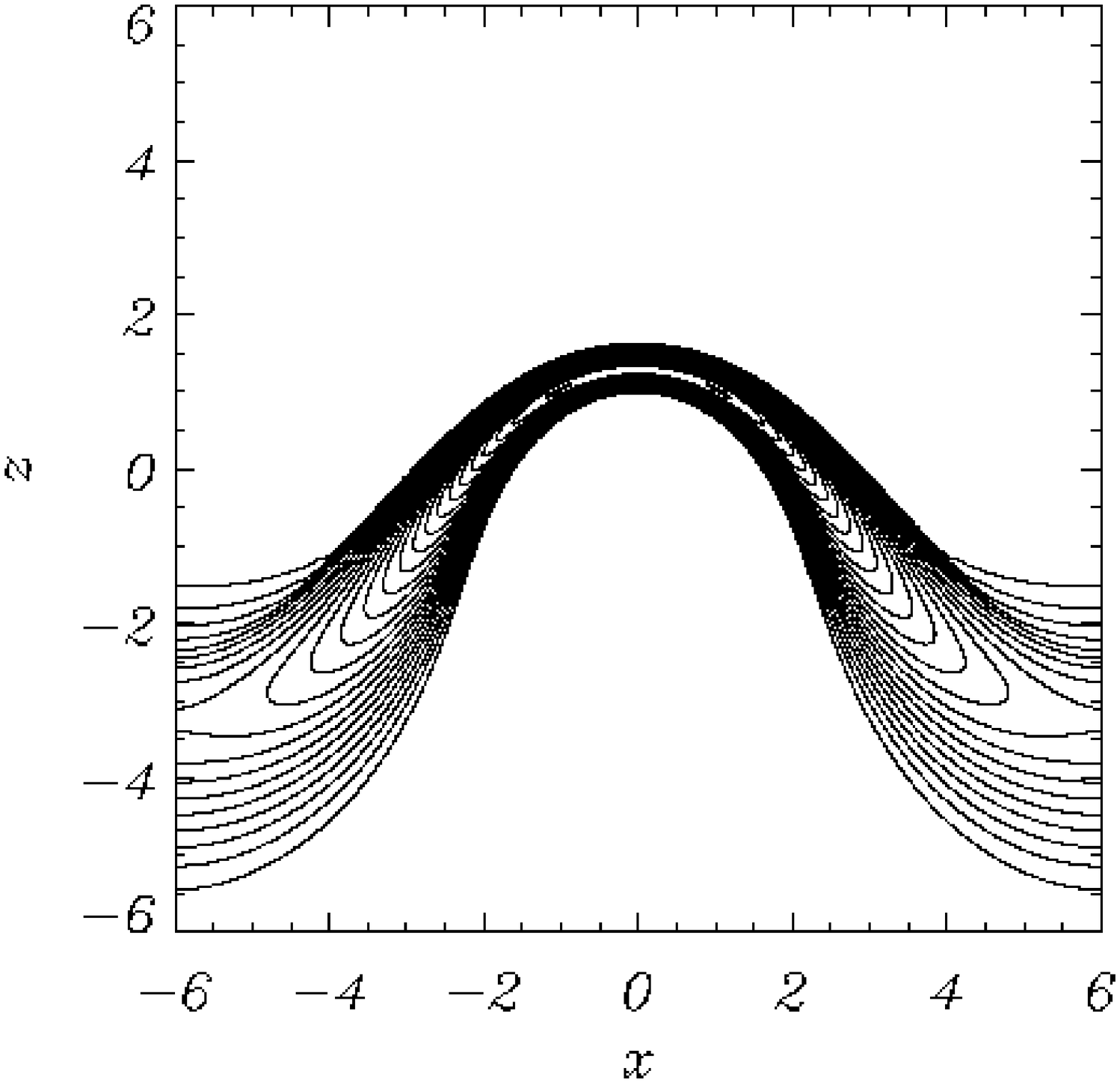}\\
\includegraphics[width=2.0in]{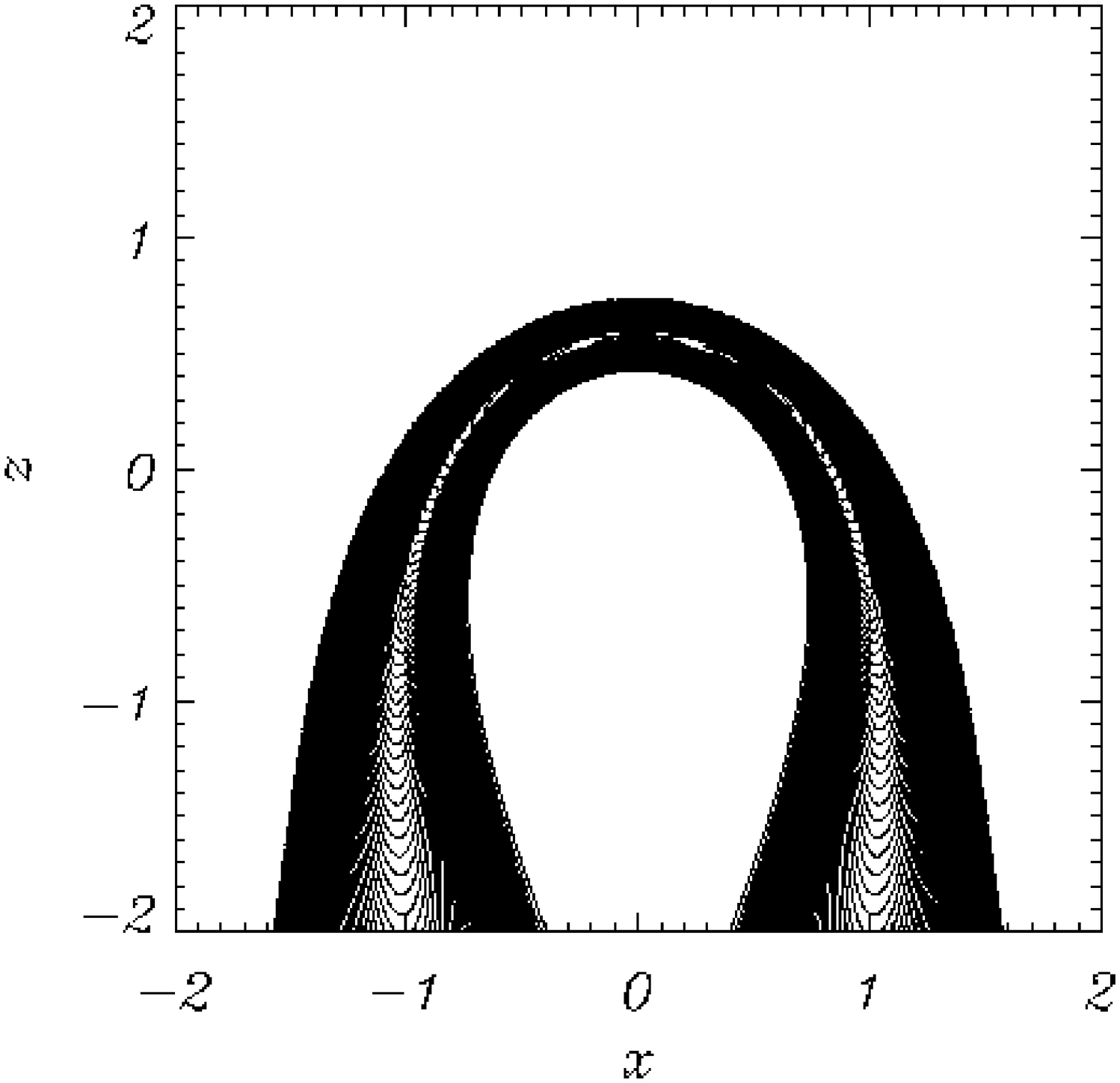}
\hspace{0.15in}
\includegraphics[width=2.0in]{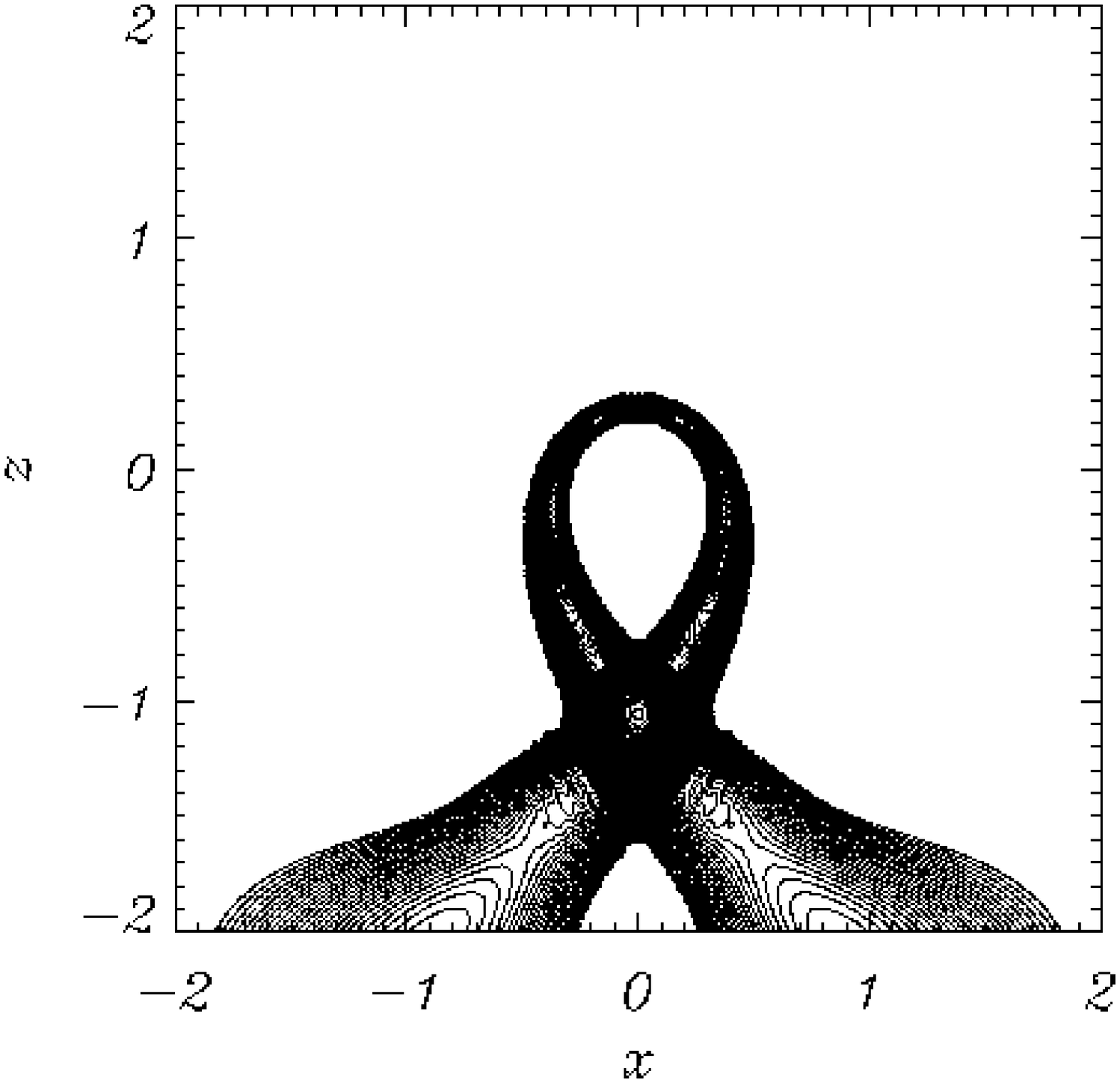}
\caption{Contours of $V$ for a fast wave sent in from the upper boundary and its resultant propagation at times $(a)$ $t$=1.0, $(b)$ $t$=1.8, $(c)$ $t=$2.6  and $(d)$ $t$=3.4, labelling from top left to bottom right. Here, $\omega = 2 \pi$ and time is measured in units of $ \left( \mu \rho _0 \right) ^ {1/2} / B _0 $. The axes have been rescaled in these lower two subfigures to draw attention to the central behaviour. The number of points in each direction is $1600 \times 1600$.}
\label{figuretwo}
\end{center}
\end{figure*}

We solve the linearised MHD equations for the fast wave, namely equations (\ref{fastalpha}), numerically using a two-step Lax-Wendroff scheme. The numerical scheme is run in a box with $-6 \leq x \leq 6$ and $-6 \leq z \leq 6$ and we initially consider a single wave pulse coming in from the top boundary.  For the single wave pulse, the boundary conditions were set such that:
\begin{eqnarray*}
\qquad V(x, 6) &=& \left\{\begin{array}{cl}
{\sin { \omega t } } & {\mathrm{for} \; \;0 \leq t \leq \frac {\pi}{\omega} } \\
{0} & { \mathrm{otherwise} }
\end{array} \right. \; , \\
\; \frac {\partial V } {\partial x } | _{x=-6} &=& 0 \; , \quad \frac {\partial V } {\partial x }  | _{x=6} = 0 \; , \quad \frac {\partial V } {\partial z }  | _{z=-6}  = 0 \; .
\end{eqnarray*}

Tests show that the central behaviour is largely unaffected by these choices of side and bottom boundary conditions. The other boundary conditions on the perturbed magnetic field follow from the remaining equations and the solenodial condition, $\nabla \cdot {\mathbf{B} _1} =0 $.

We find that the linear, fast magnetoacoustic wave travels towards the neighbourhood of the X-point and bends around it. Since the Alfv\'{e}n speed, $v_A \left( x,z \right)$, is spatially varying, different parts of the wave travel at different speeds, and it travels faster the further it is away from the origin (i.e. the further away a point is from the origin, the greater in magnitude $v_A \left( x,z \right)$ is). So the wave demonstrates \emph{refraction} and this can be seen in Figure \ref{figuretwo}. A similar refraction phenomenon was found by \cite{Nakariakov1995}. It is this refraction effect that wraps the wave around the null point and it is this that is the key feature of fast wave propagation.

Since the Alfv\'{e}n speed drops to zero at the null point, the wave never reaches there, but the length scales (this can be thought of as the distance between the leading and trailing edges of the wave pulse) rapidly decrease, indicating that the current (and all other gradients) will increase. As a simple illustration, consider the wavefront as it propagates down the $z$ axis at $x=0$. Here the vertical velocity is $v_z = \frac {d z} {d t} = -z $. Thus, the start of the wave is located at a position $z_s=6 e^{-t}$, when the wave is initally at $z=6$. If the end of the wave leaves $z=6$ at $t=t_0$ then the location of the end of the wave is $z_e=6 e^{t_0 - t}$. Thus, the distance between the leading and trailing edge of the wave is $\delta z = 6 \left( e^{t_0} -1 \right) e^{-t} $ and this decreases with time, suggesting that all gradients will increase exponentially.

\begin{figure*}[htb]
\begin{center}
\includegraphics[width=2.4in]{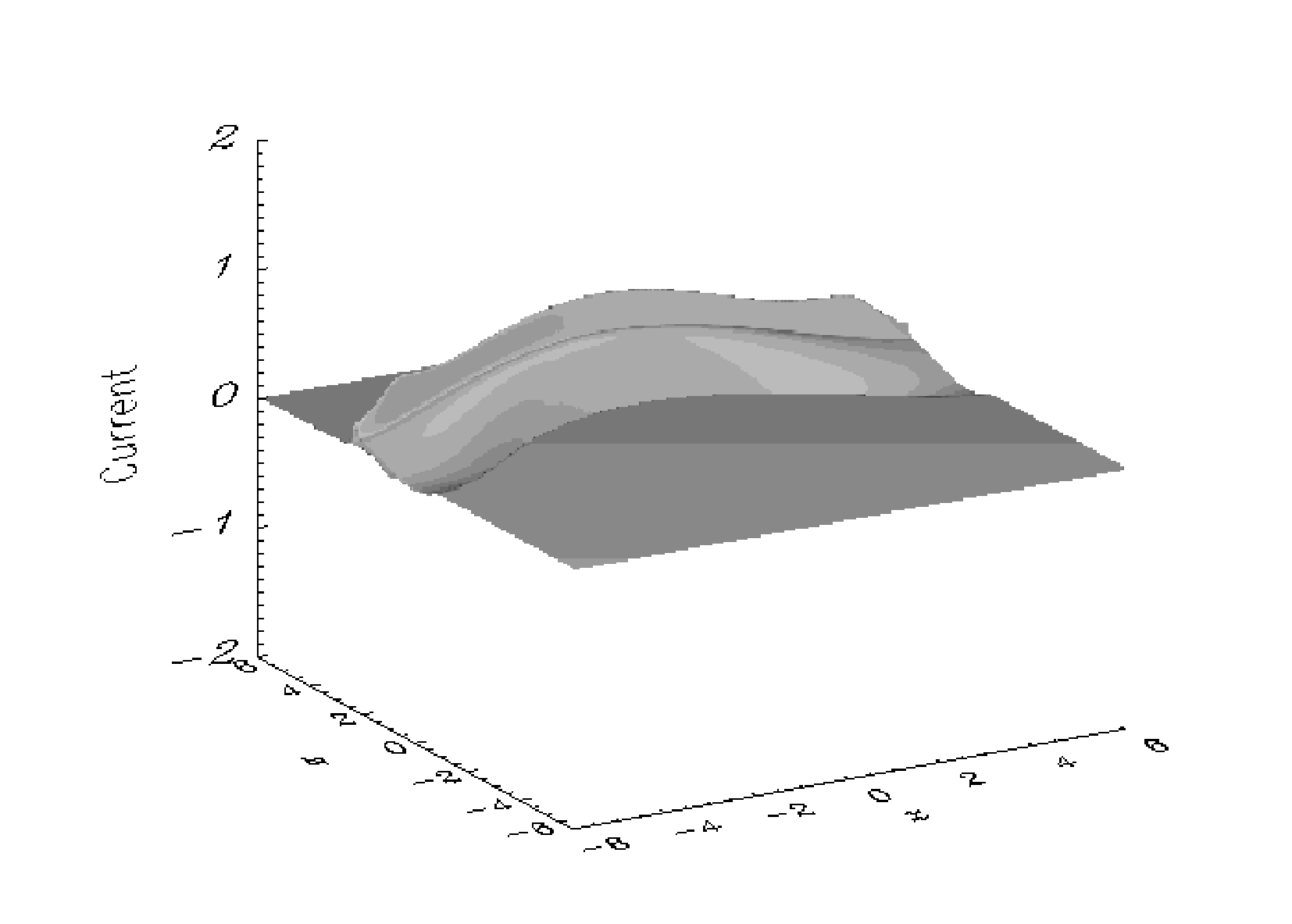}
\hspace{0.4in}
\includegraphics[width=2.4in]{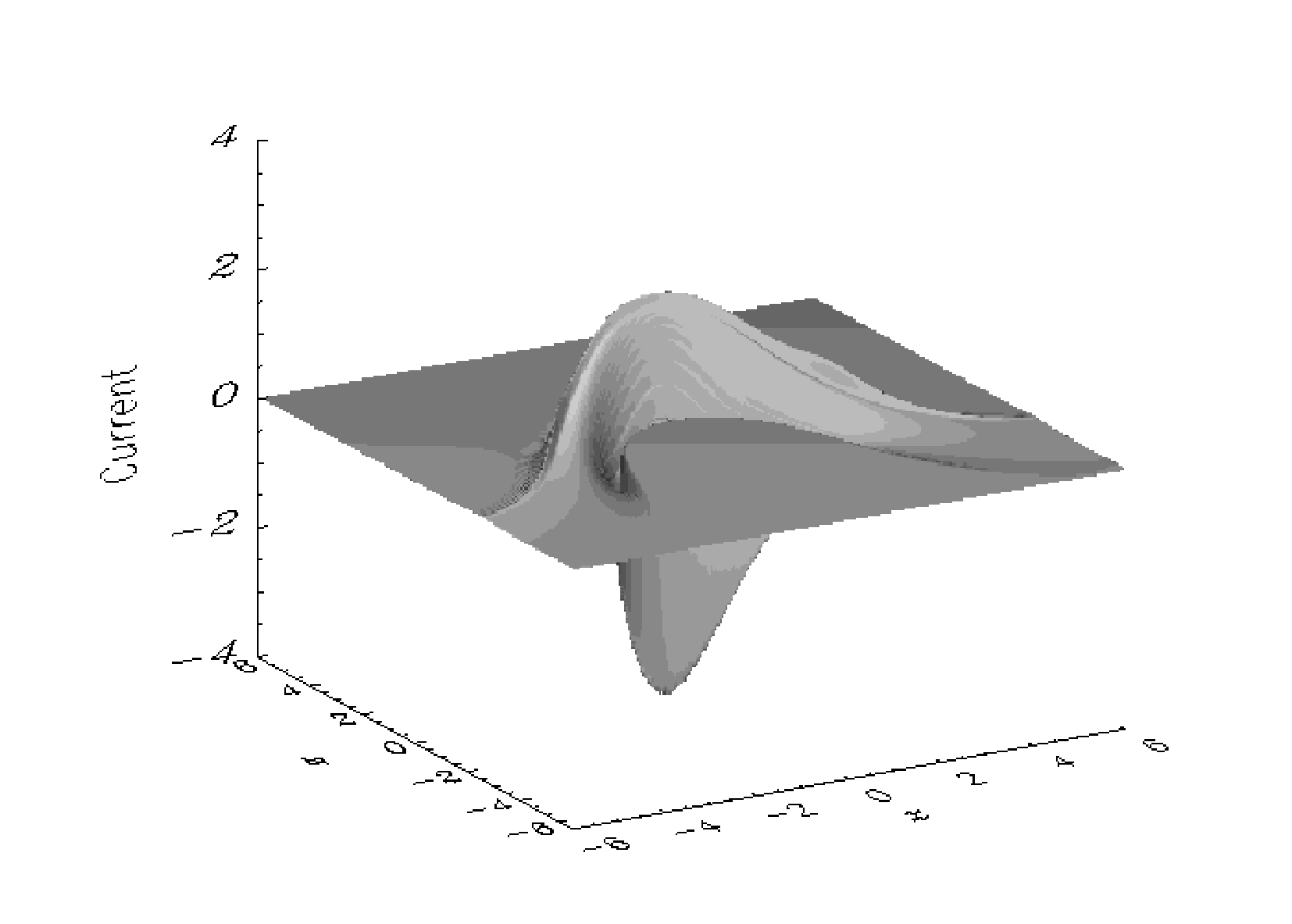}\\
\includegraphics[width=2.4in]{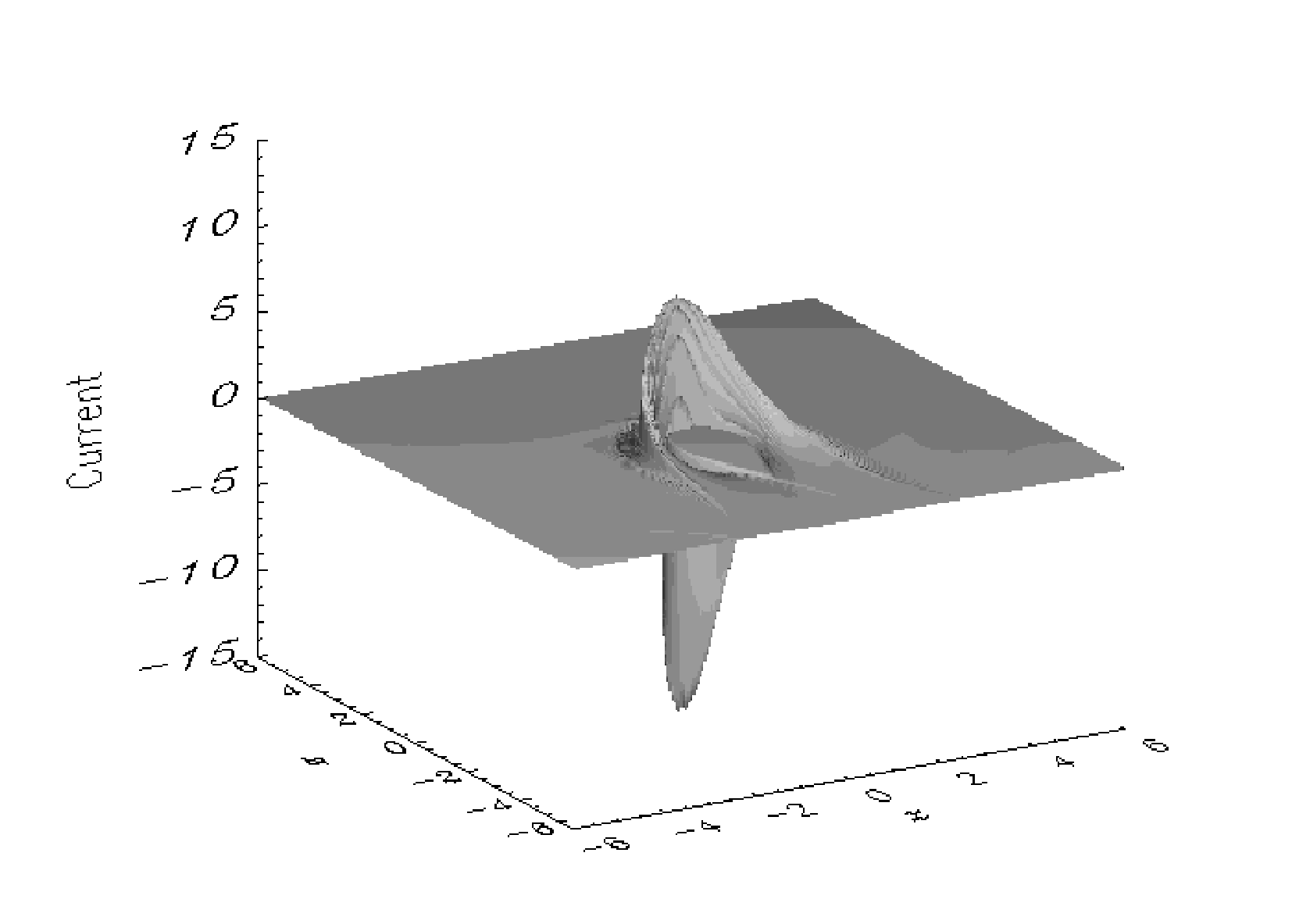}\\
\includegraphics[width=2.4in]{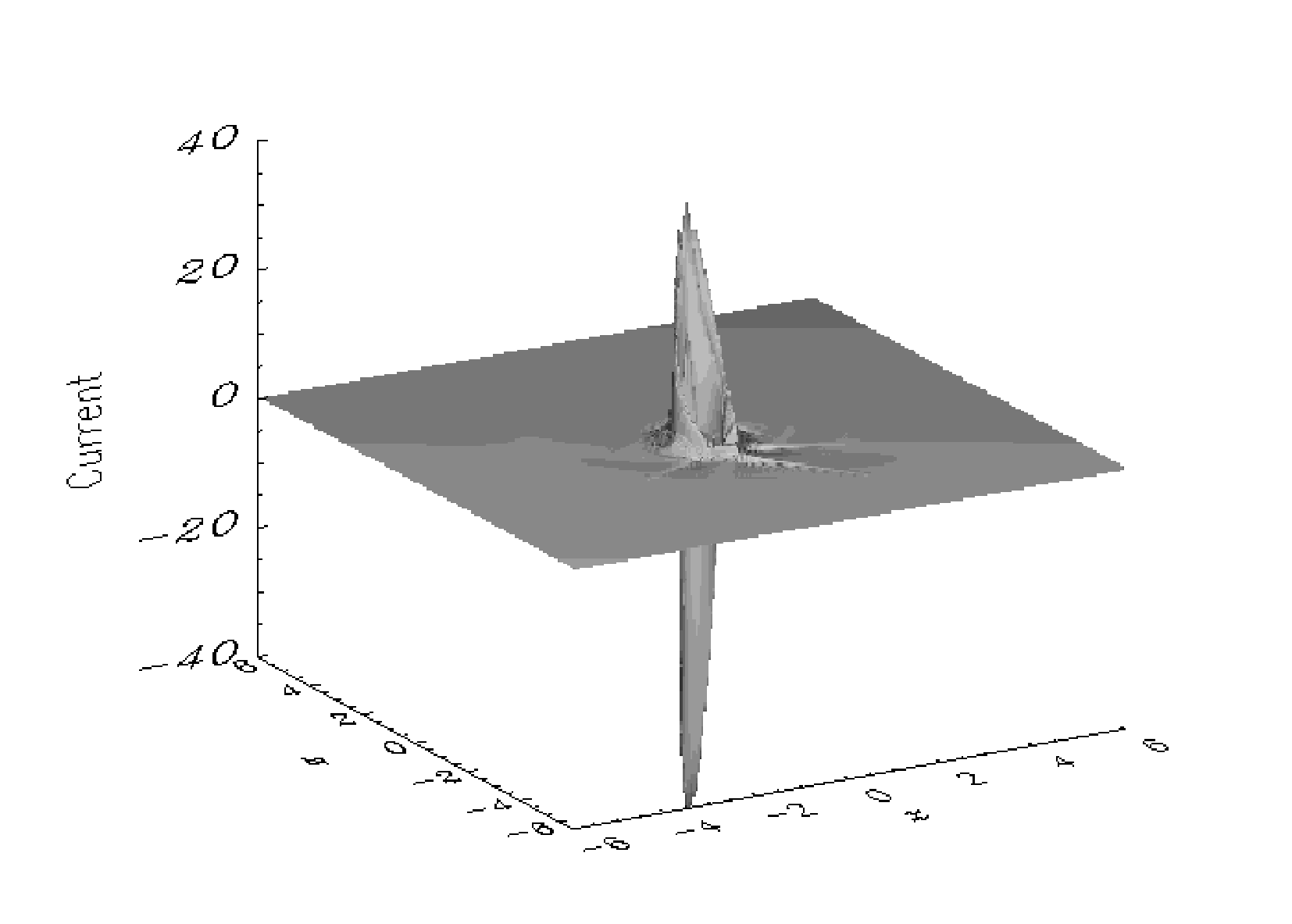}
\hspace{0.4in}
\includegraphics[width=2.4in]{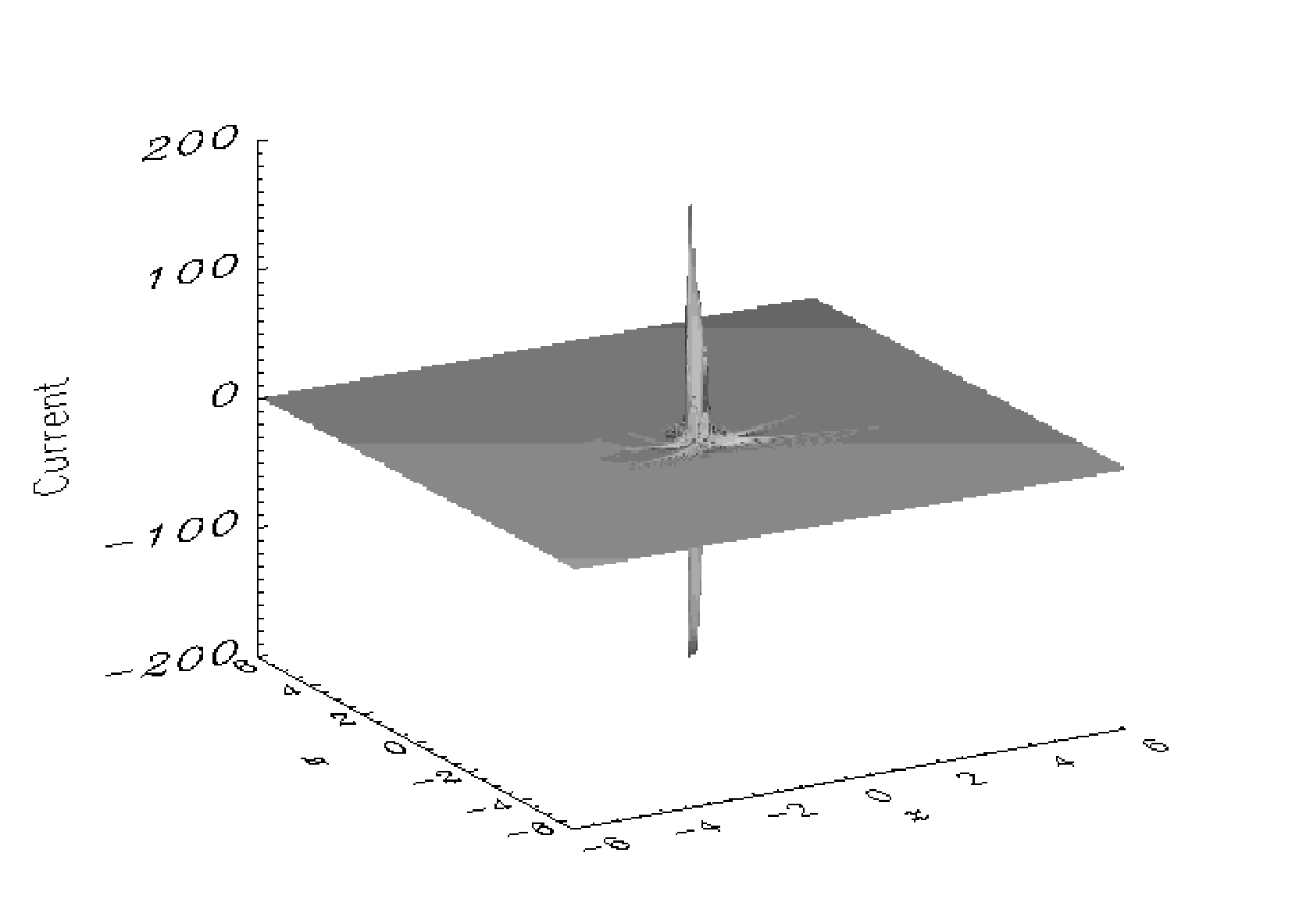}
\caption{Shaded surfaces showing the build up of current at times $(a)$ $t$=1.0, $(b)$ $t$=1.8, $(c)$ $t$=2.6, $(d)$ $t$=3.4 and $(e)$ $t$=4.2, labelling from top left to bottom right. Note the change in vertical scale.}
\label{figurethree}
\end{center}
\end{figure*}

\subsection{Current}\label{sec:3.2}

\begin{figure*}[htb]
\begin{center}
\includegraphics[width=2.4in]{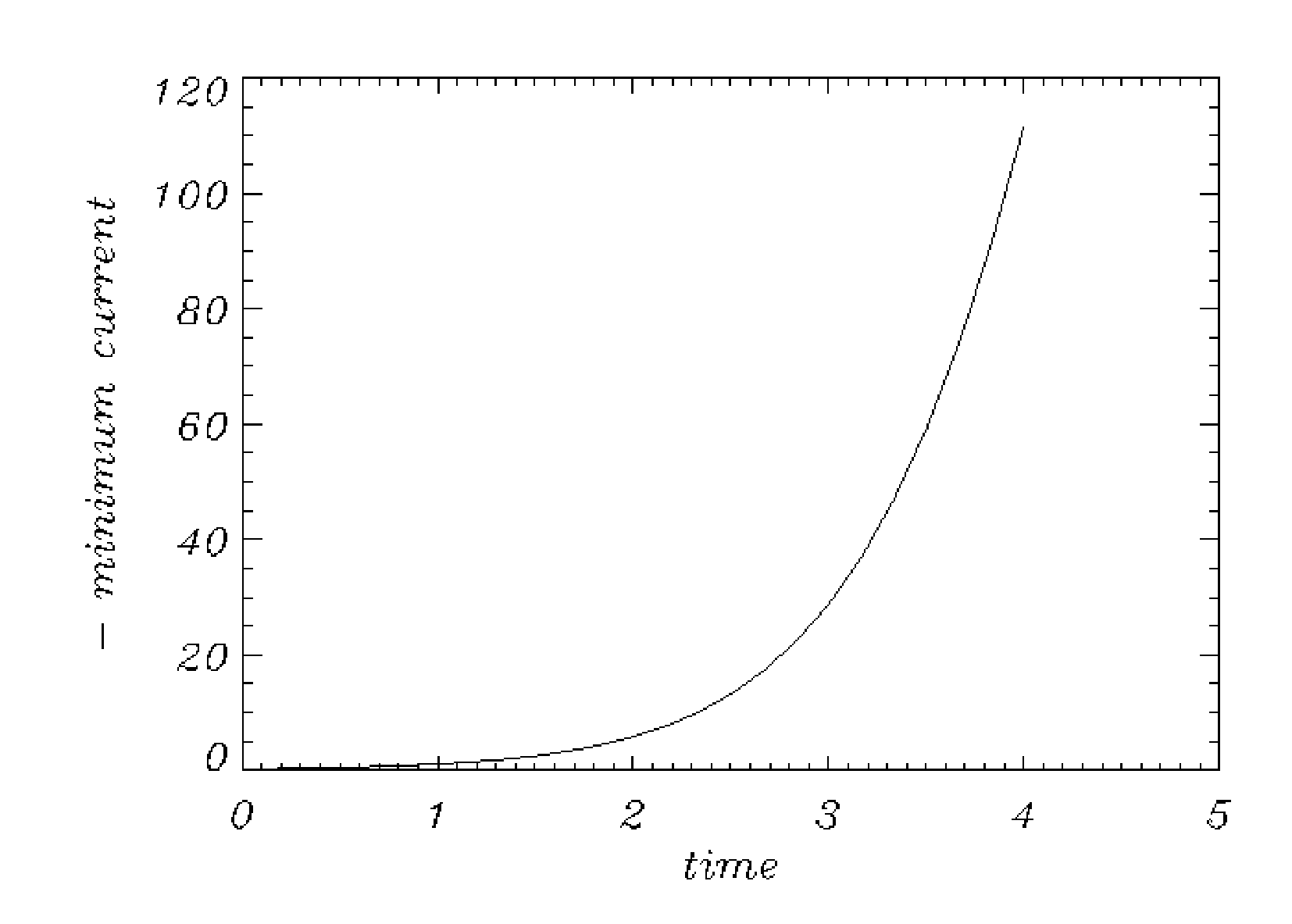}
\hspace{0.4in}
\includegraphics[width=2.4in]{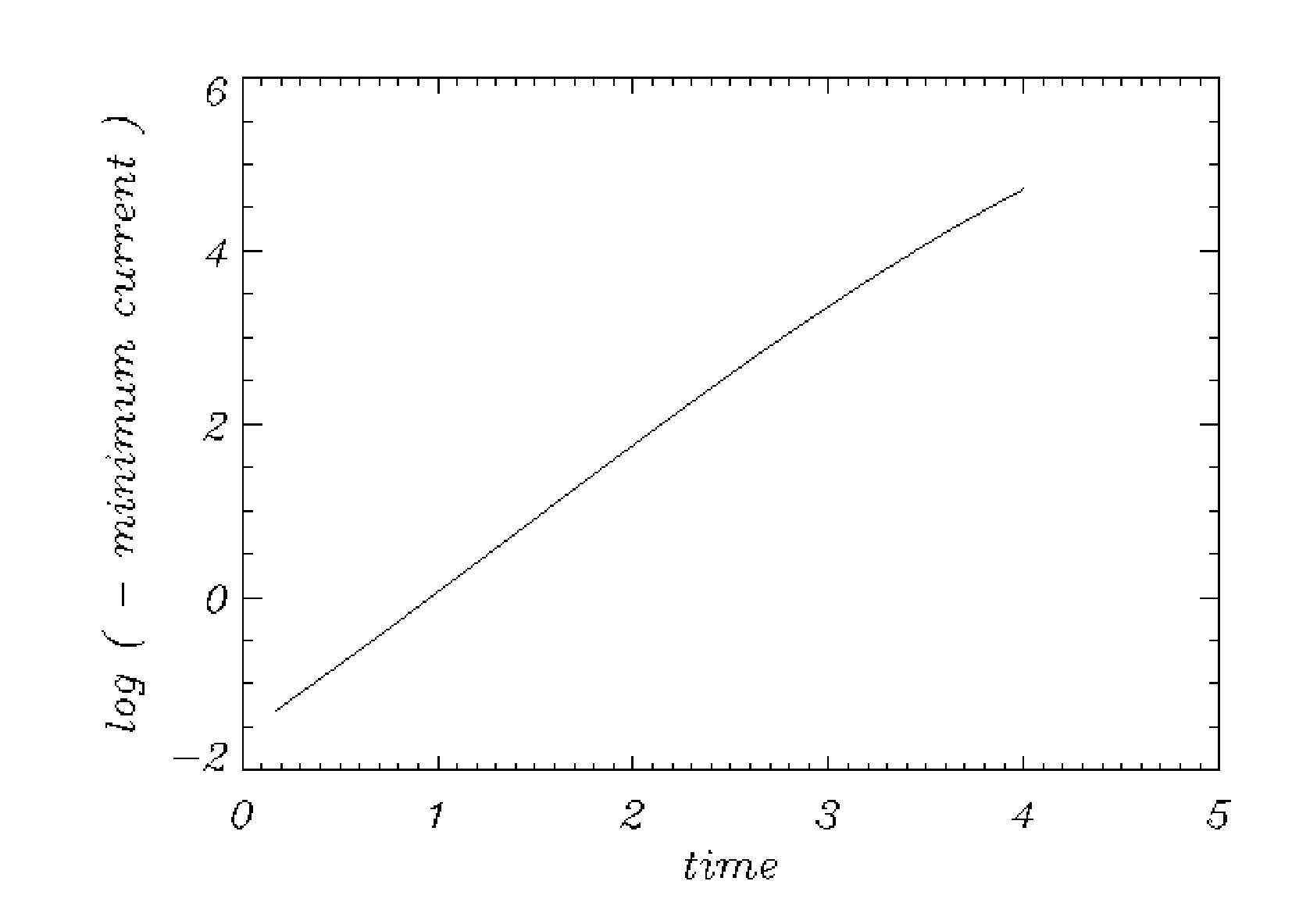}
\caption{ ( - minimum current) against time elapsed (left), log ( - minimum current) against time elapsed (right). The slope of the line between $t=1$ and $t=3$ is $1.65$.}
\label{figurefive}
\end{center}
\end{figure*}

Since we have a changing {perturbed} magnetic field whose gradients are increasing in time, we have a build up of current density. Simulations show that there is a large current accumulation at the neutral point (Figure \ref{figurethree}) and that this build up is exponential in time (Figure \ref{figurefive}), in keeping with our discussion on the thickness of the wave pulse. Figure \ref{figurefive} shows the maximum of the absolute value of the current at each time (left) and a plot of the logarithm of it against time (right).

This build-up of current is extremely important since it implies that resistive dissipation will eventually become important, regardless of the size of the resistivity, and will convert the wave energy into heat. In fact, the exponential growth of the current means that the time for magnetic diffusion to become important will depend only on $\log{\eta}$. Thus, refraction of the wave focuses the majority of the wave energy at the null point. This key result will be discussed further in the conclusions. Note that the topology of the current accumulation seems to be approaching that of a current line. The current line comes from the collapse of the width of the fast wave as it approaches the null. Note that while the current grows exponentially in time, the velocity remains finite in magnitude.

As an aside, note that the line in Figure \ref{figurefive} (right) starts to deviate from a straight line after (approximately) $t=3.5$. Recall the discussion above concerning the decrease in width of the pulse, where the distance between the leading and trailing edge of the wave was shown to be $\delta z = 6 \left( e^{t_0} - 1 \right) e^{-t}$. This exponential decrease in length scales means that we will always run out of numerical resolution for the simulation in Figure \ref{figuretwo}; for any topology we want to resolve, for which $\delta z$ will be of the form $\frac {C}{N}$ (where $N$ is the number of grid points and $C$ is a constant of inverse proportionality), we see that a solution is improperly resolved when $e^t \geq \left(  e^{ t_0 } -1 \right) N / C$, i.e. $t \geq \log {N} + \log { \left(  e^{ t_0 } -1 \right) } - \log {C}$. Thus, because of this logarithmic dependence on $N$, we will only be able to resolve the solution up until a certain time and increasing the number of grid points will not substantially improve this. Thus, the line in Figure \ref{figurefive} tails off due to a lack of numerical resolution.

\subsection{Analytical results}\label{sec:3.1}

\begin{figure*}[t]
\begin{center}
\includegraphics[width=2.0in]{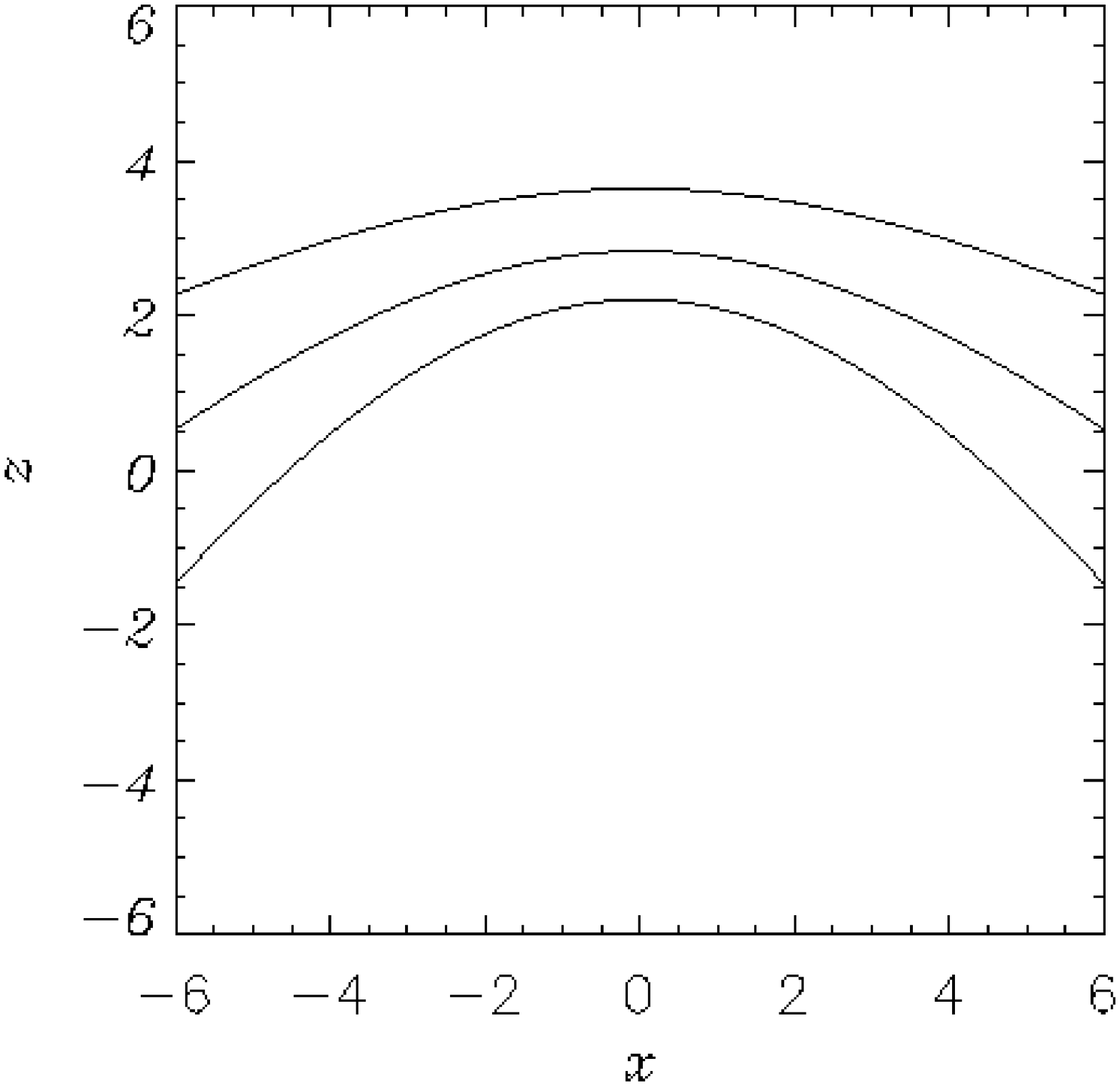}
\hspace{0.15in}
\includegraphics[width=2.0in]{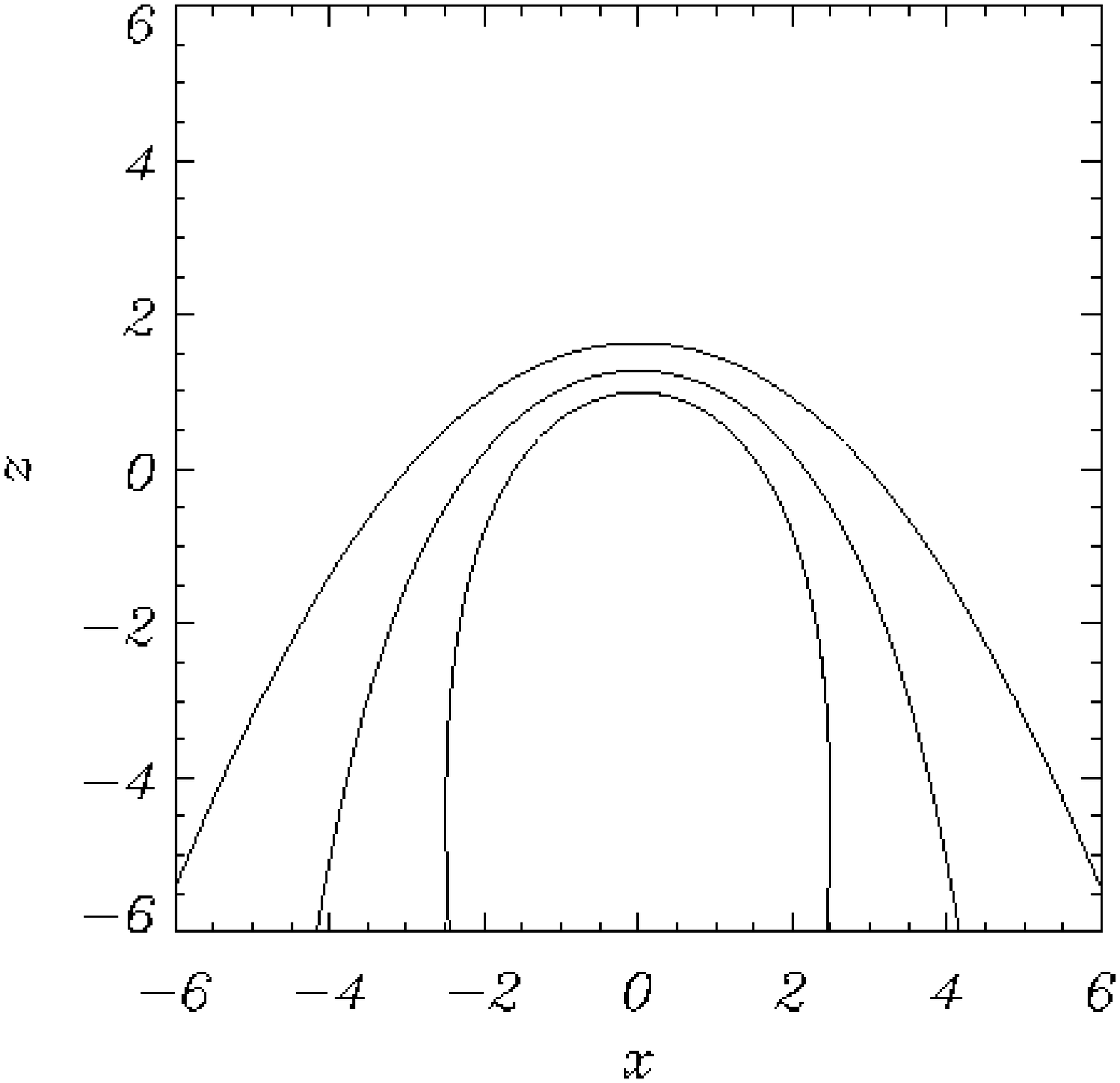}\\
\includegraphics[width=2.0in]{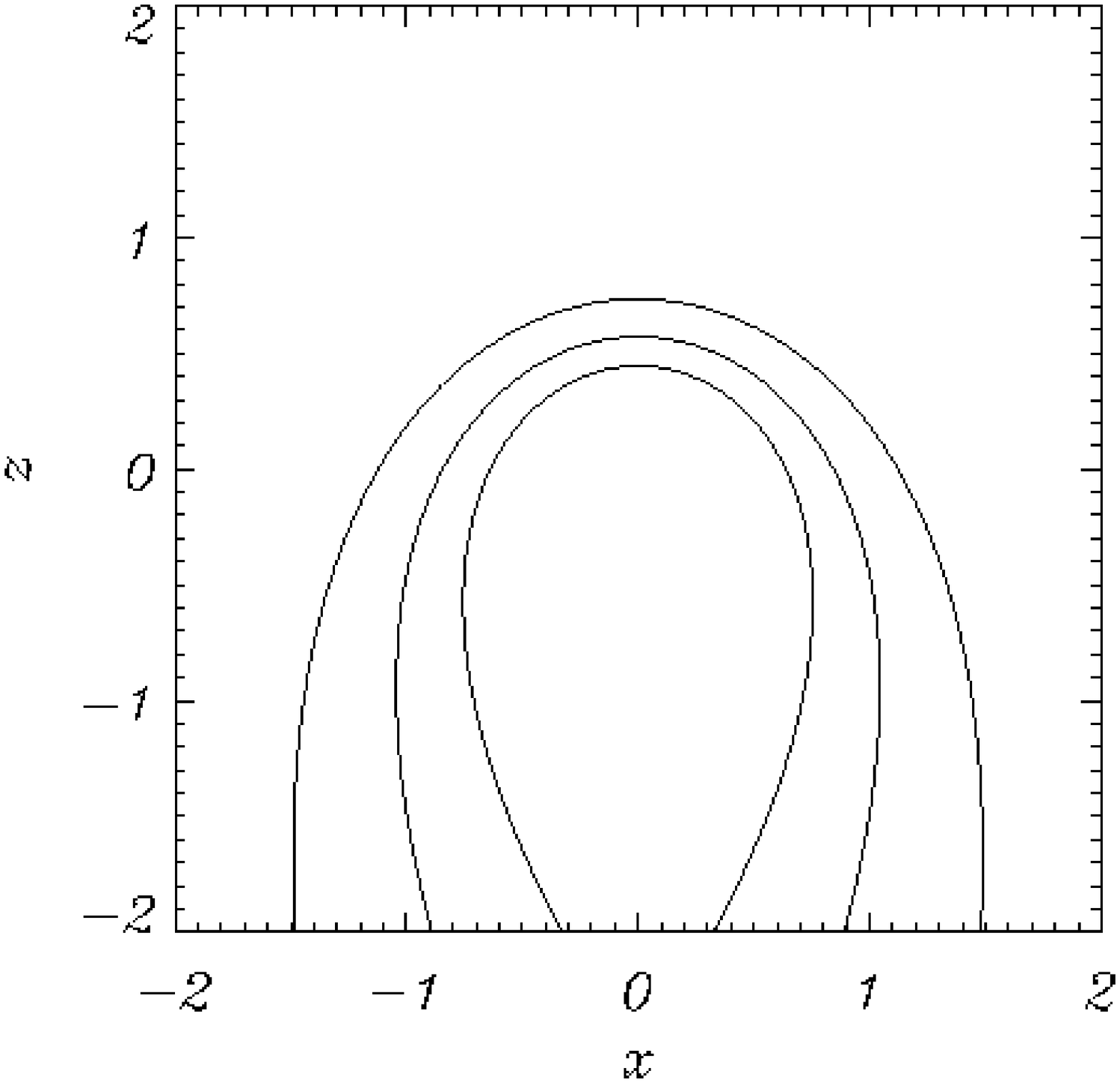}
\hspace{0.15in}
\includegraphics[width=2.0in]{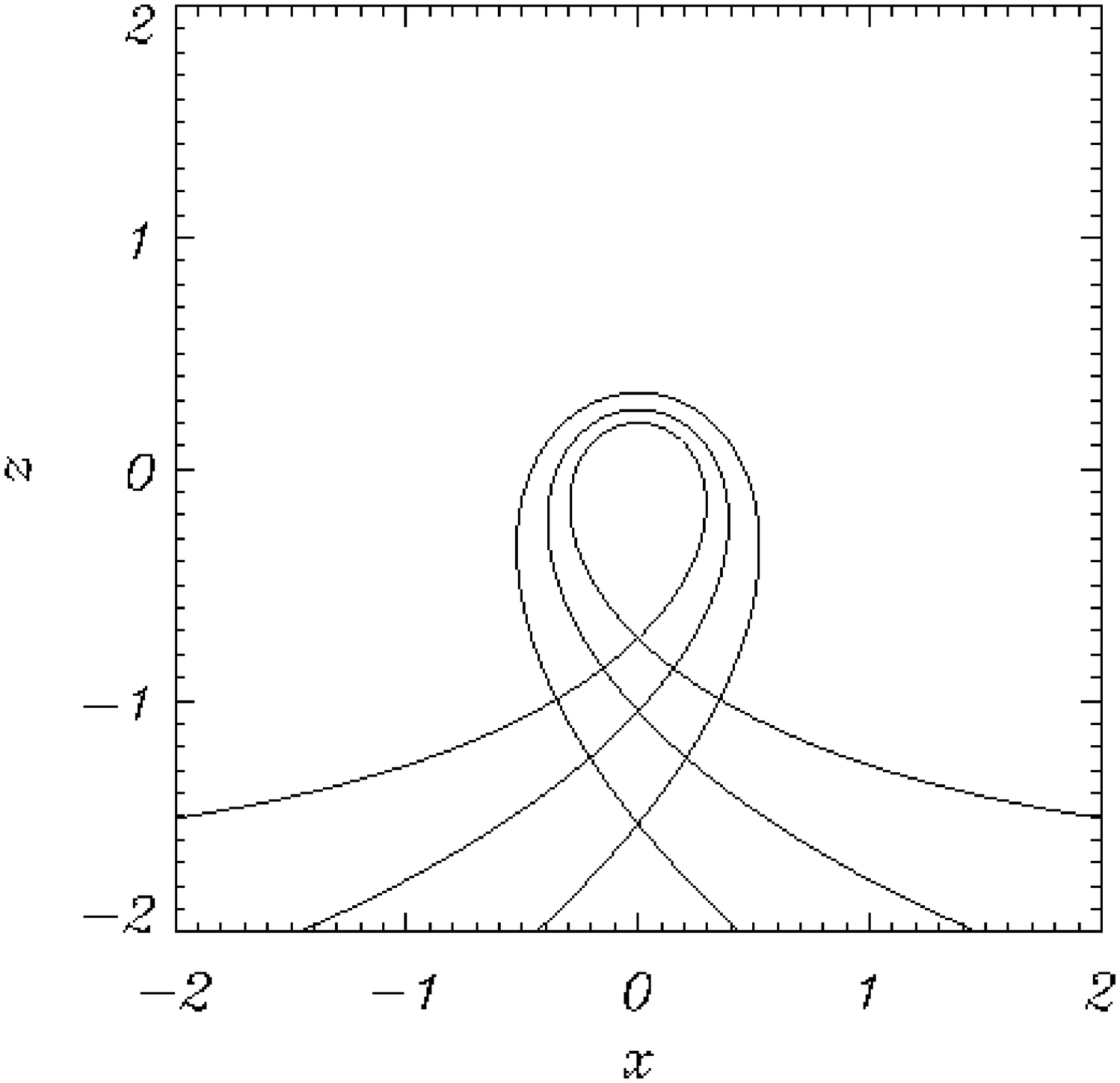}
\caption{Plots of WKB solution for a wave sent in from the upper boundary and its resultant positions at times $(a)$ $t$=1.0, $(b)$ $t=1.8$, $(c)$ $t=2.6$ and $(d)$ $t=3.4$, labelling from top left to bottom right. The lines represent the front, middle and back edges of the wave, where the pulse enters from the top of the box. Note the axes change in $(c)$ and $(d)$.}
\label{figurefour}
\end{center}
\end{figure*}

\begin{figure*}
\begin{center}
\includegraphics[width=2.0in]{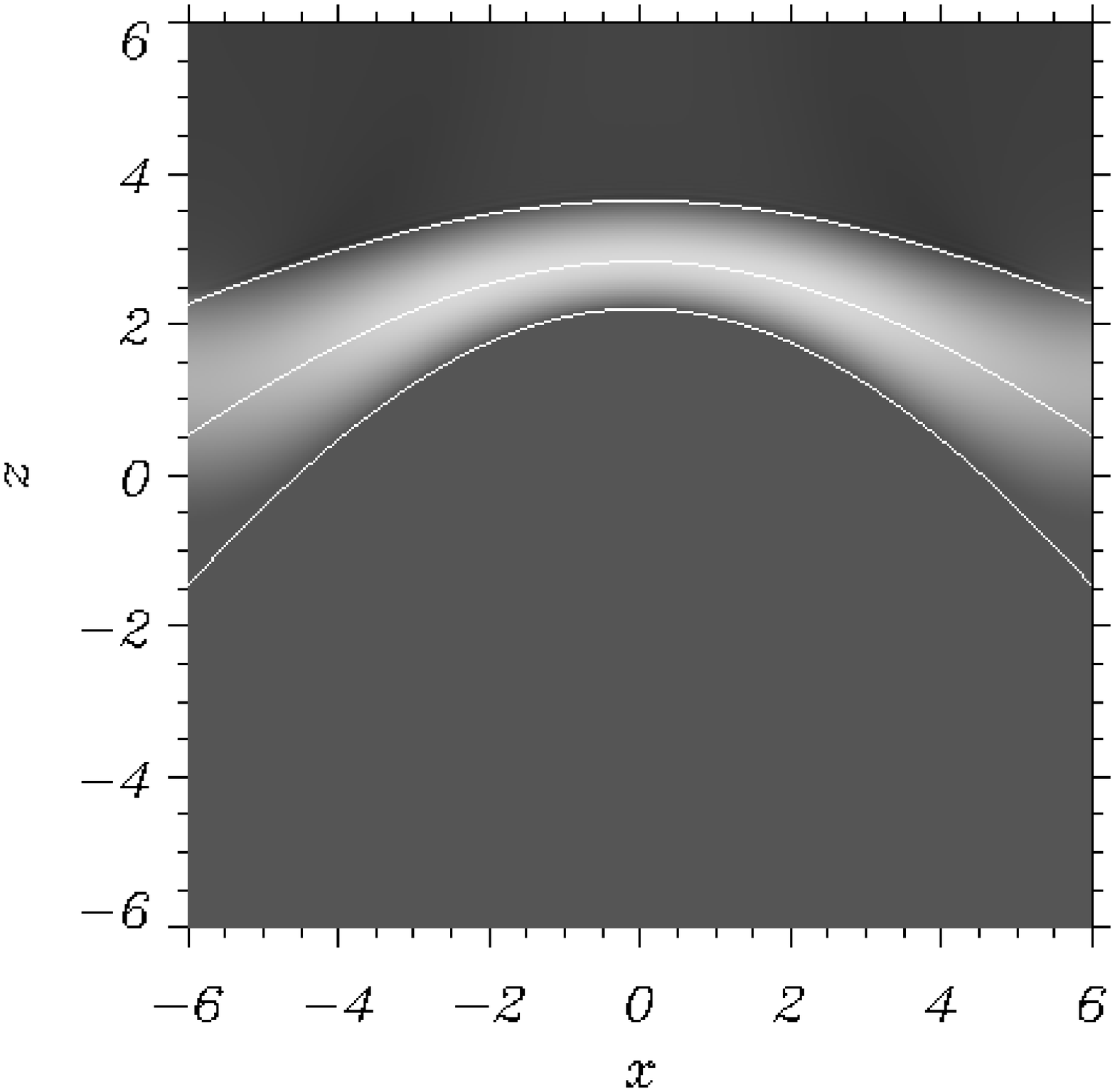}
\hspace{0.15in}
\includegraphics[width=2.0in]{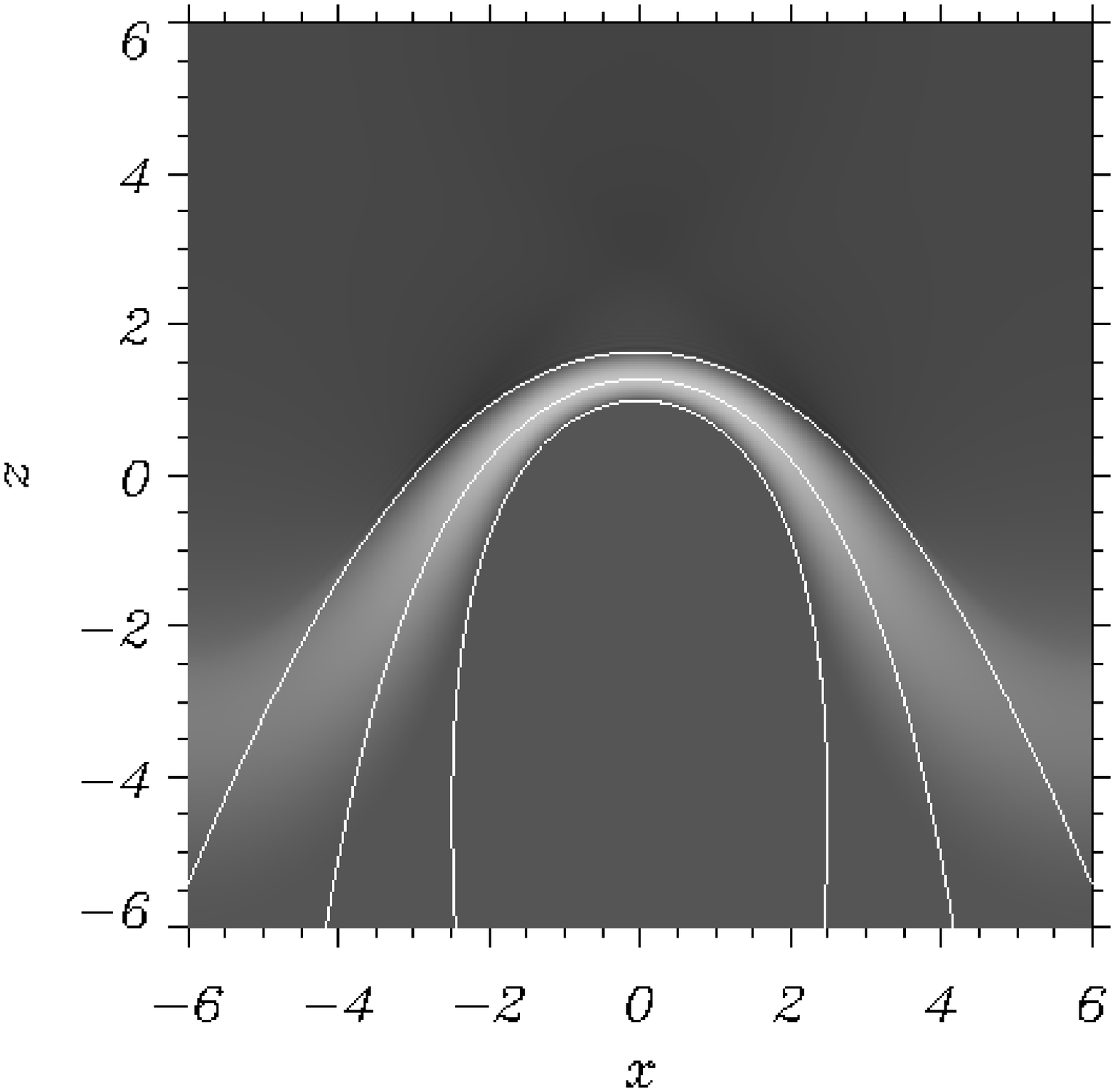}\\
\includegraphics[width=2.0in]{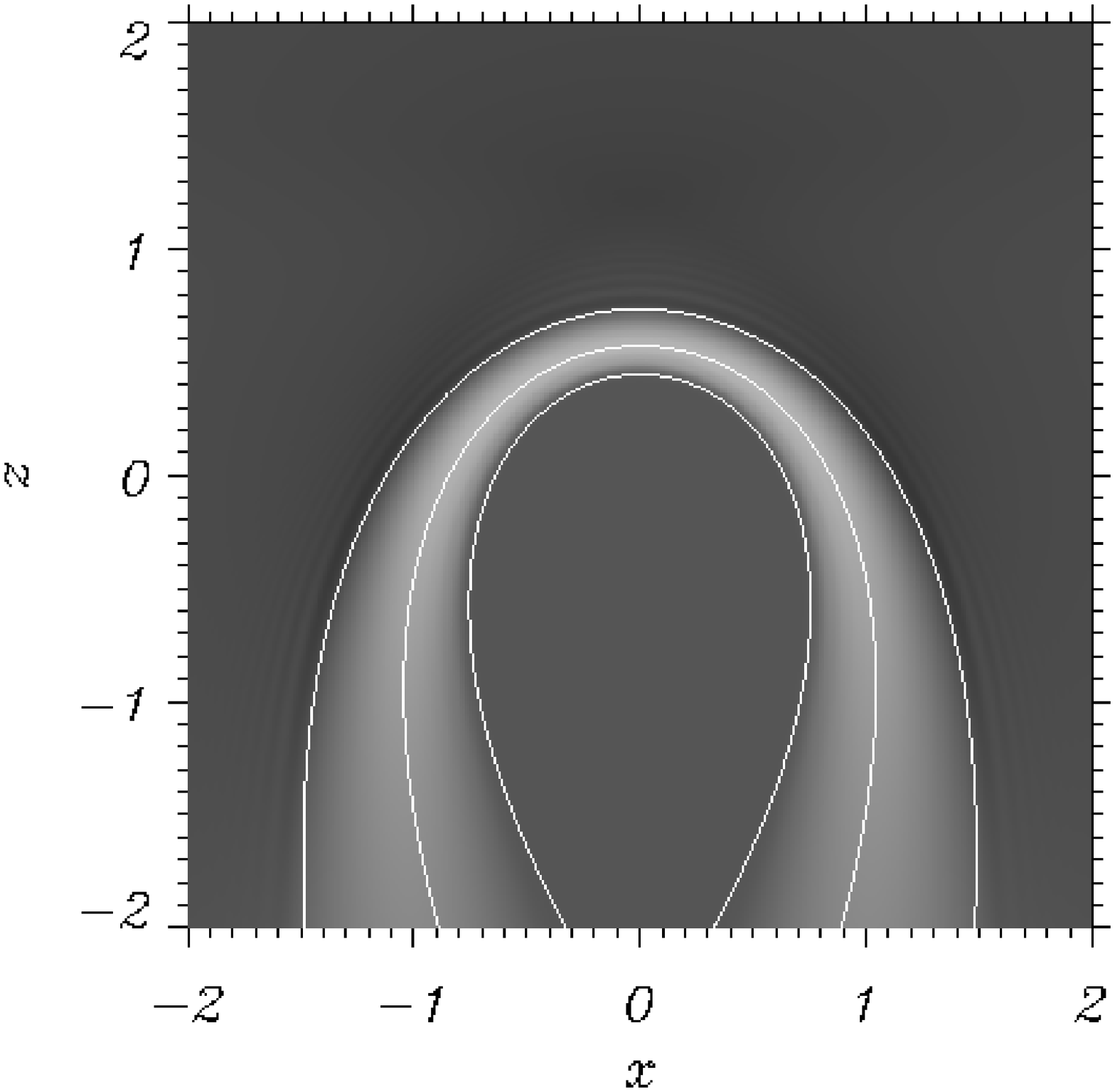}
\hspace{0.15in}
\includegraphics[width=2.0in]{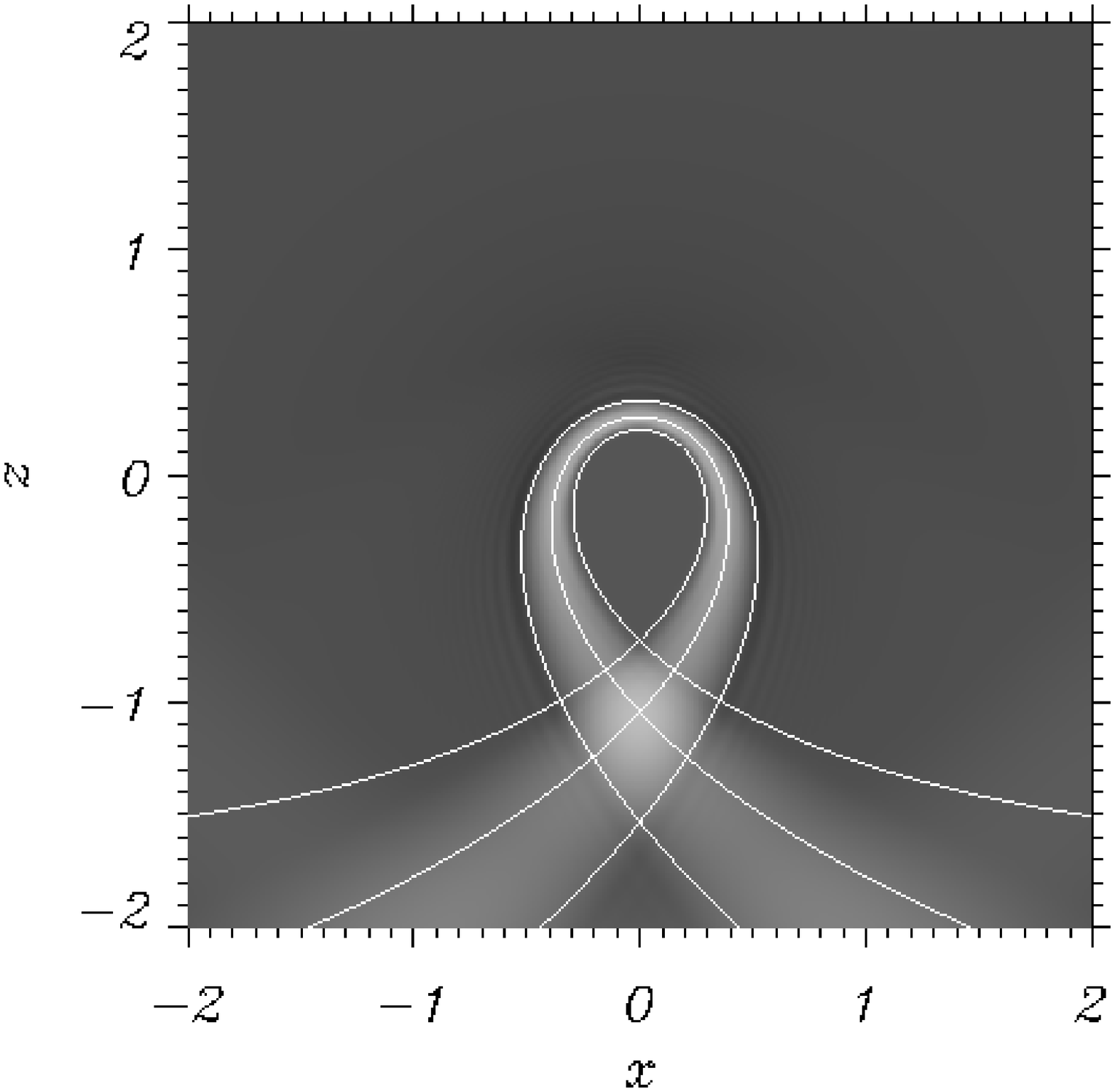}
\caption{Comparison of numerical simulation and analytical solution at times $(a)$ $t$=1.0, $(b)$ $t=1.8$, $(c)$ $t=2.6$ and $(d)$ $t=3.4$, labelling from top left to bottom right. As before, the lines are the front, middle and back edges of the analytical solution.}
\label{figurealpha}
\end{center}
\end{figure*}

We can approximately solve equation (\ref{fastbeta}) for the fast wave to gain more insight into the current build-up observed in the numerical simulations. Substituting $V = e^{i \phi (x,z) } \cdot e^{-i \omega t}$ into (\ref{fastbeta}) and assuming that $\omega \gg 1 $ (WKB approximation), leads to a first order PDE of the form $\mathcal{F} \left( x,z,\phi,\frac {\partial \phi } {\partial x}, \frac {\partial \phi } {\partial z} \right)=0$. Applying the method of characteristics, we generate the solution:
\begin{eqnarray*}   
\qquad \phi &=& -\omega ^2 s  \; ,\\
x &=& \left[ x_0 \cos { \left( \frac {Ax_0}{z_0} s \right) }  + z_0   \sin { \left( \frac {Ax_0}{z_0} s \right) } \right] e^{-As} \; ,\\ 
z &=& \left[   z_0 \cos { \left( \frac {Ax_0}{z_0} s \right) } - x_0 \sin { \left( \frac {Ax_0}{z_0} s \right) } \right] e^{-As} \; ,
\end{eqnarray*}
where $s$ is some parameter along the characteristic, $x_0$ is a starting point distinguishing between different characteristic curves, $z_0$ is a second starting point ($z_0=6$ in our simulations), $\omega$ is the frequency of our wave and $A$ is a constant such that $A =  {z_0 \; \omega } / { \sqrt{ x_0 ^2 + z_0 ^2 } } $.
Figure {\ref{figurefour}} shows constant $\phi$ at four different values of the parameter $s$. Constant $\phi$ can be thought of as defining the position of the leading edge of the wave pulse, i.e. with this choice of $s$, the WKB solution represents the front of the wave. $s$ is comparable to $t$ and so the subfigures  can be directly compared to Figure {\ref{figuretwo}}. The agreement between the analytic model and the leading edge of the wavefront is very good, as seen in an overplot of a contour and our WKB solution in Figure {\ref{figurealpha}}. Note that there is a difference between the side and bottom boundary conditions of the simulations and the analytical model.

We can also use our WKB approximation to predict the current density build up. The current density is given by $j = \nabla ^2 V$ which from above can be approximated by:
\begin{eqnarray*}
\qquad j = \nabla ^2 V &=& - \frac {-\omega ^2 V}{x^2+z^2} \; \:  = - \frac {-\omega ^2  e^{2As} V}{x_0^2+z_0^2} \\
&=& -A^2 e^{2As} V = -A^2 e^{2As} e^{i \phi} e^{-i \omega t} \\
&=& -A^2 e^{2As} e^{- i \omega ^2 s } e^{-i \omega t} \; . 
\end{eqnarray*}
By considering the modulus of this result, we can see that $|j|$ will grow exponentially with $s$, i.e. as  $e^{2As}$, and that the current will be negative (in agreement with figure \ref{figurefive}). To compare with our simulations, we first note that the time in our numerical scheme, $t$, is related to our parameter $s$ by $t=\omega s$. Now, the exponent $2As$ in the exponential is equal to ${12 \omega s} / {\sqrt{x_0^2 + z_0^2}}$  and so along $x_0=0$ and at $z_0=6$ ($x_0=0$ is where the maximum current of our numerical simulations  occured) this is simply $2t$. The slope of the numerical experiment in figure \ref{figurefive} between $t=1$ and $t=3$ is $1.65$. This agreement between the analytical and numerical current density build up is quite good, considering that the WKB solution is only valid for a harmonic wavetrain with $\omega \gg 1$.

\subsection{Asymmetric Pulse}\label{sec:3.3}

\begin{figure*}[htb]
\begin{center}
\includegraphics[width=2.0in]{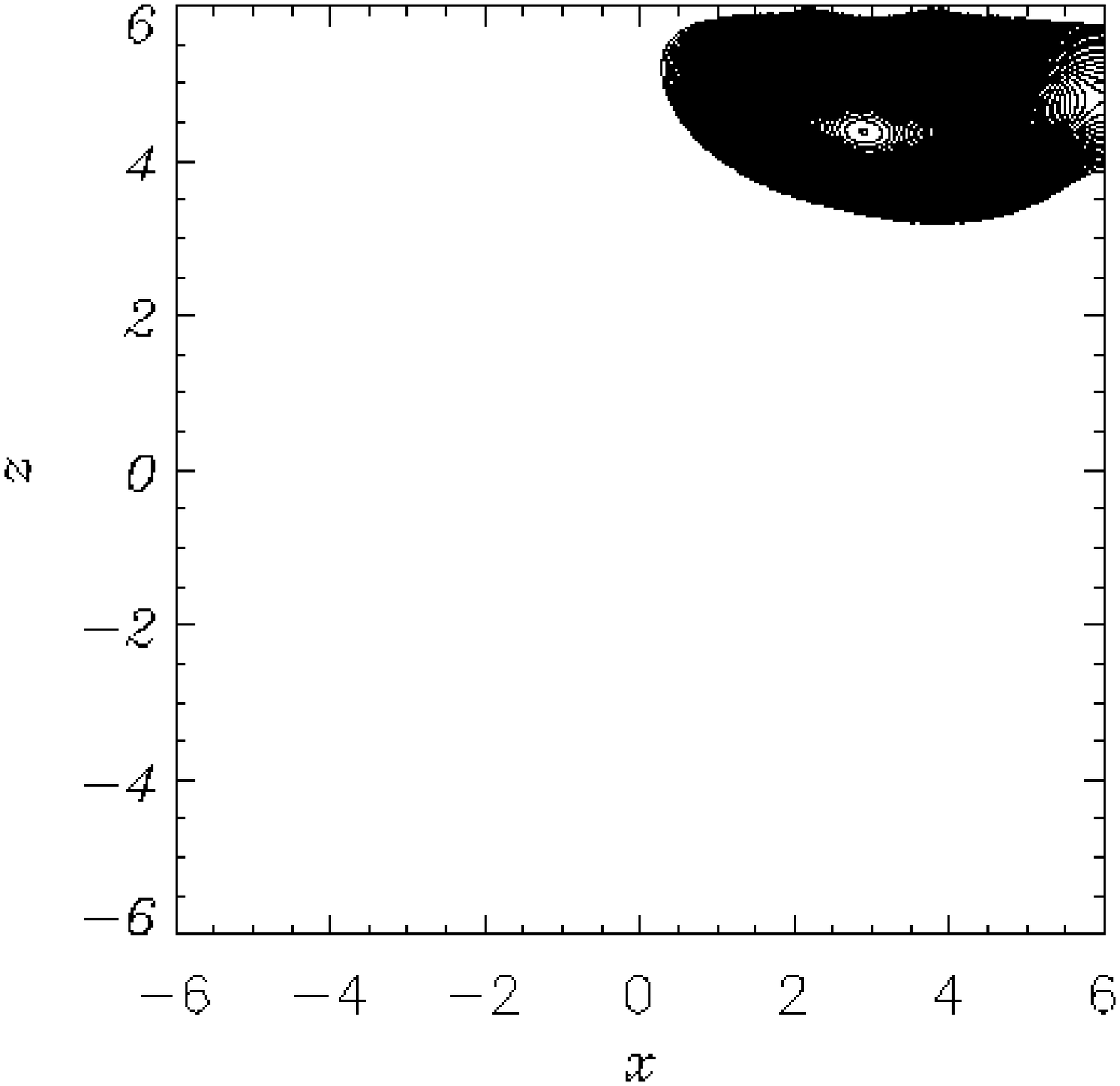}
\hspace{0.15in}
\includegraphics[width=2.0in]{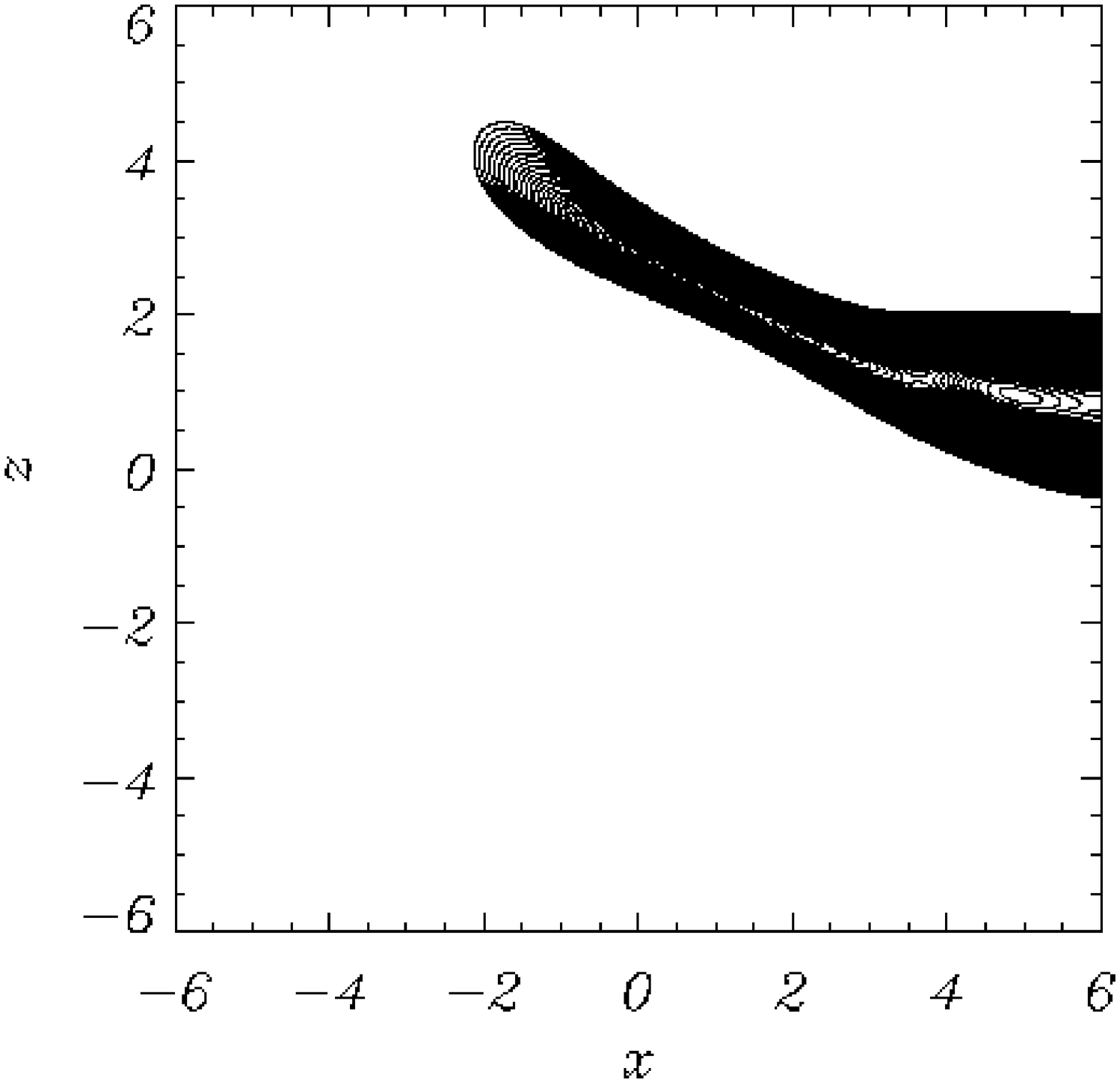}
\hspace{0.15in}
\includegraphics[width=2.0in]{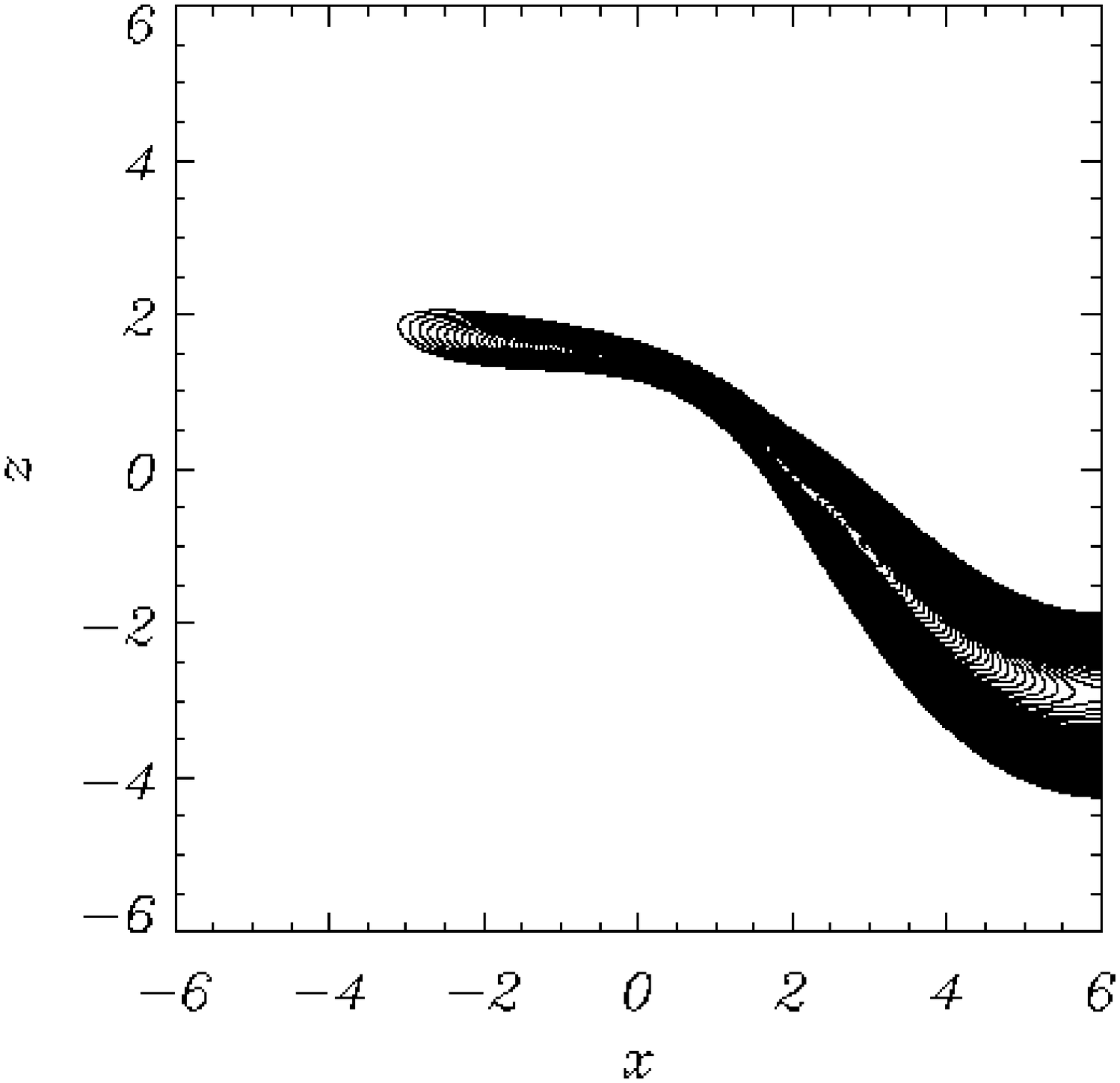}\\
\includegraphics[width=2.0in]{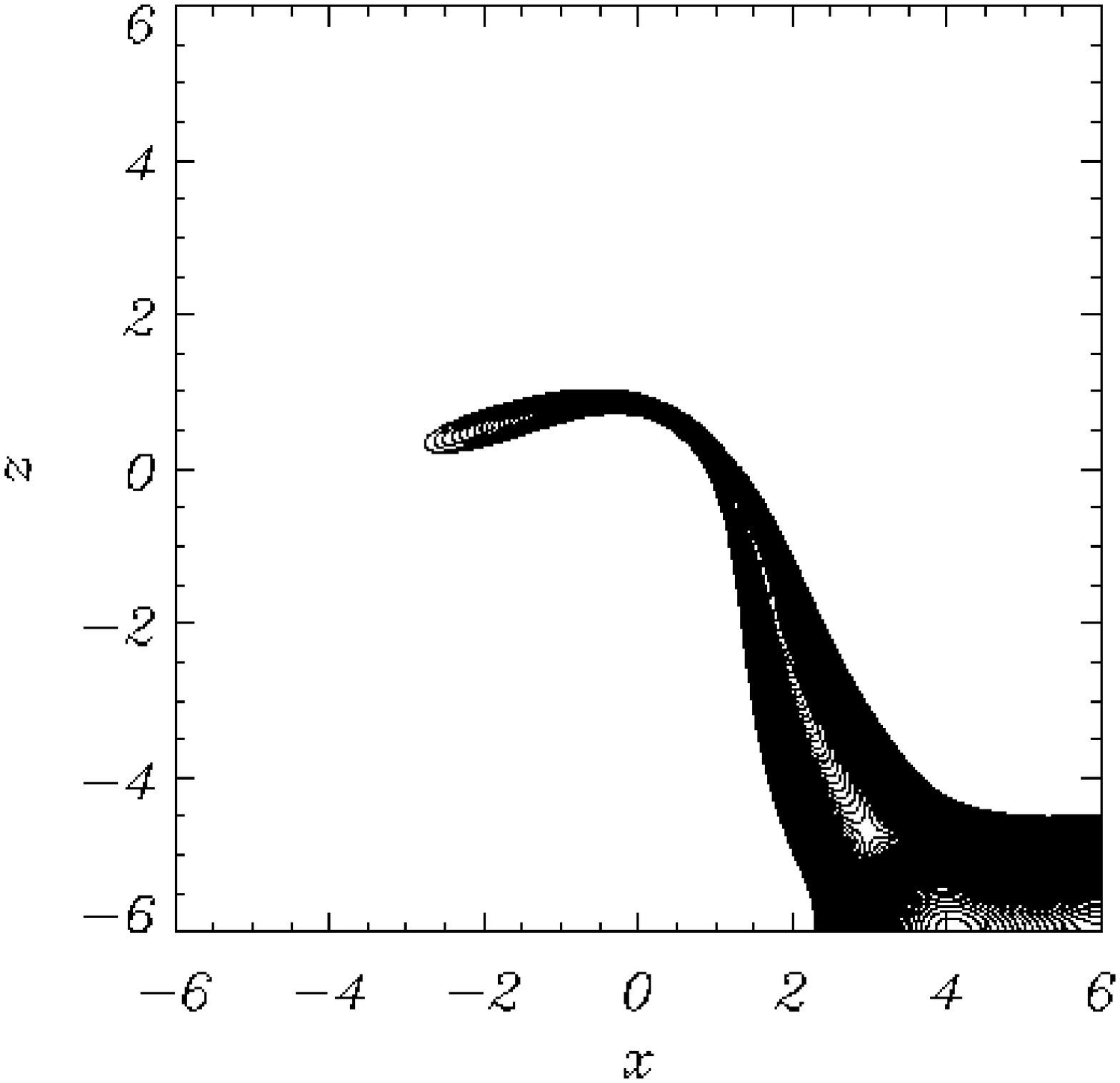}
\hspace{0.15in}
\includegraphics[width=2.0in]{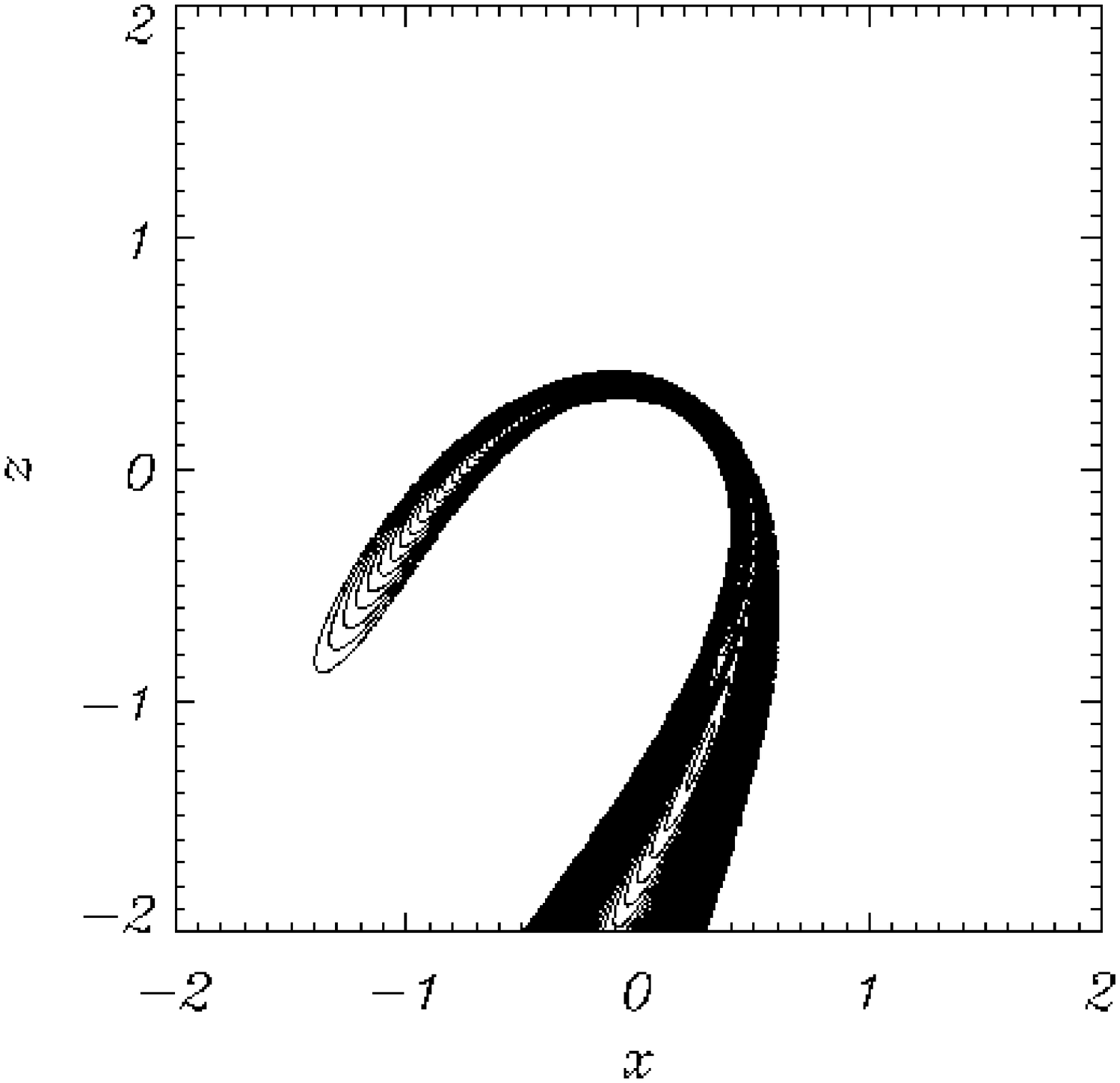}
\hspace{0.15in}
\includegraphics[width=2.0in]{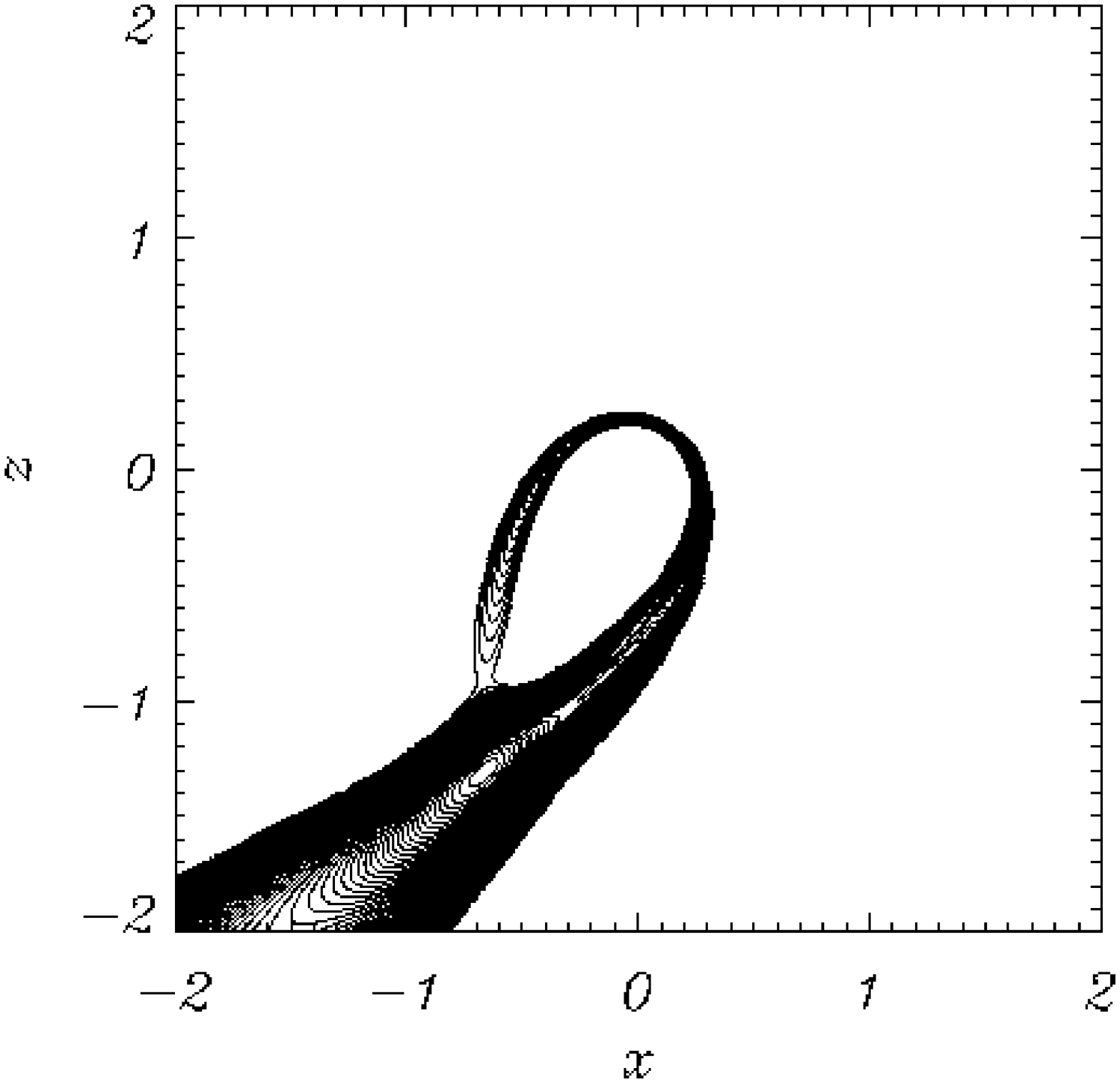}
%\vspace{1cm}
\caption{Contours of $V$ for a fast wave sent in from upper boundary in accordance with equation (\ref{lambda}) with $\omega = 2 \pi$ , and its resultant propagation at times $(a)$ $t$= 0.5, $(b)$ $t$= 1.17, $(c)$ $t=$ 1.83,  $(d)$ $t$= 2.33, $(e)$ $t$= 3.17 and $(f)$ $t$= 3.67, labelling from top left to bottom right. The axes have been scaled down in the last two subfigures to draw attention to the central behaviour. The number of grid points in each direction is $ 1200 \times 1200$.}
\label{figurethirteen}
\end{center}
\end{figure*}

As mentioned in the introduction, MHD waves in the neighbourhood of null points have been investigated before in cylindrical coordinates. In this paper however, we can readily consider asymmetric pulses and how they propagate in the neighbourhood of coronal null points. Hence, to demonstrate the wrapping around the null phenomena further, we consider the same numerical experiment above but with a pulse being fed in only on one part of the upper boundary. This pulse took the form;
\begin{eqnarray}
V = \sin {\left( \omega t \right) } \left[ 1+\cos{  \pi \left( x-3 \right)  } \; \right] \; \left\{ \begin{array}{cl}
{2 \leq x \leq 4} &  {, \: z=6} \\
{0 \leq t \leq \frac {\pi}{\omega}} & {~}
 \end{array} \right. \label{lambda}
\end{eqnarray}
 The propagation can be seen in Figure \ref{figurethirteen}. Again, it is important to note the majority of the wave energy tends to accumulate at the coronal null point; this is where wave heating would occur. The growth of the current density occurs in a similar manner to that demonstrated in Figure \ref{figurethree}. However, it is important to note that the more the pulse is displaced to one side, the more the boundary conditions will influence the subsequent evolution.

\section{Alfv\'en waves}\label{sec:4}

\begin{figure*}[htb]
\begin{center}
\includegraphics[width=2.0in]{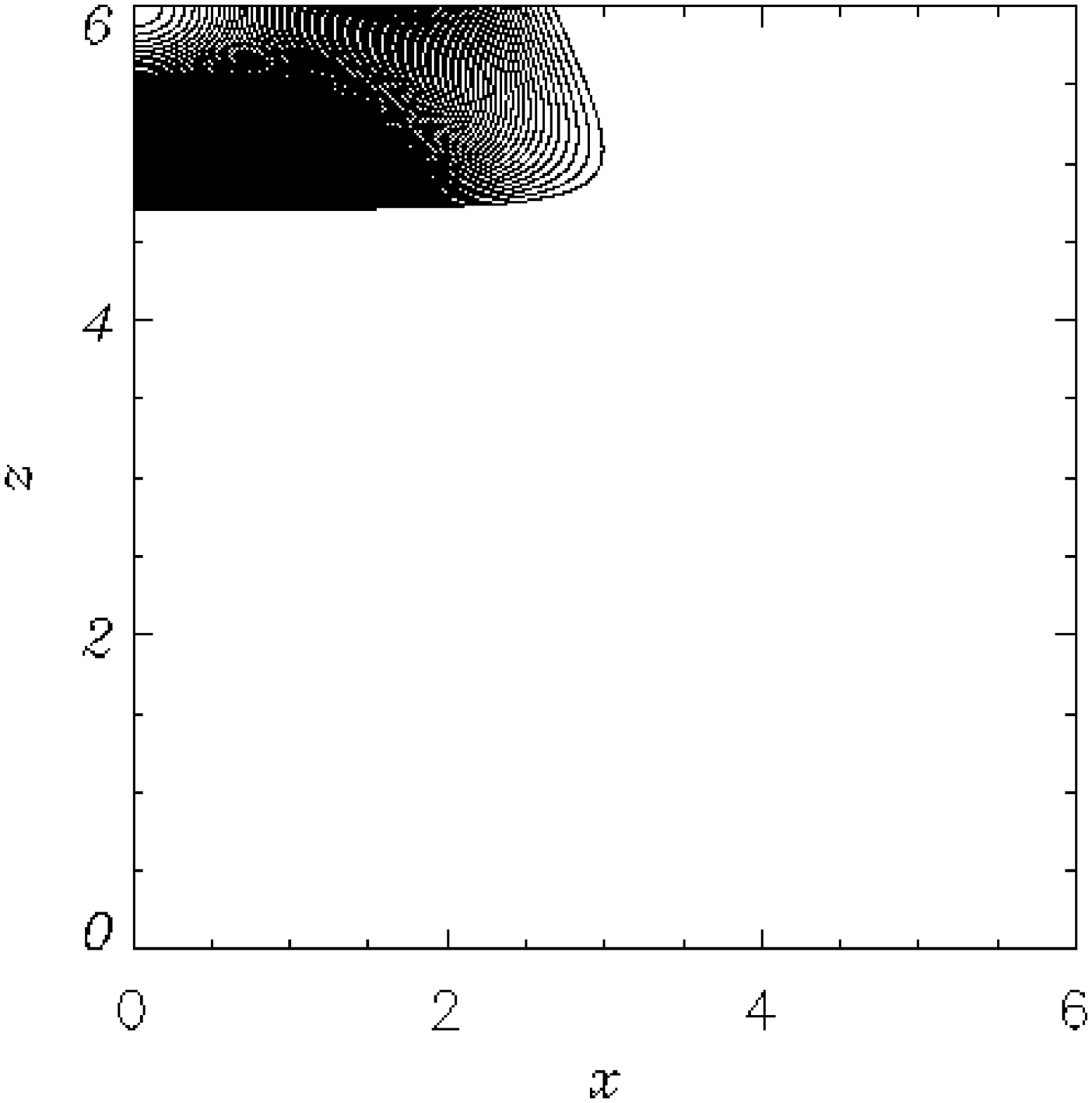}
\hspace{0.15in}
\includegraphics[width=2.0in]{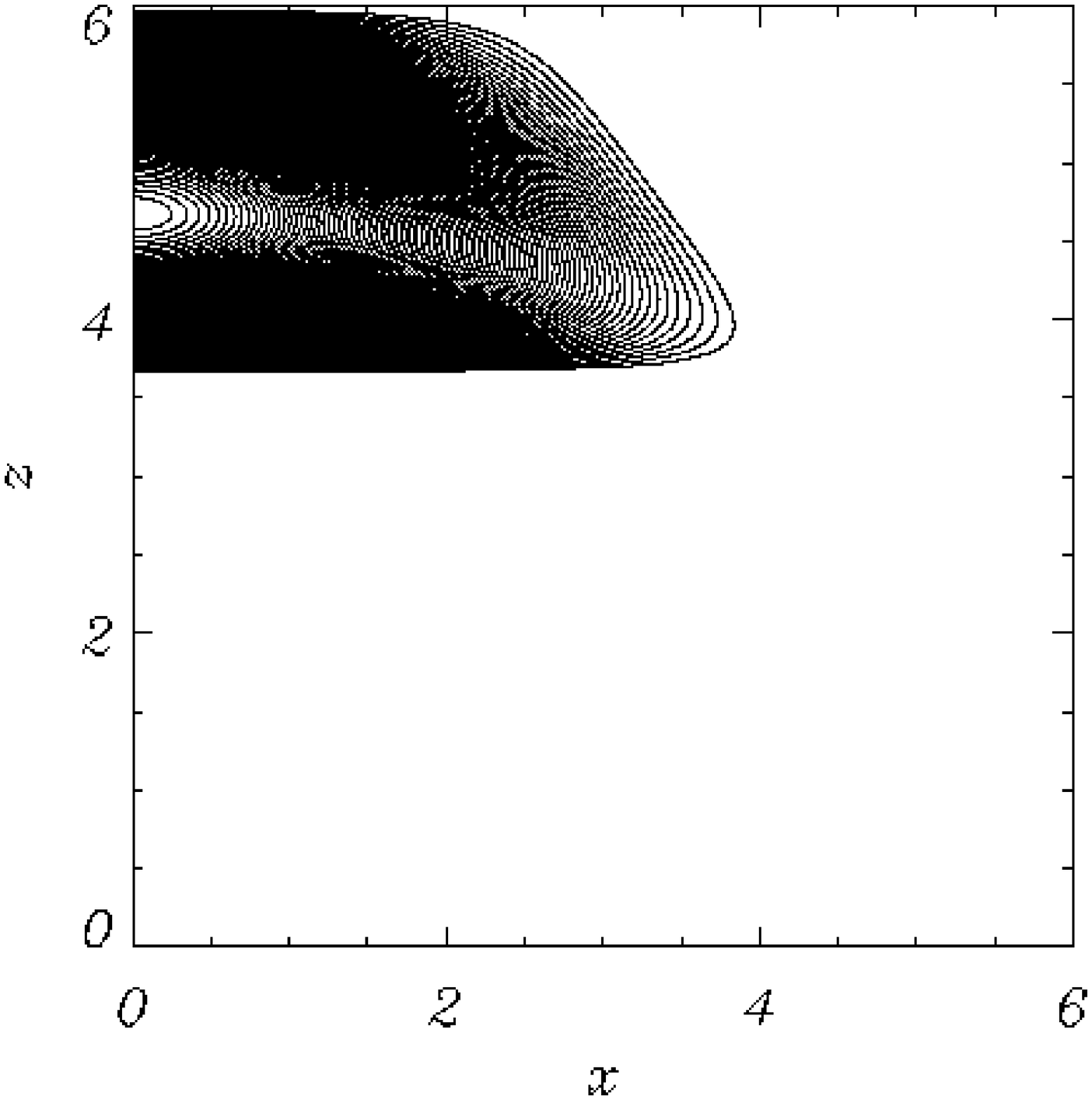}
\hspace{0.15in}
\includegraphics[width=2.0in]{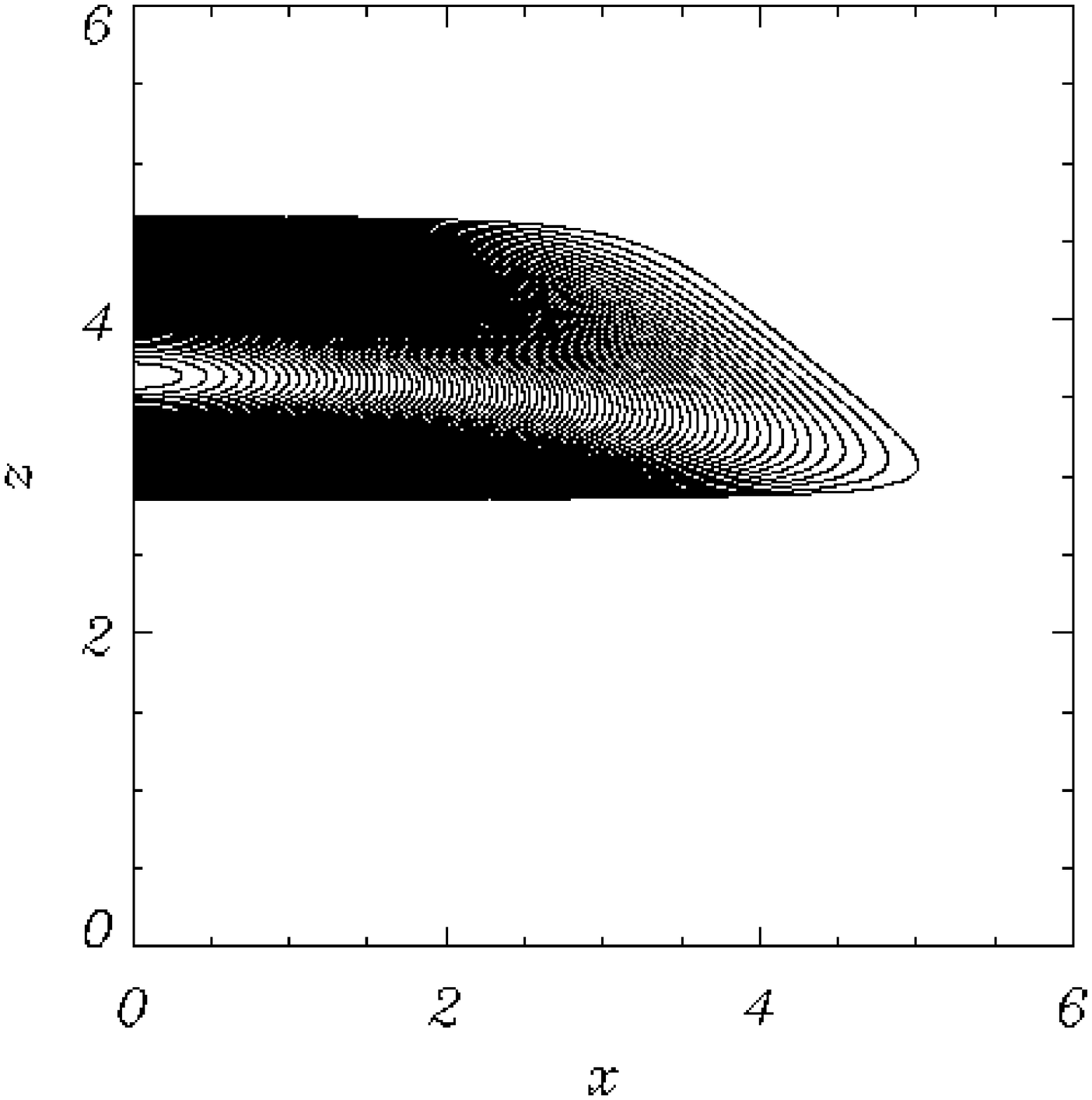}\\
\includegraphics[width=2.0in]{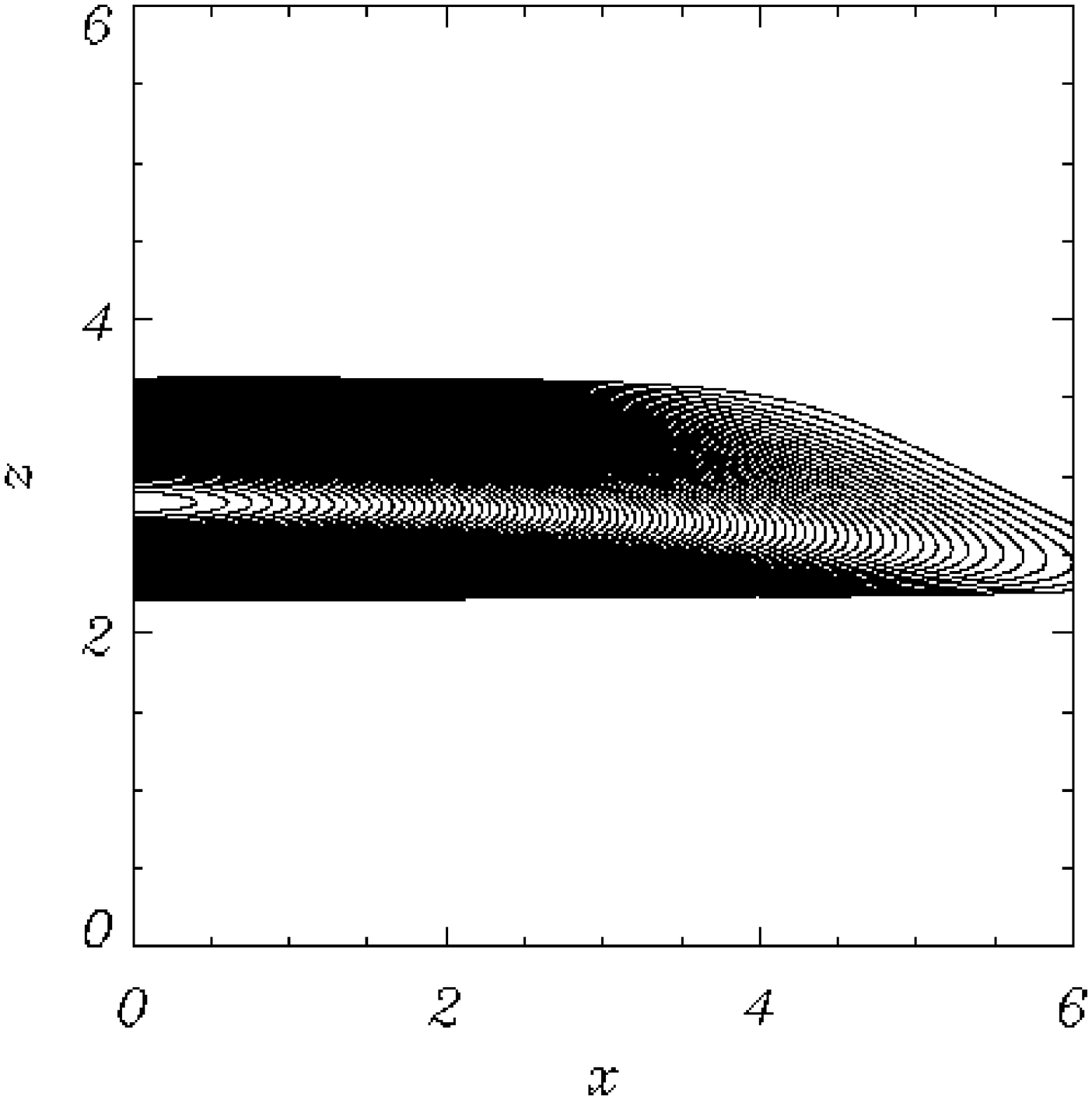}
\hspace{0.15in}
\includegraphics[width=2.0in]{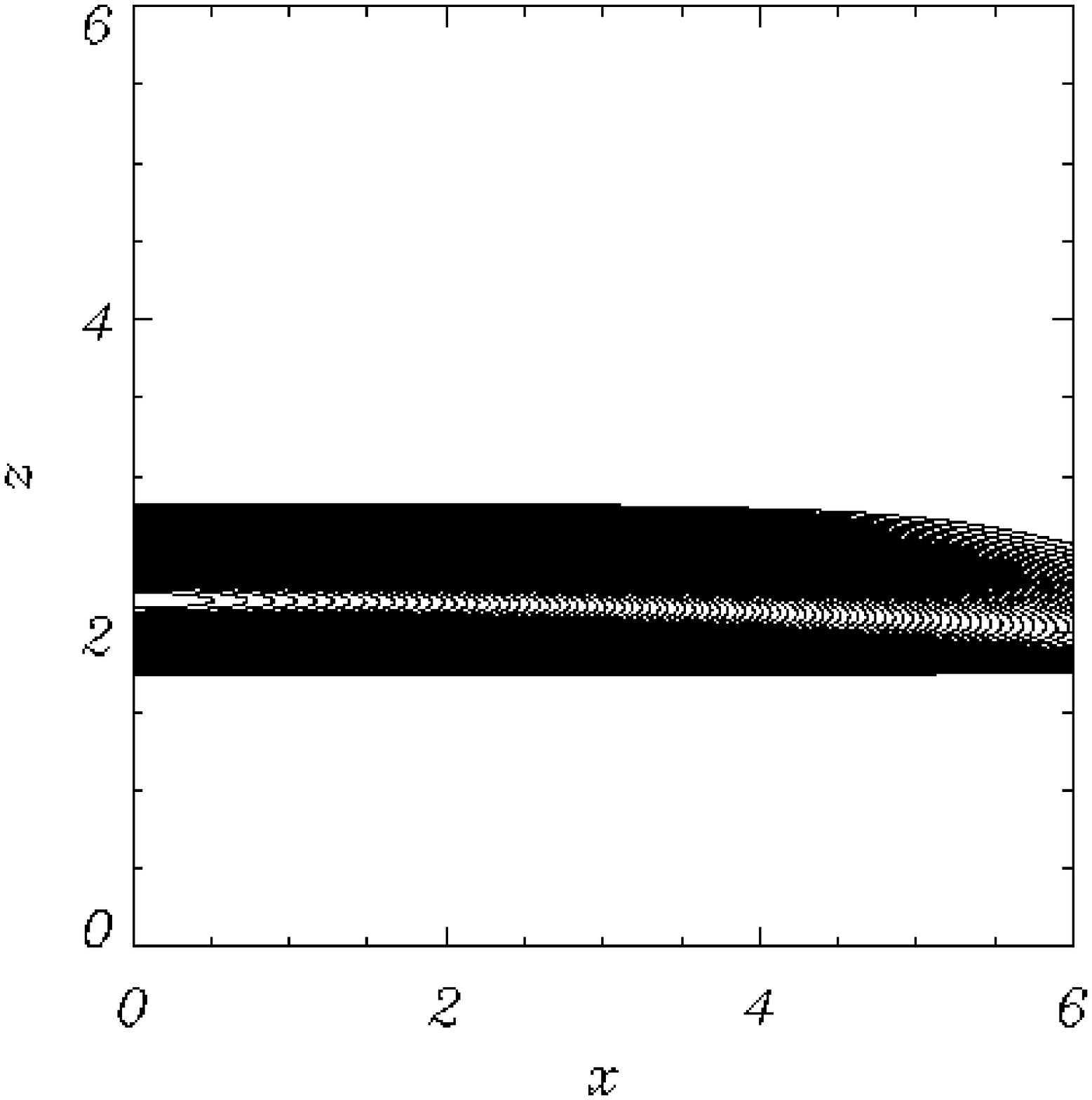}
\hspace{0.15in}
\includegraphics[width=2.0in]{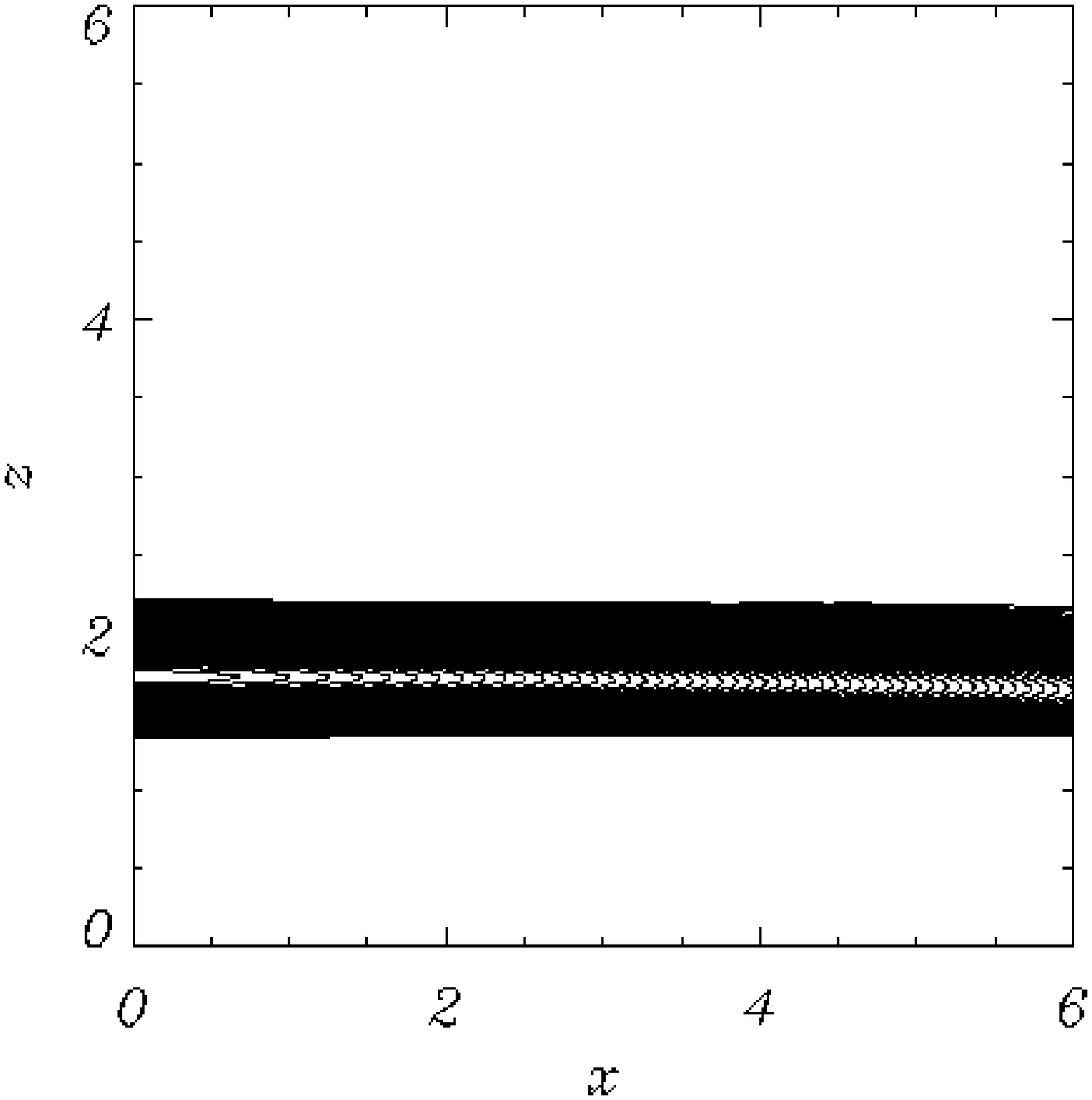}
\caption{Contours of $v_y$ for an Alfv\'en wave sent in from upper boundary for $0\leq x \leq 3.0 $ and its resultant propagation at times $(a)$ $t$=0.25, $(b)$ $t$=0.5, $(c)$ $t$=0.75, $(d)$ $t=$1.0, $(e)$ $t$=1.25 and $(f)$ $t$=1.5, labelling from top left to bottom right.}
\label{figureten}
\end{center}
\end{figure*}

The equations describing the behaviour of the Alfv\'en wave, equations (\ref{alfvenalpha}), were solved numerically using the same two-step Lax-Wendroff scheme. We initially consider a box ($0 \leq x \leq 6$, $0 \leq z \leq 6$) with a single wave pulse coming in across half of the top boundary ($0\leq x\leq 3$). We chose such a pulse because, as shown in Figure \ref{figureten}, the Alfv\'en wave spreads out along the field lines as it propagates and we found that this choice of boundary condition illustrated this effect much clearer. The full boundary conditions were;
\begin{eqnarray*}
\begin{array}{cl}
{v_y(x, 6) = \sin { \omega t } \left( 1 + \cos \frac {\pi x }{3} \right) } & {\mathrm{for} \; \; \left\{\begin{array}{c}  {0 \leq x \leq 3} \\ {0 \leq t \leq \frac {\pi}{\omega} } \end{array}\right. } \\
{\frac {\partial v_y }{\partial z} | _{z=6} = 0 } & { \mathrm{otherwise} }
\end{array} \; ,  \\
\frac {\partial v_y } {\partial x } | _{x=6} =0 \; , \qquad \frac {\partial v_y } {\partial x }  | _{x=0} = 0 \; , \qquad \frac {\partial v_y } {\partial z }  | _{z=0}  = 0 \; .
\end{eqnarray*}
Tests show that the central behaviour is unaffected by these choices. The other boundary conditions follow from the remaining equations and the solenodial condition, $\nabla \cdot {\mathbf{B} _1} =0$. Note that we have used a slightly different inital pulse to those in the fast wave investigation. This is because the field lines (see Figure \ref{figureone}) leave the box and we know that the Alfv\'en wave follows the field lines. Hence, we made this choice of inital pulse in case reflections from the side boundaries influenced the subsequent evolution.

We found that the linear Alfv\'en wave travels down from the top boundary and begins to spread out, following the field lines. As the wave approaches the lower boundary (the separatrix), it thins but keeps its original ampitude. The wave eventually accumulates very near the separatrix; defined by the $x$ axis. This can be seen in Figure \ref{figureten}.

\subsection{Analytical results}\label{sec:4.1}

The Alfv\'en equation we have to solve takes the form:
\begin{eqnarray*}
\qquad \frac {\partial ^2 v_y } {\partial t^2 } = \left( x \frac {\partial }{\partial x} - z \frac {\partial }{\partial z } \right) ^2 v_y \; ,
\end{eqnarray*}
and this can be solved using the method of characteristics. Let $\frac {\partial }{\partial s} =  \left( x \frac {\partial }{\partial x} - z \frac {\partial }{\partial z } \right) $ and comparing the original equation with $\frac {\partial v_y}{\partial s} =   \frac {\partial x}{\partial s} \frac {\partial v_y}{\partial x} +  \frac {\partial z}{\partial s} \frac {\partial v_y}{\partial z }$ leads to:
\begin{eqnarray}
\qquad x = x_0 e^s \; , \quad z = z_0 e^{-s} \; \label{exponential}, 
\end{eqnarray}
where $x_0$ and $z_0$ are the starting positions of our characteristics. In our simulation, $z_0=6$. Thus, our characteristic equation, $ \frac {\partial ^2 v_y } {\partial t^2 } =\frac {\partial ^2 v_y } {\partial s^2 }$ can be solved with a D'Alembert solution such that: 
\begin{eqnarray}
\qquad v_y = \mathcal{F} \left( x_0 \right)  \mathcal{G} \left( t - s  \right) \;.
\end{eqnarray}
In order to compare these analytical results with the numerical results above, we substitute the same initial conditions into the D'Alembert solution, i.e. $\mathcal{F} \left( x_0 \right) = 1+\cos {\left(\frac { \pi x_0}  {3} \right)}$ and $\mathcal{G} \left( t  \right) = \sin {\left( \omega t \right)}$ to get the analytical solution for $v_y$, namely:
\begin{eqnarray}
\qquad v_y (x,z,t) = \left[ 1+\cos {\left( \frac {\pi x z } {18} \right)} \right] \sin { \omega \left( t + \log { \frac{z}{6} } \right) } \nonumber\\
 \mathrm{for} \; \left\{ \begin{array}{c} {0 \leq t + \log { \frac {z}{6} } \leq {\frac {\pi}{\omega}}} \\ {0 \leq  \frac {\pi x z } {6} \leq 3} \end{array} \right.  \; .
\end{eqnarray}
The agreement between the analytical and numerical results is excellent (the contours essentially lie on top of each other), even though the analytical solution does not satisfy the numerical boundary conditions.

Furthermore, we use our analytical solution to calculate $b_y$, $j_x$ and $j_z$:
\begin{eqnarray}
\qquad b_y &=&  \; -  \; \left[ 1+\cos {\left( \frac {\pi x z } {18} \right)} \right] \sin { \omega \left( t + \log { \frac {z}{6} } \right) } \; , \\
j_x  &=& {\frac {2 \pi}{z}\left[ 1+\cos {\left( \frac {\pi x z } {18} \right)} \right] \cos { \omega \left( t + \log { \frac{z}{6} } \right) } }
\nonumber\\
 && \qquad { -  \frac {\pi x }{18}\sin {\left( \frac {\pi x z } {18} \right) }\sin { \omega \left( t + \log { \frac{z}{6} } \right) }}  \; \label{jx} , \\
j_z  &=&  \frac {\pi z }{18} \sin { \left( \frac {\pi x z } {18}  \right) } \sin { \omega \left( t + \log { \frac{z}{6} } \right) } \; \label{jz} \\
&& \qquad \qquad \mathrm{all~for} \; \left\{ \begin{array}{c} {0 \leq { t + \log { \frac {z}{6} } } \leq \frac {\pi}{\omega} } \\ {0 \leq  {\frac {\pi x z } {6} } \leq 3} \end{array} \right. \; \nonumber .
\end{eqnarray}

\subsection{Current}\label{sec:4.2}

\begin{figure*}[b]
\begin{center}
\includegraphics[width=2.4in]{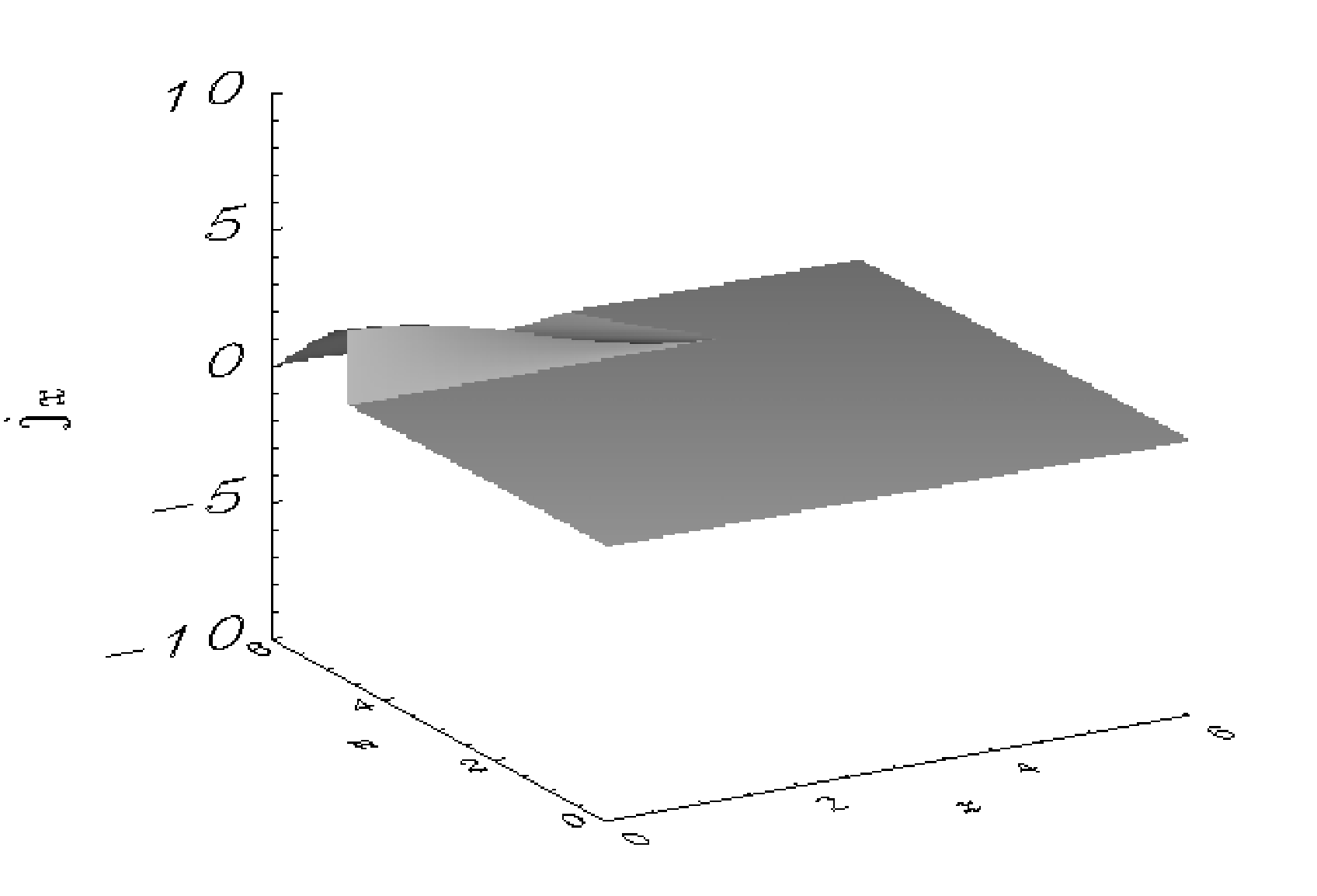}
\hspace{0.15in}
\includegraphics[width=2.4in]{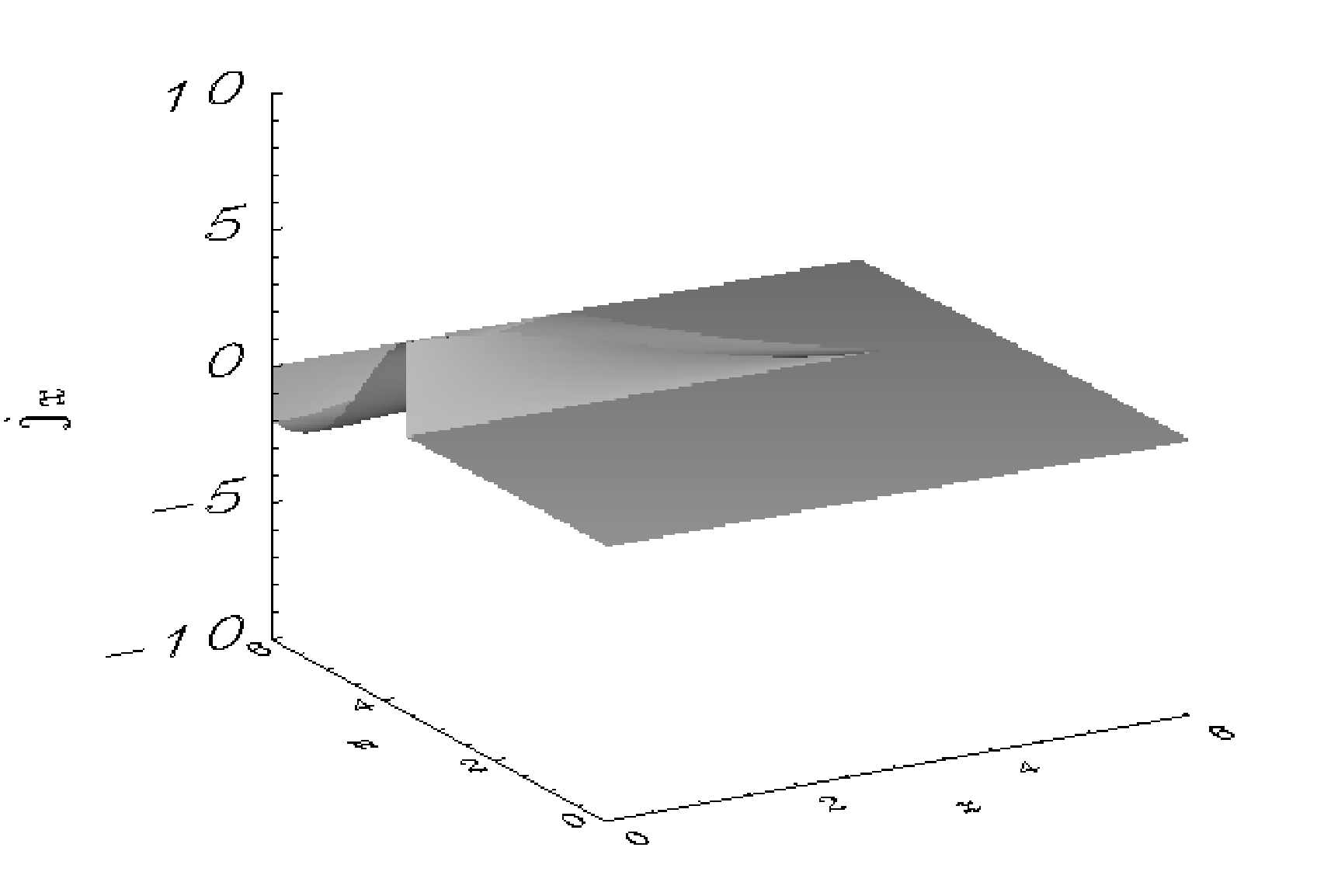}\\
\includegraphics[width=2.4in]{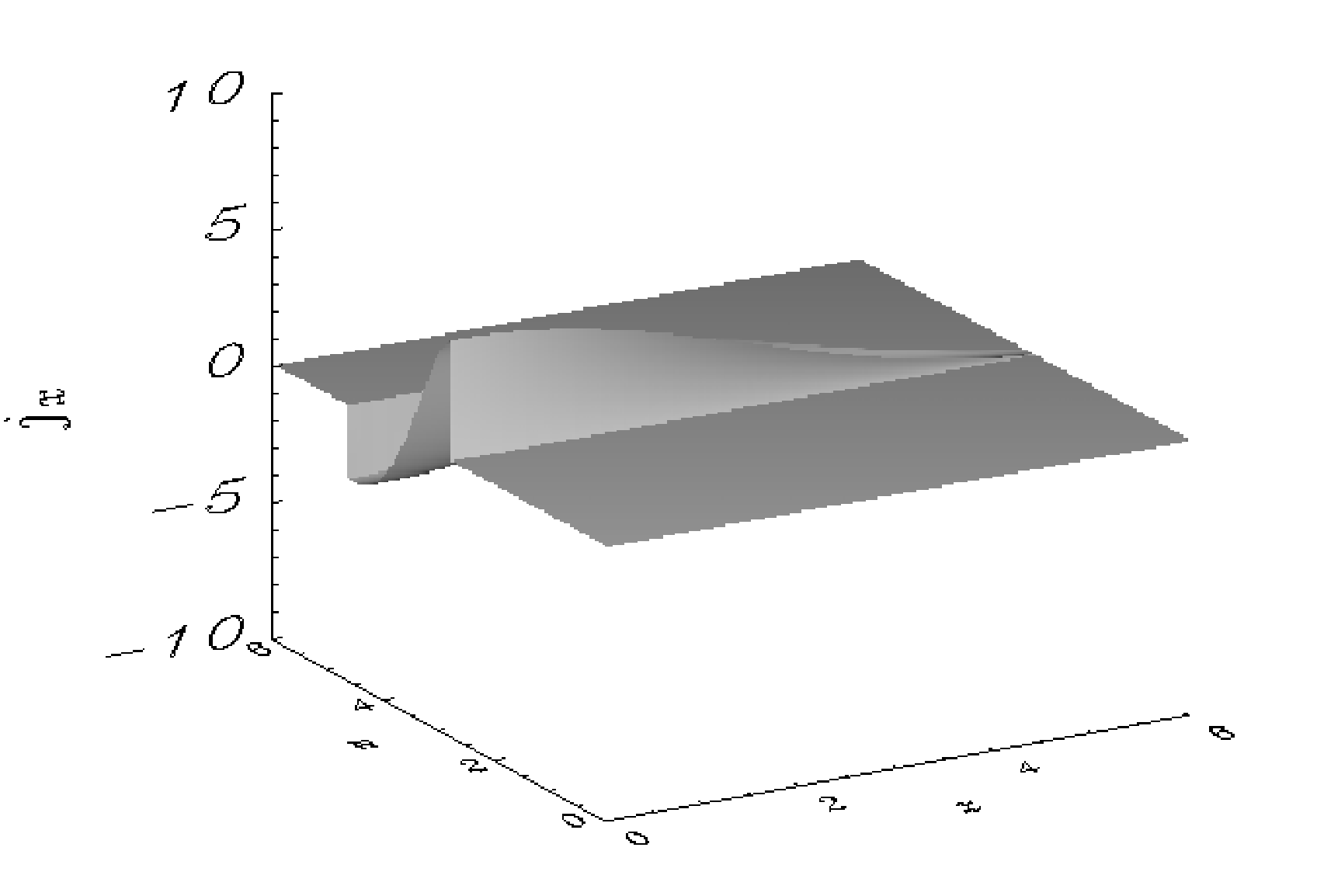}
\hspace{0.15in}
\includegraphics[width=2.4in]{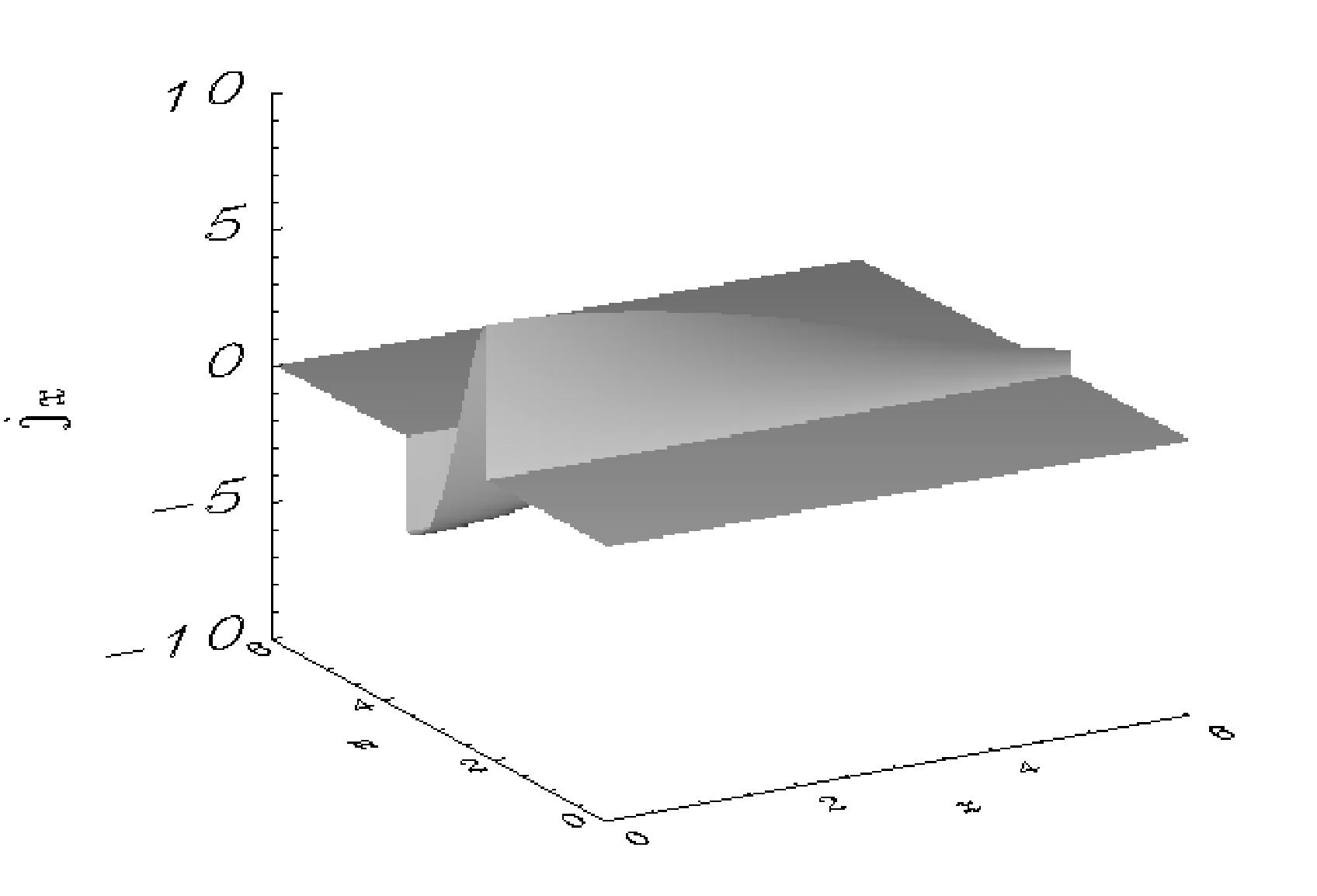}\\
\includegraphics[width=2.4in]{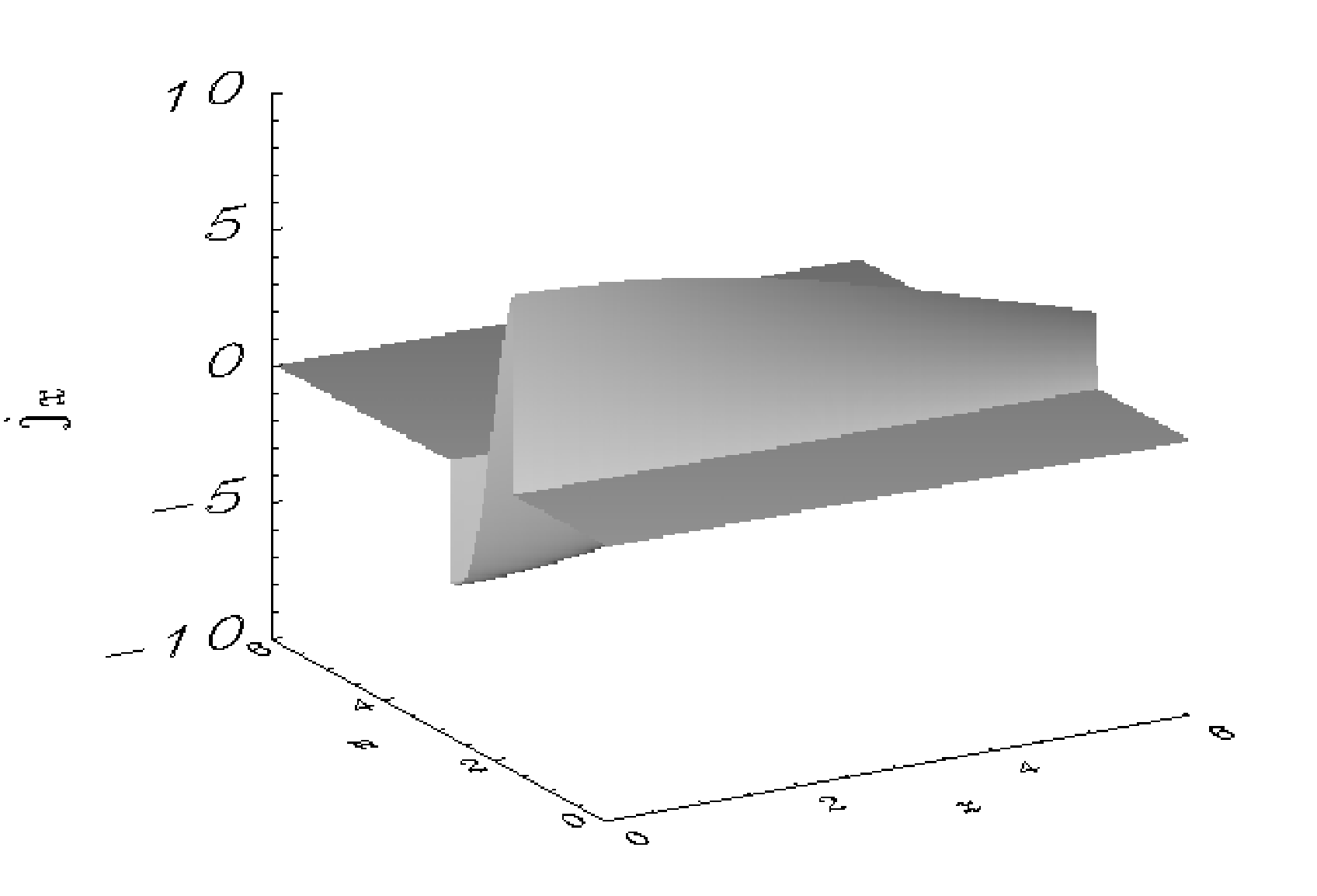}
\hspace{0.15in}
\includegraphics[width=2.4in]{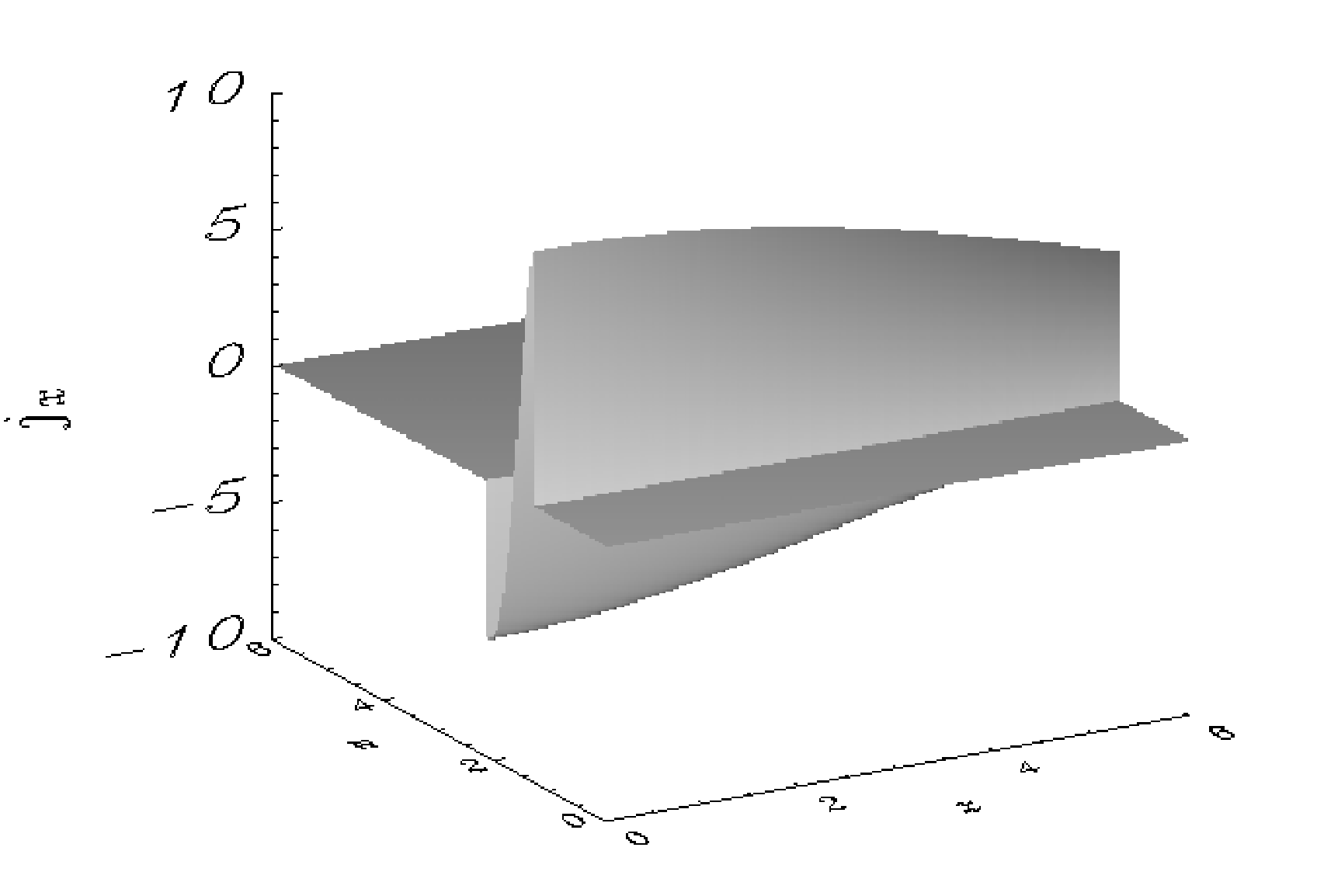}
\caption{Surfaces of $j_x$ at times $(a)$ $t$=0.25, $(b)$ $t$=0.5, $(c)$ $t$=0.75, $(d)$ $t=$1.0, $(e)$ $t$=1.25 and $(f)$ $t$=1.5, labelling from top left to bottom right.}
\label{figureeleven}
\end{center}
\end{figure*}

\begin{figure*}[b]
\begin{center}
\includegraphics[width=2.4in]{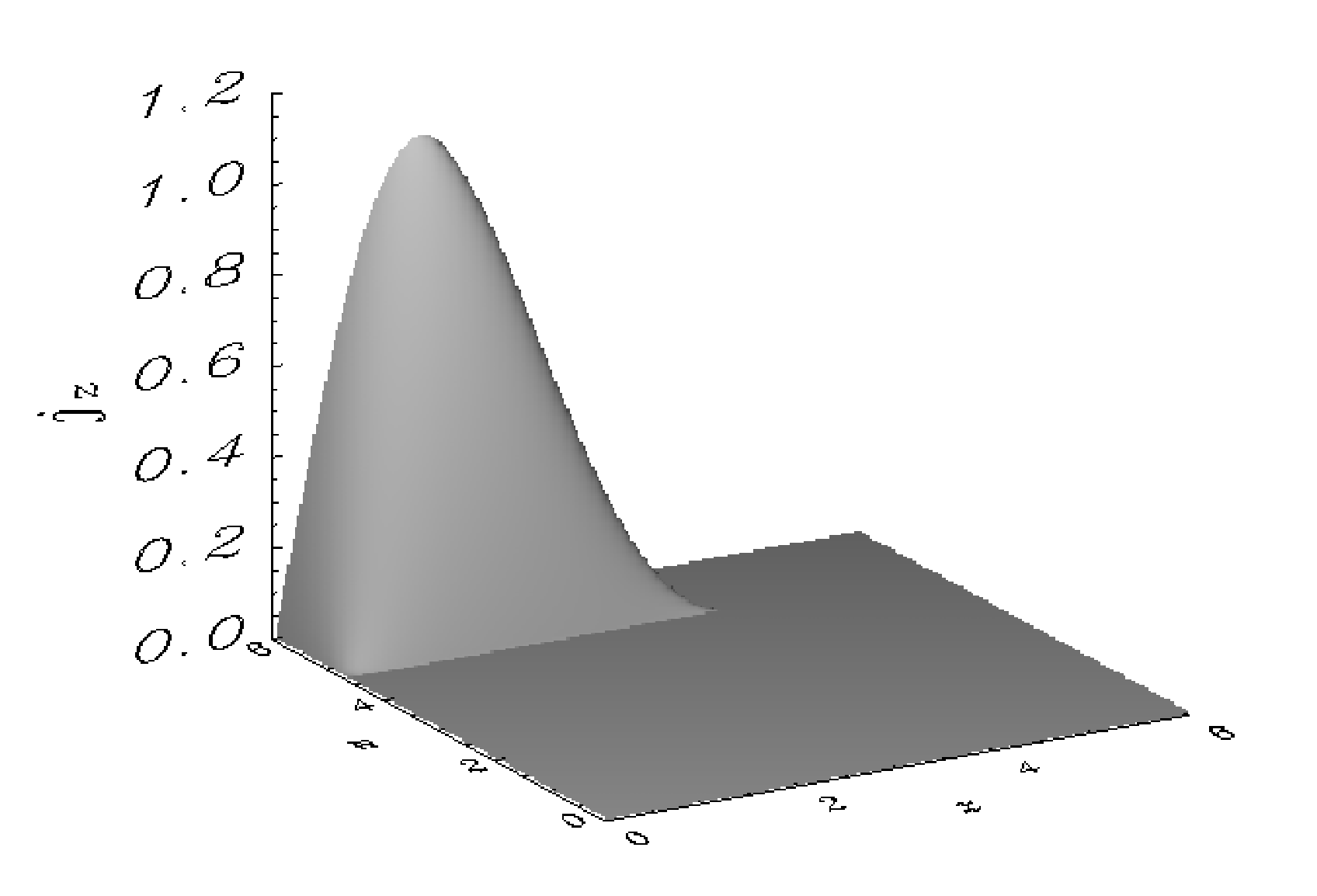}
\hspace{0.15in}
\includegraphics[width=2.4in]{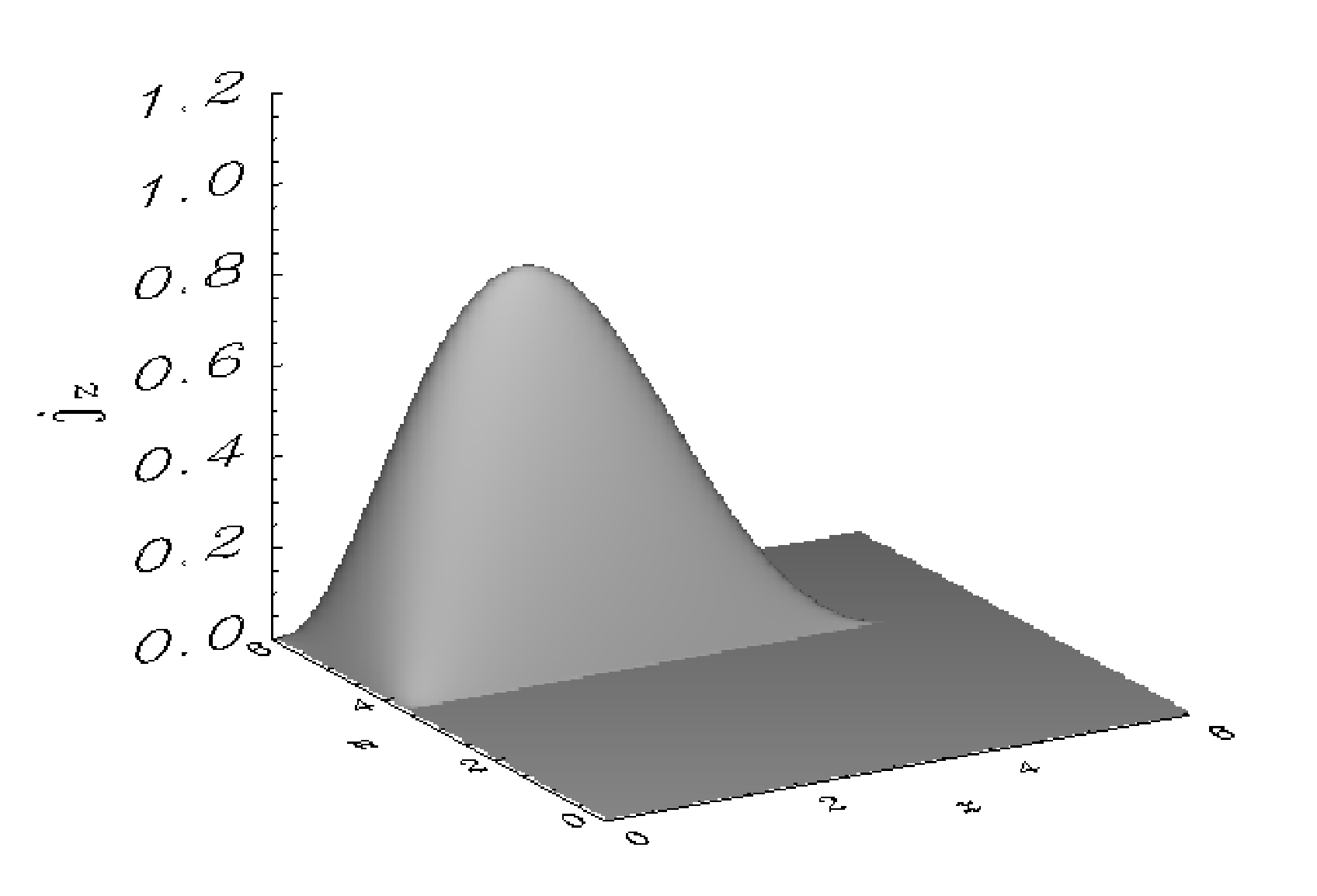}\\
\includegraphics[width=2.4in]{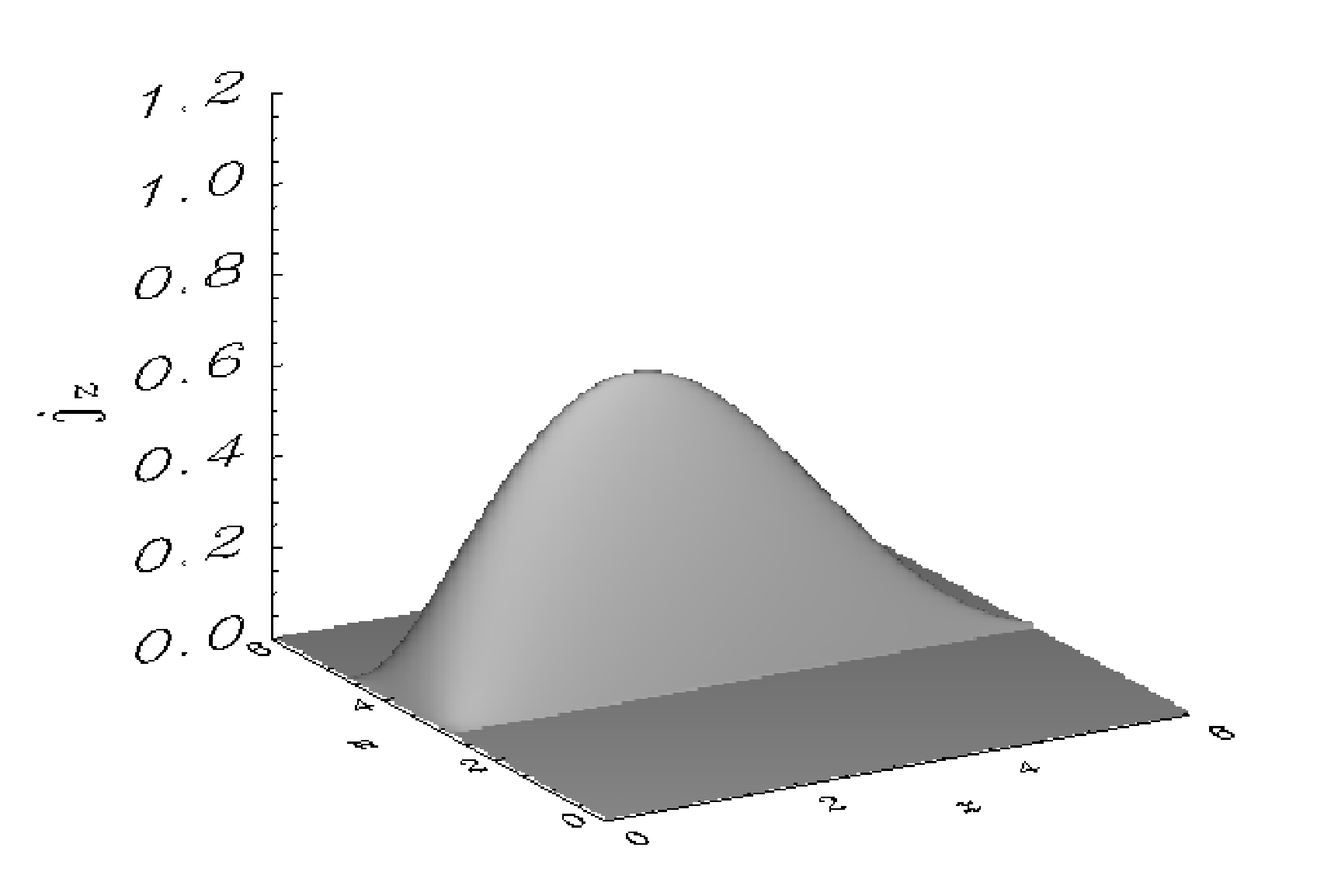}
\hspace{0.15in}
\includegraphics[width=2.4in]{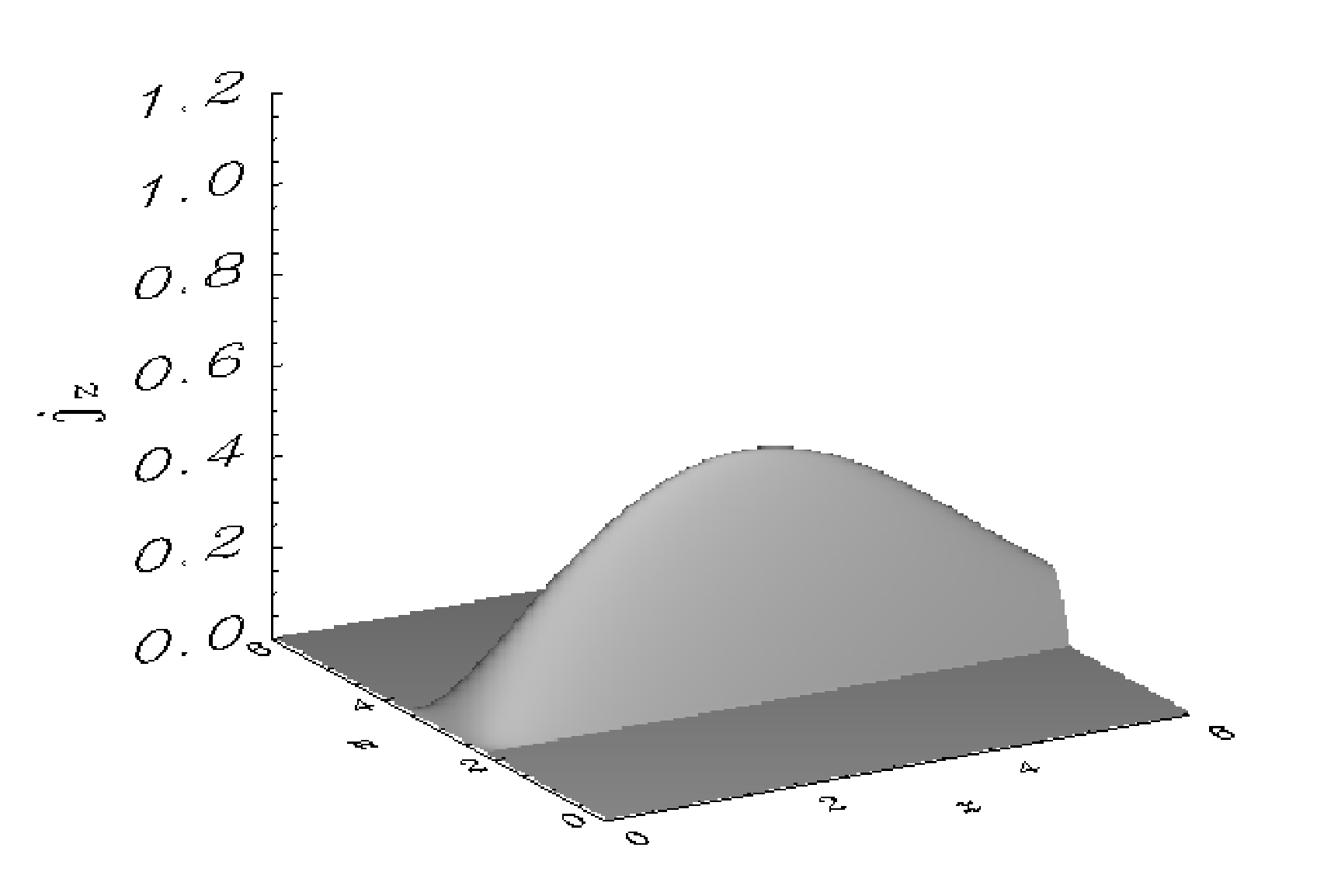}\\
\includegraphics[width=2.4in]{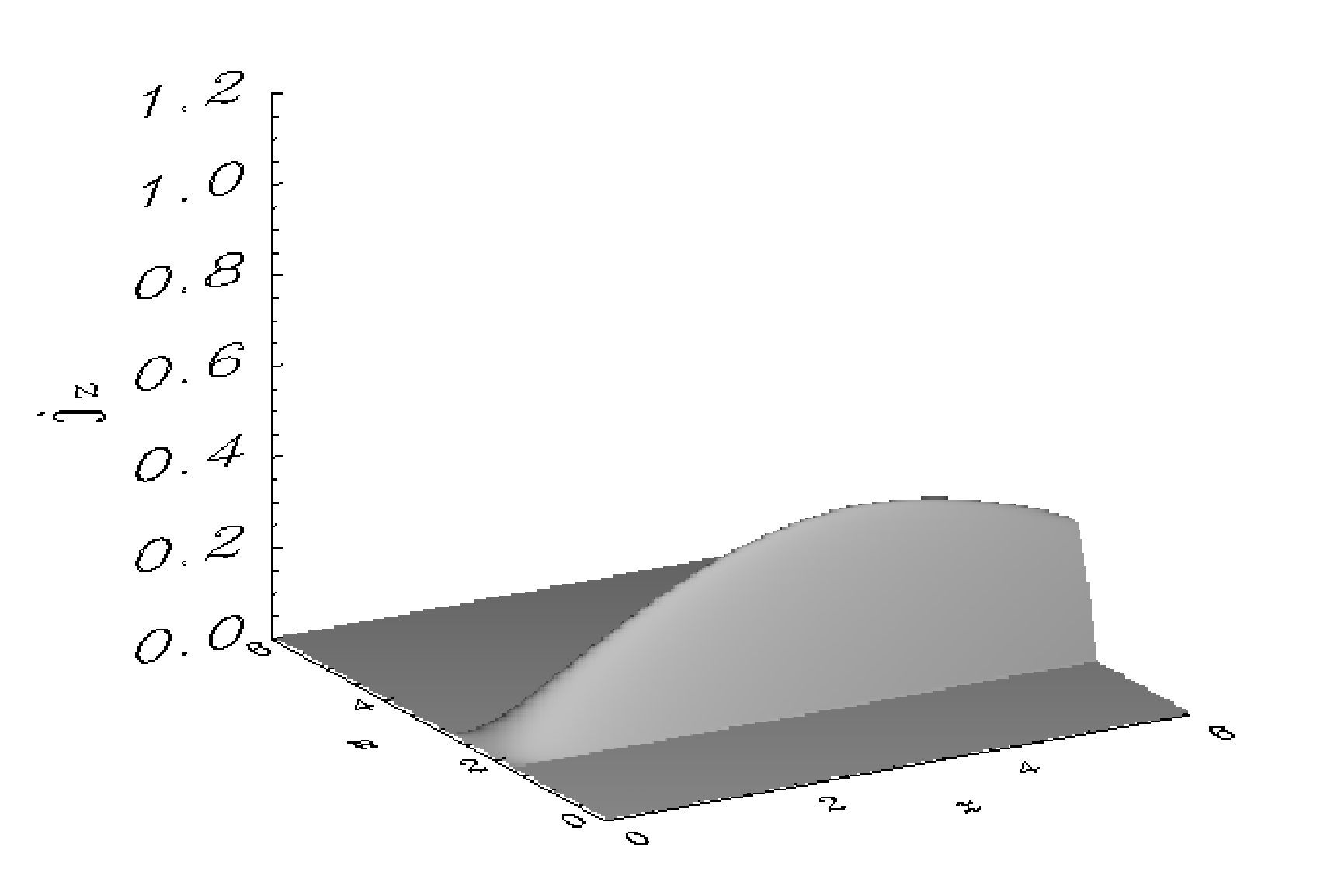}
\hspace{0.15in}
\includegraphics[width=2.4in]{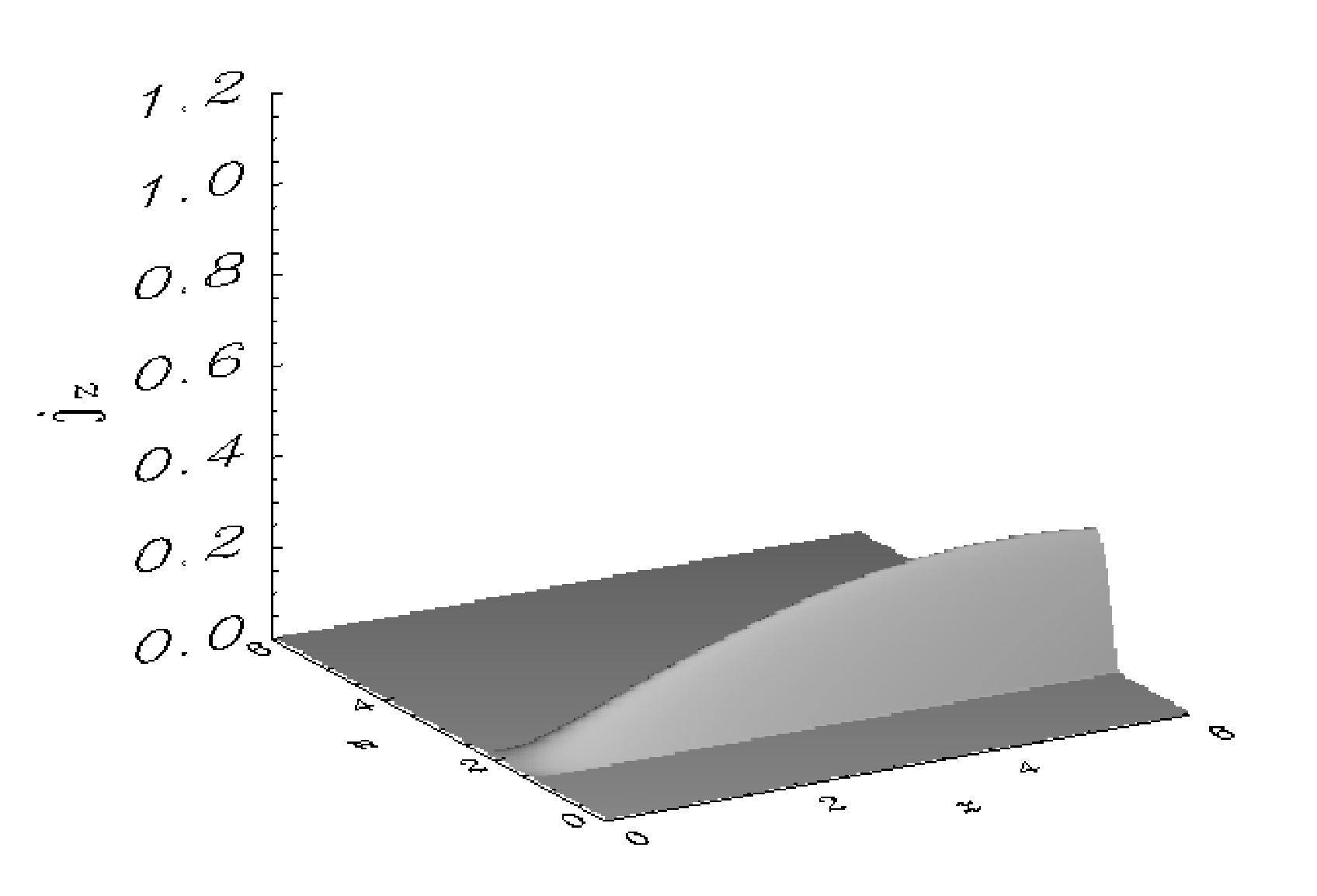}
\caption{Surfaces of $j_z$ at times $(a)$ $t$=0.25, $(b)$ $t$=0.5, $(c)$ $t$=0.75, $(d)$ $t=$1.0, $(e)$ $t$=1.25 and $(f)$ $t$=1.5, labelling from top left to bottom right.}
\label{figuretwelve}
\end{center}
\end{figure*}

As in the fast wave case, we have a spatially varying perturbed magentic field and so current is forming, given by $\frac {1}{\mu} \left( \nabla \times {\mathbf{B}} _1 \right) = (j_x,0,j_z)$. In the Alfv\'en case only $j_x = - \frac {1}{\mu} \frac {\partial b_y }{\partial z}$ and $j_z = \frac {1}{\mu} \frac {\partial b_y }{\partial x}$ are present. The evolution of the current can be seen in Figures \ref{figureeleven} and \ref{figuretwelve}.

\begin{figure*}
\begin{center}
\includegraphics[width=2.4in]{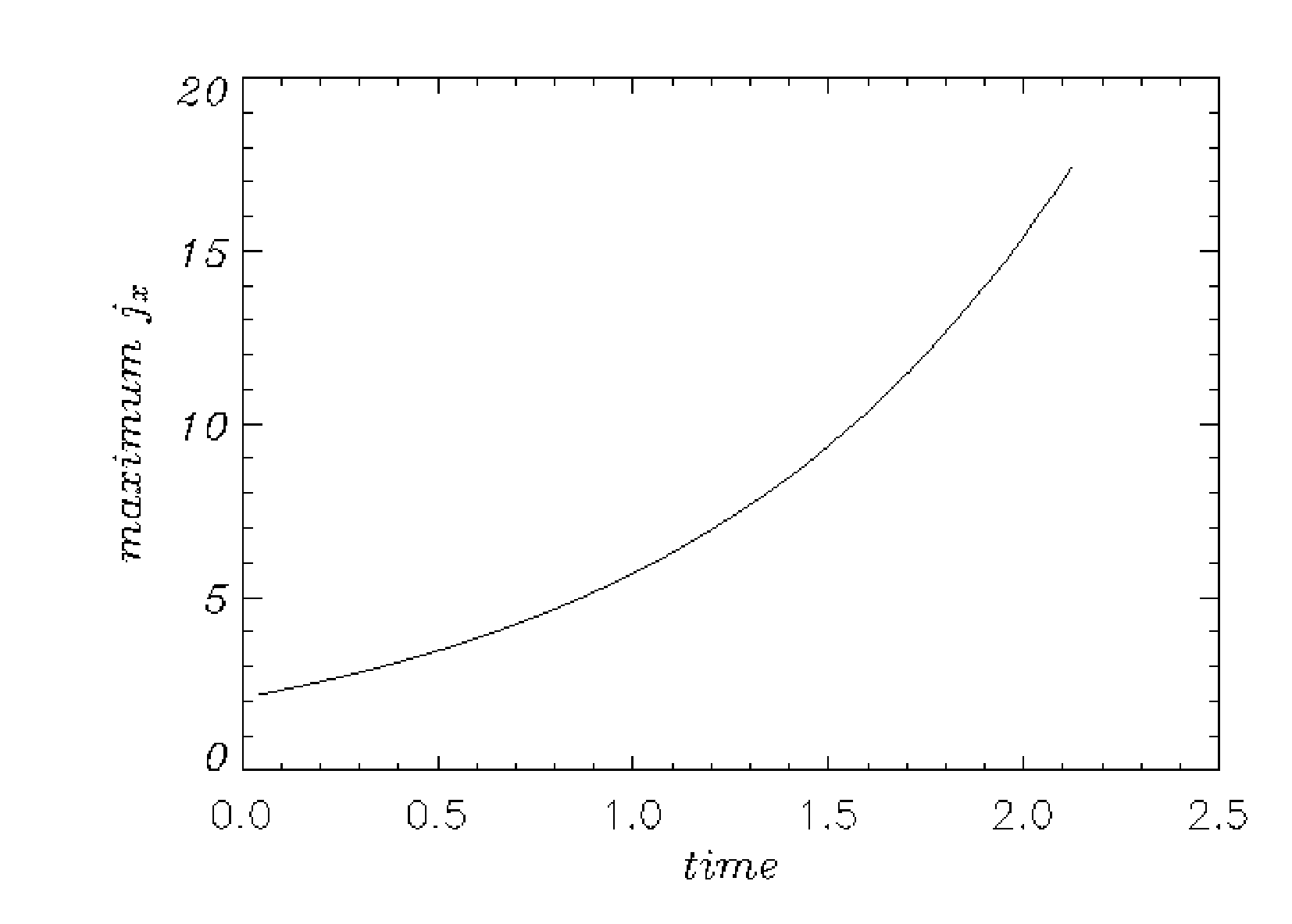}
\hspace{0.4in}
\includegraphics[width=2.4in]{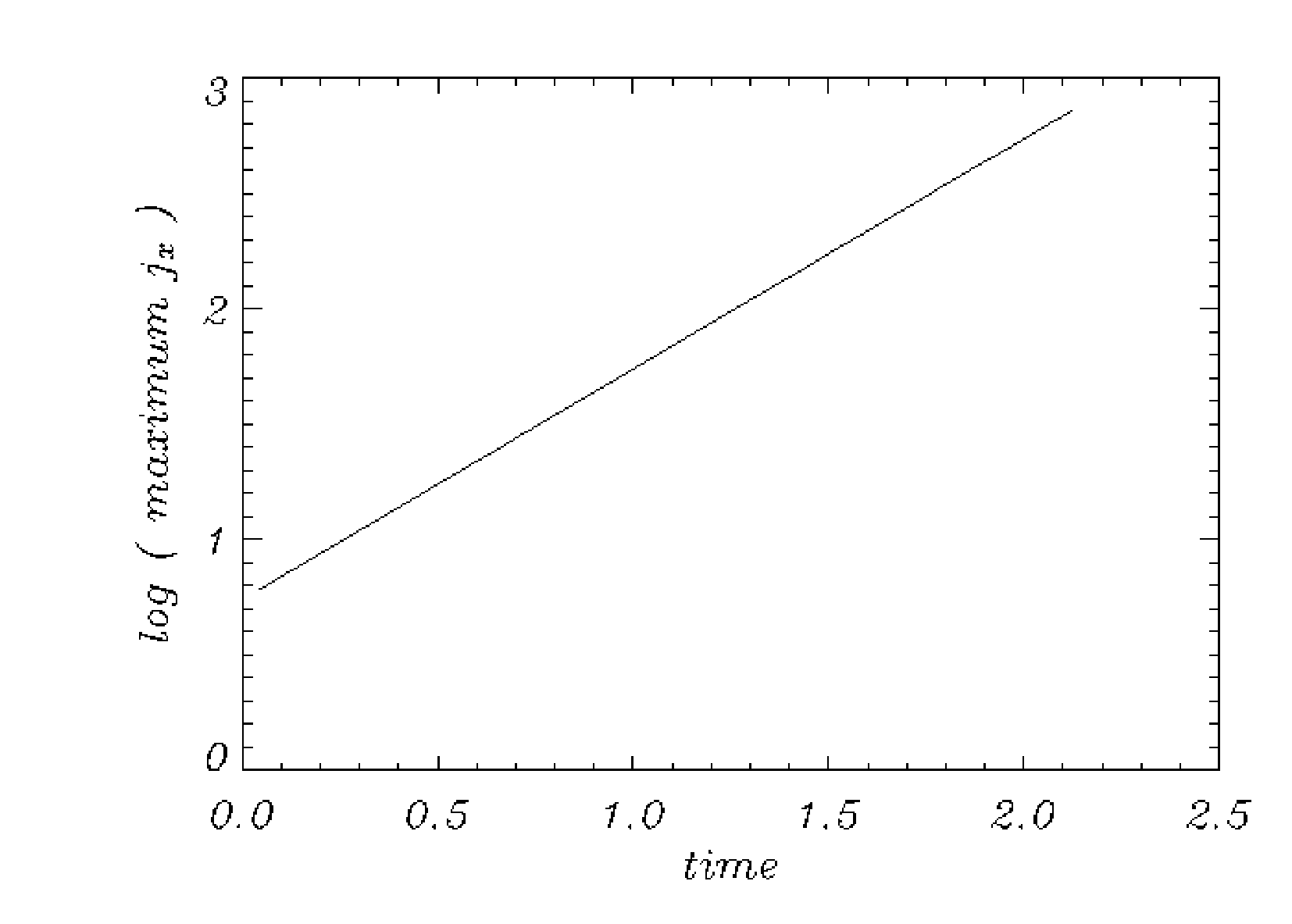}
\caption{ Maximum $j_x$ against time elapsed (left), log ( maximum $j_x$) against time elapsed (right). The slope of the line is $+1.0$.}
\label{jxgrowth}
\end{center}
\end{figure*}

From Figure \ref{figureeleven}, we see that $j_x$ spreads out along the field lines, accumulating along the separatrix; $z=0$. $j_x$ takes a discontinuous form;  this is due to our choice of initial conditions and this is confirmed by the analytical solution. $j_x$ also increases in time. From equation {\ref{jx}}, we see that $j_x$ grows like $\frac {1}{z}$ and thus, in accordance with equation {\ref{exponential}}, this means $j_x$ grows like $e^s$, i.e. grows as $e^t$ (since $s=t + \textrm{constant}$ in the D'Alembert solution). The behaviour of the maximum of $j_x$ with time can be seen in Figure \ref{jxgrowth}; the slope of the line is $+1.0$ in agreement with our analytical solution.

\begin{figure*}
\begin{center}
\includegraphics[width=2.4in]{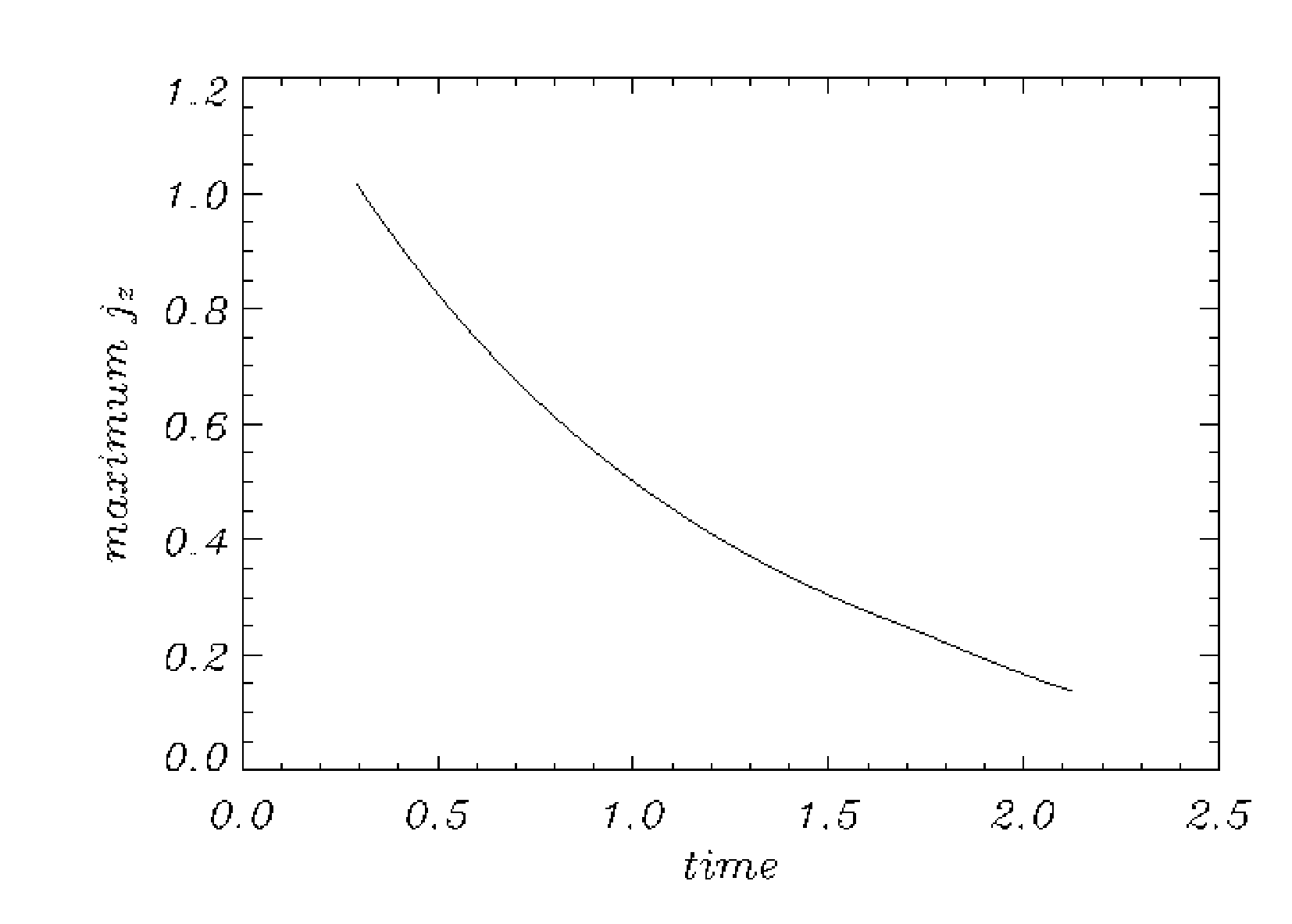}
\hspace{0.4in}
\includegraphics[width=2.4in]{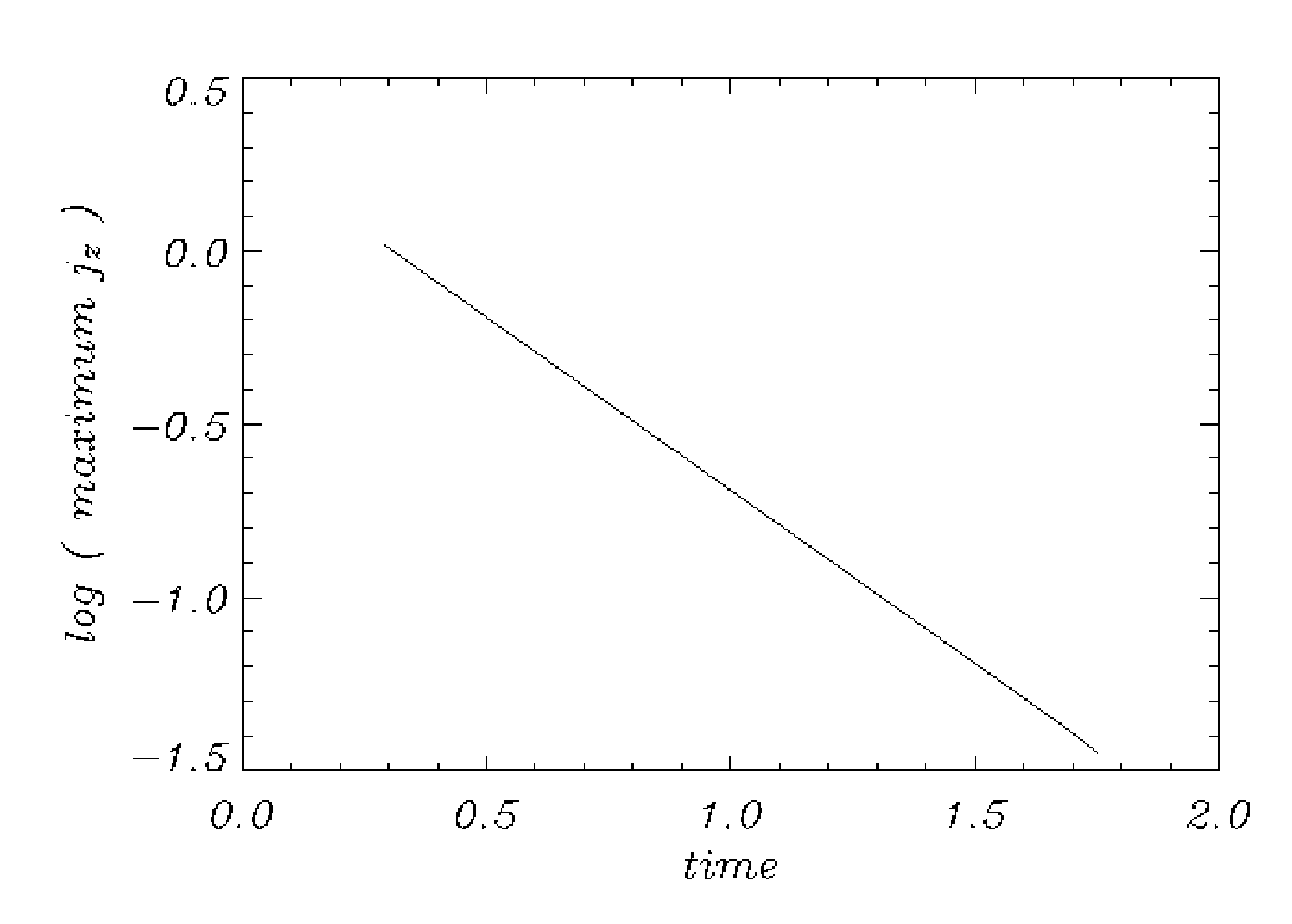}
\caption{ Maximum $j_z$ against time elapsed (left), log ( maximum $j_z$) against time elapsed (right). The slope of the line is $-1.0$.}
\label{jzgrowth}
\end{center}
\end{figure*}

The behaviour of $j_z$ can be seen in Figure \ref{figuretwelve}. $j_z$ takes the spatial form of the inital pulse and, in the same way as $v_y$, spreads out along the fieldlines. It decays in amplitude as it approaches the separatrix. The analytical solution, equation \ref{jz}, shows that it behaves like $z$ (and $z$ is going to zero). Thus, according to equation {\ref{exponential}}, $j_z$ decays as $e^{-t}$. The behaviour of the maximum of $j_z$ with time can be seen in Figure \ref{jzgrowth}. In this case, the slope of the line is $-1.0$ in agreement (again) with our analytical solution.

Hence, the Alfv\'en wave causes current density to build up along the separatrix.

\section{Conclusions}\label{sec:5}

This paper describes the start of an investigation into the nature of MHD waves in the neighbourhood of null points. From the work explained above, it has been seen that when a fast magnetoacoustic wave propagates near a magnetic X-type neutral point, the wave wraps itself around the null point due to refraction (at least in two dimensions). It has also been seen that this behaviour causes a large current density to accumulate at the null and simulations have shown that this build up is exponential in time, although the exponential growth in this linear simulation will be modified by non-linearities. We also note that for the set of disturbances investigated here, there is no evidence of the X-point collapsing; rather, the current density seems to form a spike. However, it is clear that the refraction of the wave focusses the energy of the incident wave towards the null point. As seen from both the numerical work and analytical approximations, the wave continues to wrap around the null point, again and again. The physical significance of this is that any fast magnetoacoustic disturbance in the neighbourhood of a neutral point will be drawn towards the region of zero magnetic field strength and focus all of its energy at this point. Hence, this is where the build up of current will occur and energy will be dissipated. Experiments are being carried out to extend the analysis to multiple null points and three dimensions but, if the results transfer, then null points should effectively trap and dissipate the energy contained in fast magnetoacoustic waves. Therefore, wave heating will naturally occur at coronal null points.

The numerical experiments and analytical work described above were all conducted using the ideal MHD equations. However, we can make some comments about the addition of resistivity into the model. For the fast magnetoacoustic wave, all the current density accumulates at the null point and appears to form a null line. Hence, no matter how small the value of the resistivity is, if we include the dissipative term, then eventually the $\eta \nabla ^2 {\mathbf{B}} _1$ term in equation (\ref{eq:2.5}) will become non-negligible and dissipation will become important. In addition, since $\nabla ^2 {\mathbf{B}} _1$ grows exponentially in time, the diffusion terms become important in a time that depends on $\log{\eta}$; as found by \cite{CraigWatson1992} and \cite{CraigMcClymont1993}. This means that linear wave dissipation will be very efficient. Thus, we deduce that null points will be the locations of wave energy deposition and preferential heating.

In the case of the Alfv\'en wave, the results show that the wave propagates along the field lines, accumulating on the separatrix and hence, due to symmetry, along the separatrices. The wave also thins and stretches along the separatrices. The current $j_x$ increases and accumulates along the separatrix, whilst $j_z$ decays away. This is seen in both the analytical and numerical work.

Now consider the effect of including resistivity. Considering $\nabla ^2 \mathbf{B} {_1}$, this can be split into $\frac {\partial}{\partial x} \frac {\partial b_y}{\partial x}+\frac {\partial}{\partial z} \frac {\partial b_y}{\partial z}$$= \frac {\partial j_z }{\partial x} + \frac {\partial j_x }{\partial z}$. As shown in the appendix, $  \frac {\partial j_z }{\partial x}$ decays away (exponentially) but $  \frac {\partial j_x }{\partial z}$  increases (exponentially). We have also seen that the current accumulates along the separatrices. Hence, eventually (due to its exponential increase), the resistive term, $\eta \frac {\partial ^2 b_y }{\partial z^2} = \eta \frac {\partial j_x }{\partial z}$, will become important, no matter how small the value of $\eta$. Hence, all the Alfv\'en wave energy will be dissipated along the separatrices. This is a different behaviour to that of the fast wave in the sense that the two wave types deposit all their wave energy at different areas (along the separatrices as opposed to the null point), although the phenomenon of depositing wave energy in a specific area is common to both.

One of the next steps to be taken is to extend the model to 3D. Galsgaard \emph{et al.} (2003) performed a similar analysis of MHD wave propagation for a particular disturbance of a symmetric 3D null. Their findings show a close link to our own and support the possibility of transferring our results to 3D.

Another of the steps to be taken is to investigate the effect of pressure on the system. This has been investigated by \cite{CraigWatson1992} with cylindrical symmetry and they find that the rapid current growth is halted. The most obvious effect of including a finite $\beta$ is the introduction of slow magnetoacoustic waves. Fast waves can now pass through the null point (as we would now have a non-zero sound speed) and thus perhaps take wave energy away from that area. The exact nature depends on the choice of boundary conditions. In a simple manner, if the $\beta$ has a value of $\beta_0 \ll 1$ at, say, $x^2+z^2=1$, the finite pressure effects will become important once $x^2+z^2=2 \beta$. From our WKB solution, with $z_0=6$ and $x_0=0$, the gas pressure will become non-negligible when $t \simeq -\frac {1}{2} \log {\beta _0} $. This is not true for Alfv\'en waves since they are unaffected by the finite $\beta$.

Finally, the validity of the linearisation is questionable once the perturbed velocity becomes comparable to the magnitude of the local Alfv\'en speed. In a similar argument to the finite $\beta$ case, the linearisation is valid until $t \simeq - \log {M_A} $, where $M_A$ is the initial Alfv\'en Mach number. Once the Alfv\'en Mach number exceeds unity, the fast wave is likely to shock and the wave energy will still be dissipated but this time in the shock. A more rigorous analysis requires detailed numerical simulations of the non-linear MHD equations and will be investigated in the future.

%---------------------------------------------------

\section*{Acknowledgements}

J. A. McLaughlin acknowledges financial assistance from the Particle Physics and Astronomy Research Council (PPARC) and the helpful comments of the referee.

\section*{Appendix}\label{sec:appendixA}

\begin{figure*}[tb]
\begin{center}
\includegraphics[width=2.4in]{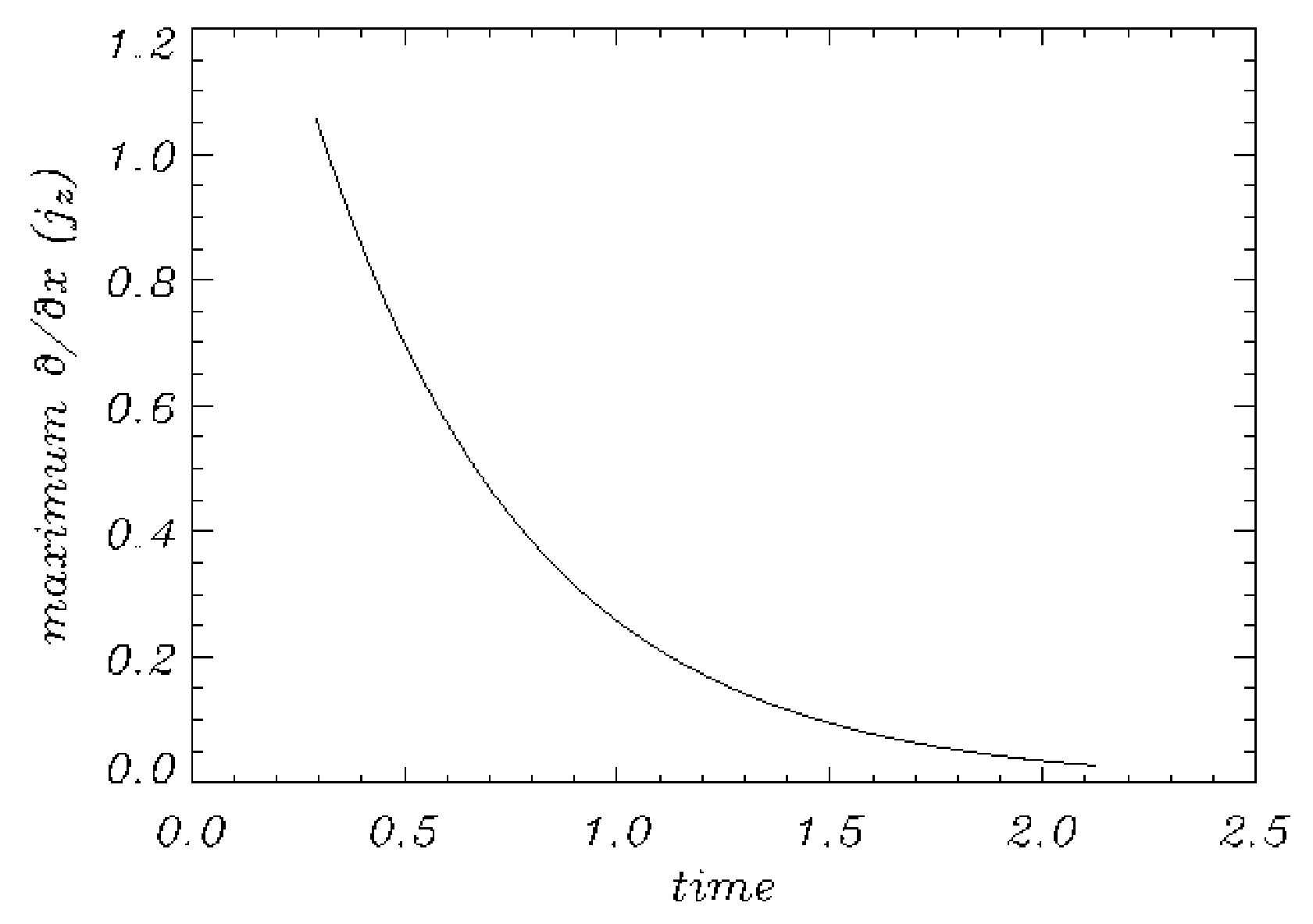}
\hspace{0.2in}
\includegraphics[width=2.4in]{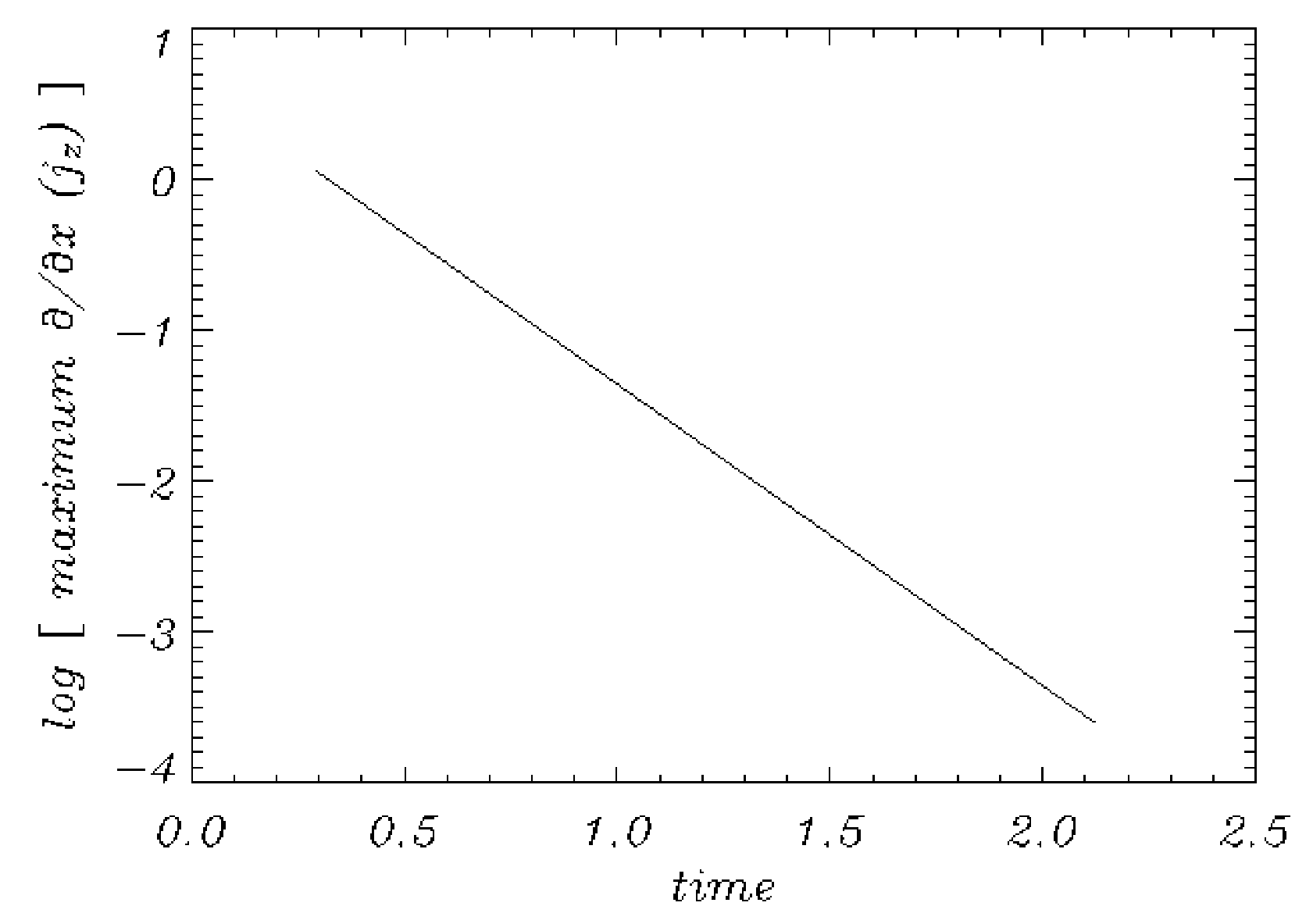}
\caption{Plot of maximum $\frac {\partial}{\partial x} j_z$ against time (left), plot of $\log \mathrm{maximum} {\frac {\partial}{\partial x} j_z }$ against time (right). The slope of the line is $-2$.} 
\label{figuregrowthjz}
\end{center}
\end{figure*}

\begin{figure*}[tb]
\begin{center}
\includegraphics[width=2.4in]{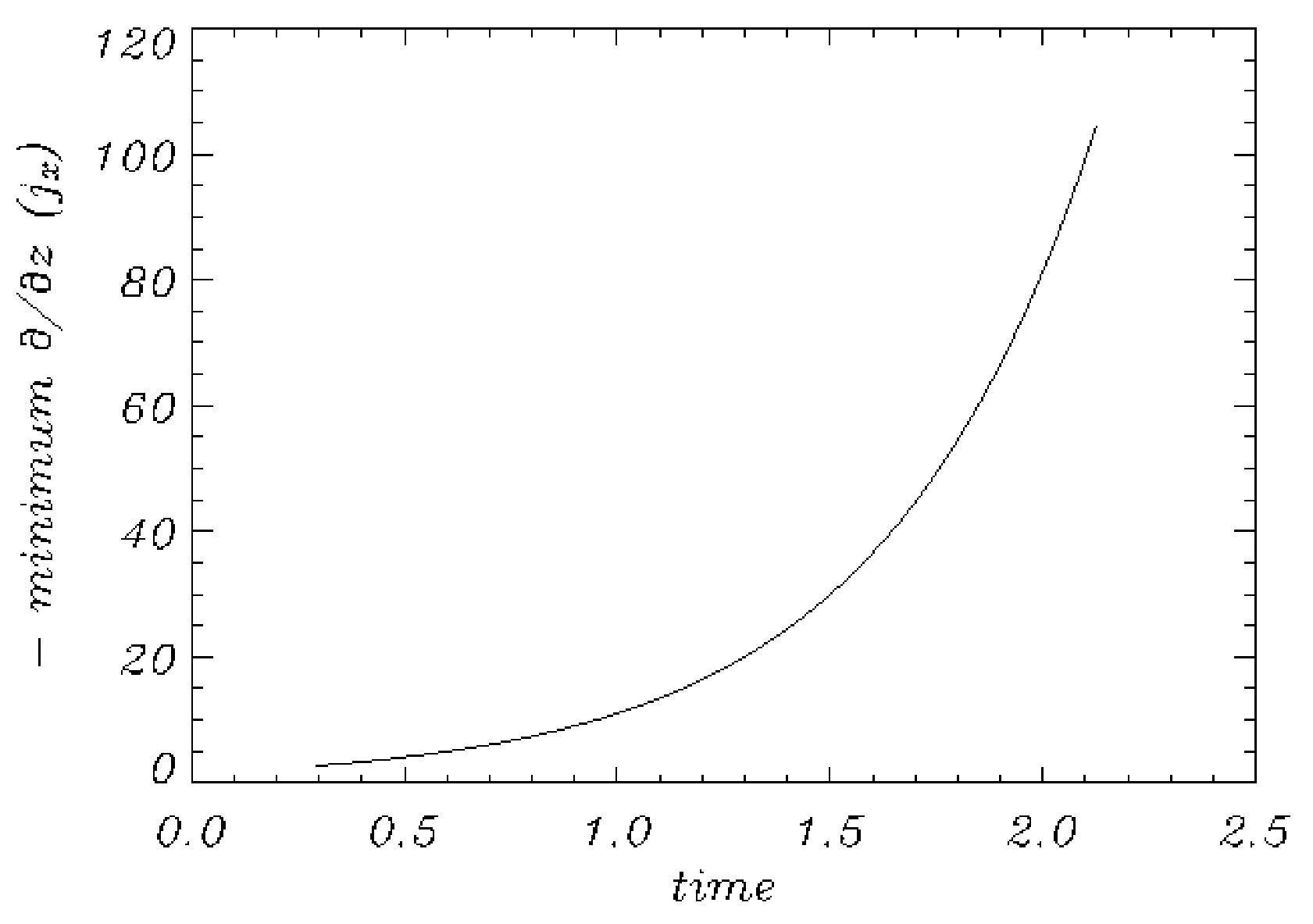}
\hspace{0.2in}
\includegraphics[width=2.4in]{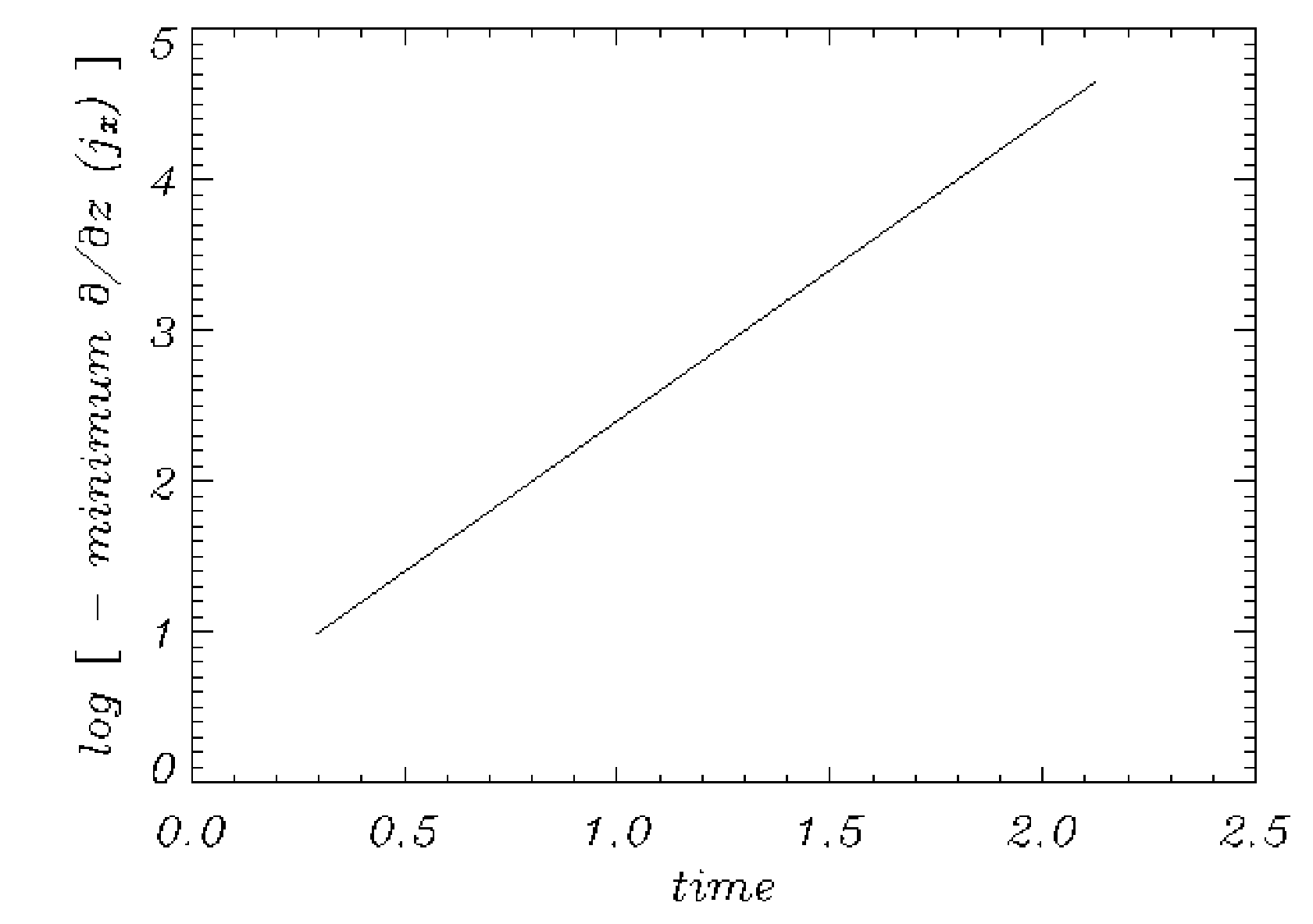}
\caption{Plot of minimum $- \frac {\partial}{\partial z} j_x$ against time (left), plot of $\log { - \mathrm{minimum} \frac {\partial}{\partial z} j_x }$ against time (right). The slope of the line is $+2$.}
\label{figuregrowthjx}
\end{center}
\end{figure*}

Consider the induction equation for the Alfv\'en wave:
\begin{eqnarray*}
\qquad \frac {\partial b_y}{\partial t} = \left(x \frac {\partial}{\partial x} - z \frac {\partial}{\partial z} \right) v_y + \eta  \frac {\partial}{\partial x} j_z - \eta  \frac {\partial}{\partial z} j_x
\end{eqnarray*}
In order for the resistive terms to become important, they have to become comparable to the convective term; $\left(x \frac {\partial}{\partial x} - z \frac {\partial}{\partial z} \right) v_y$. So we have to calculate the behaviour of $ \frac {\partial}{\partial x} j_z$ and the behaviour of $ \frac {\partial}{\partial z} j_x$ over time. From Figure \ref{figuregrowthjz}, we see that the maximum of $ \frac {\partial}{\partial x} j_z$ decays exponentially over time. From Figure \ref{figuregrowthjx}, we see that the maximum of $ - \frac {\partial}{\partial z} j_x$ increase exponentially over time. Hence, the resistive terms invloving $\frac {\partial}{\partial z} j_x$ will eventually become important, no matter how small the value of the resistivity, whereas the resistive terms invloving $ \frac {\partial}{\partial x} j_z$ will become negligible.

\end{document}